\documentclass[longauth]{aa}  

\usepackage{graphicx}
\usepackage{txfonts}	
\usepackage{amsmath}	
\usepackage{longfigure}
\usepackage{longtable}
\usepackage{subfig}
\usepackage[colorinlistoftodos]{todonotes}
\usepackage[export]{adjustbox}
\usepackage{makecell}
\usepackage{orcidlink}
\usepackage{placeins}

\usepackage[dvipsnames]{xcolor}
\hypersetup{colorlinks=true,
linkcolor=RoyalBlue,
citecolor=RoyalBlue,
filecolor=magenta,
urlcolor=RoyalBlue,
pdfpagemode=FullScreen}

\definecolor{Gold}{HTML}{F2AB27}
\definecolor{Silver}{HTML}{C0C0C0}
\definecolor{Bronze}{HTML}{B8422D}

\newcommand\encircle[1]{%
  \tikz[baseline=(X.base)] 
    \node (X) [draw, shape=circle, inner sep=0] {\strut #1};}

\newcommand{\gold}{\textcolor{Gold}{\bf \encircle{g}}}
\newcommand{\silver}{\textcolor{Silver}{\bf \encircle{s}}}
\newcommand{\bronze}{\textcolor{Bronze}{\bf \encircle{b}}}

\newcommand{\sigmalt}{$\sigma_{\rm lt}\,$}

\newcommand{\znirspecerrA}[2]{$z_{\rm NIRSpec} = #1 \pm #2$}

\newcommand{\onlyvalueA}[2]{$#1 \pm #2$}
\newcommand{\onlyvalueB}[3]{$#1^{+#2}_{-#3}$}
\newcommand{\Halpha}{H$\alpha\ $}

\newcommand{\csl}[1]{$\sigma/m_{\chi} < {#1}\, \mathrm{cm}^2~\mathrm{g}^{-1}$}

\def\cse{$\sigma/m_{\chi}\ $}

\defcitealias{Richard21}{R21}
\defcitealias{cha25}{C25}
\defcitealias{Benavides23}{B23}
\defcitealias{Sarrouh25}{Sarrouh \& Asada et al. (2026)}

\begin{document}

   \title{Mapping dark matter in the Bullet Cluster using JWST imaging and spectroscopy}

\author{G. Rihtar\v{s}i\v{c}
          \inst{1}
          \fnmsep\thanks{gregor.rihtarsic@fmf.uni-lj.si}
          \and
          M. Brada\v{c}
          \inst{1,2}
          \and
         G. Desprez
         \inst{3}
         \and
          A. Harshan
         \inst{4,5}
         \and
         N. S. Martis
          \inst{1}
          \and
          C. J. Willott
          \inst{6}
          \and 
          Y. Asada
        \inst{7,8,9}
        \and 
         G. T. E. Sarrouh
         \inst{7}
          \and
          C. Cornil-Ba\"{\i}otto
          \inst{10}
          \and
         A. Biviano
          \inst{11,12}
        \and 
        D. Clowe
        \inst{13}
        \and
        A. H. Gonzalez
         \inst{14}
         \and
        C. Jones
        \inst{15}
         \and
        J. Judež
         \inst{1}
         \and
         S. Y. Kim
         \inst{16}
         \and
         B. C. Lemaux
         \inst{17,2}
         \and
        M. Lombardi
         \inst{18,19}
         \and
         D. Marchesini
         \inst{20}
         \and
         M. Markevitch
         \inst{21}
         \and
         V. Markov
         \inst{1}
         \and 
         G. Noirot
         \inst{22}
         \and 
         A. H. G. Peter
         \inst{23}
         \and
         S. W. Randall
         \inst{15}
         \and
         A. Robertson
         \inst{16}
         \and 
         M. Sawicki
         \inst{24}
         \and 
         R. Tripodi
         \inst{25}
          }

   \institute{Faculty of Mathematics and Physics, Jadranska ulica 19, SI-1000 Ljubljana, Slovenia
    \and
    Department of Physics and Astronomy, University of California Davis, 1 Shields Avenue, Davis, CA 95616, USA
    \and
    Kapteyn Astronomical Institute, University of Groningen, P.O. Box 800, 9700AV Groningen, The Netherlands
    \and
    Kavli Institute for Cosmology, University of Cambridge, Madingley Road, Cambridge, CB3 0HA, United Kingdom
    \and
    Cavendish Laboratory - Astrophysics Group, University of Cambridge, 19 JJ Thomson Avenue, Cambridge, CB3 0HE, United Kingdom
    \and
    National Research Council of Canada, Herzberg Astronomy \& Astrophysics Research Centre, 5071 West Saanich Road, Victoria, BC, V9E 2E7, Canada
    \and
    Department of Physics and Astronomy, York University, 4700 Keele St. Toronto, Ontario, M3J 1P3, Canada
    \and
    David A. Dunlap Department of Astronomy and Astrophysics, University of Toronto, 50 St. George Street, Toronto, Ontario, M5S 3H4, Canada
    \and 
    Dunlap Institute for Astronomy and Astrophysics, 50 St. George Street, Toronto, Ontario, M5S 3H4, Canada
    \and
    Instituto de Física y Astronomía, Universidad de Valparaíso, Gran Bretaña 1111, Playa Ancha, Valparaíso 2360102, Chile
    \and 
    INAF-Osservatorio Astronomico di Trieste, via G. B. Tiepolo 11, 34143 Trieste, Italy
   \and
   IFPU-Institute for Fundamental Physics of the Universe, via Beirut 2, 34014 Trieste, Italy
    \and 
    Department of Physics and Astronomy, Ohio University, 1 Ohio University, Athens, OH 45701, USA
    \and 
    Department of Astronomy, University of Florida, Bryant Space Science Center, Gainesville, FL 32611, USA
    \and
    Center for Astrophysics, Harvard \& Smithsonian, 60 Garden Street, Cambridge, MA 02138, USA
    \and 
    Carnegie Observatories, 813 Santa Barbara Street, Pasadena, CA 91101, USA
    \and
    Gemini Observatory, NSF NOIRLab, 670 N. A'ohoku Place, Hilo, Hawai'i, 96720, USA
    \and
    University of Milan, Department of Physics, via Celoria 16, I-20133 Milan, Italy 
    \and
    INAF – OAS, Osservatorio di Astrofisica e Scienza dello Spazio di Bologna, via Gobetti 93/3, I-40129 Bologna, Italy
    \and 
    Department of Physics \& Astronomy, Tufts University, Medford, MA 02155, USA
    \and 
    NASA/Goddard Space Flight Center, Greenbelt, MD 20771, USA
    \and 
    Space Telescope Science Institute, 3700 San Martin Drive, Baltimore, Maryland 21218, USA
     \and 
    Department of Physics, Department of Astronomy, and CCAPP, The Ohio State University, 191 W. Woodruff Ave., Columbus OH 43210, USA
    \and 
    Department of Astronomy and Physics and Institute for Computational Astrophysics, Saint Mary's University, 923 Robie Street, Halifax, Nova Scotia B3H 3C3, Canada
    \and 
    INAF - Osservatorio Astronomico di Roma, Via Frascati 33, Monte Porzio Catone, 00078, Italy
             }

   \date{Received February 02, 2026; accepted April 01, 2026 }

 
  \abstract
   {The Bullet Cluster (1E 0657–56), located at a redshift of 0.296, is among the best-known merging galaxy clusters and a key laboratory for dark matter studies. Although the mass distribution in the Bullet Cluster has been modelled using an increasing number of multiply imaged galaxies, only six systems had spectroscopic redshifts published prior to this work, which are essential for system confirmation and as lens model constraints.}
   {We present an updated gravitational lens model of the Bullet cluster, obtained by combining JWST NIRCam imaging and NIRSpec spectroscopy. Our lens model has been constrained on the basis of a catalogue of 135 secure multiple images from 27 background galaxies with spectroscopic redshifts, uniformly covering both subclusters and a wide redshift range of 0.9 -- 6.7. We also provide a catalogue of 199 multiple image candidates.}
   {We modelled the cluster with the parametric lens modelling code \texttt{Lenstool}. We incorporated several large-scale halos, cluster member galaxies, intracluster gas, and group-scale halos surrounding the cluster core, motivated by spectroscopic studies of cluster member kinematics.}
   {We describe the main cluster component with a complex, elongated double-peaked distribution, along with the subcluster, modelled using a single large-scale halo coinciding closely with the brightest cluster galaxy at a projected separation of $4_{-2}^{+3}$ kpc. The uncertainty of the displacement has been improved three-fold thanks to the addition of JWST systems. The addition of group-scale substructures, roughly following the two axes of cluster assembly, improves the fit to the multiple image positions and provides a physically motivated alternative to a constant shear component. Our lens model shows the closest agreement with previous studies in aperture mass profiles at $\sim60$ kpc from the brightest cluster galaxies (BCGs), but exhibits significant differences in the detailed mass distribution as a result of different lens-modelling strategies and adopted constraints. The differences are reflected in small, but spatially coherent deviations between the new spectroscopic redshifts and redshifts predicted by earlier lens models. }
   {}

                
\titlerunning{Mapping dark matter in the Bullet Cluster using JWST imaging and spectroscopy}
\authorrunning{G. Rihtar\v{s}i\v{c} et al.}
\maketitle
\section{Introduction}

The Bullet Cluster (1E 0657–56) at a redshift of 0.296 \citep[][]{1998ApJ...496L...5T} is among the most remarkable and widely studied examples of a galaxy cluster merger. It was discovered as an extended X-ray source \citep{1995ApJ...444..532T}, which first sparked interest due to its high temperatures \citep[][]{1998ApJ...496L...5T,2000ApJ...544..686L,2002ApJ...567L..27M}. The Chandra X-ray observations by \cite{2002ApJ...567L..27M} revealed a compact, ram-pressure-stripped subcluster (i.e. the Bullet) trailing behind the cluster galaxies and featuring a bow-shock, indicating a textbook example of a high-velocity merger ($>1000\ \mathrm{km\ s^{-1}}$, \citealt{2007MNRAS.380..911S,2002ApJ...567L..27M}) in a post-collision phase. Further spectroscopic observations of cluster galaxies confirmed the merger to be observed in the plane of the sky with $\sim600\  \mathrm{km\ s^{-1}}$ line-of-sight (LOS) velocity difference between the two subclusters \citep{2002A&A...386..816B}. Due to its extraordinary properties, the Bullet Cluster has been used as a laboratory to study the hydrodynamical properties of an interacting system and dark matter (DM). By measuring the displacement of the mass peak from the X-ray peak, it has provided one of the most direct pieces of evidence for the existence of DM \citep[][]{clowe06}. Furthermore, by comparing the distribution of DM with the collisionless galaxy component, the Bullet Cluster offers a unique window into DM self-interaction properties in the high-velocity regime of particle collisions. Throughout the years, it has sparked several studies aimed at constraining the DM self-interaction cross-section \cse \citep[][]{2004ApJ...606..819M,Randall08,Robertson17}. The lack of a DM–galaxy displacement allowed \cite{Randall08} to place a tight constraint on \csl{1.25} using the lensing model of \cite{Bradac06}. More recently, \cite{cha25} (hereafter \citetalias{cha25}) used the same formalism and JWST data and reported the limit of \csl{0.2}. However, subsequent theoretical developments have shown that \cse measurements with this method are more challenging than initially assumed \citep{2014MNRAS.437.2865K,2017MNRAS.469.1414K,Robertson17}, requiring precise mass reconstruction, sensitive to asymmetries in DM distributions at large radii. The Bullet Cluster constraint from \cite{Randall08} was reevaluated and placed at \csl{2} \citep[][]{Robertson17}. 

The refinement of the \cse constraints hinges on improved theoretical modelling of the Bullet Cluster and on improved observational constraints. Since its discovery, the Bullet Cluster has been examined by several theoretical studies employing simulations \citep[e.g.][]{2007ApJ...661L.131M,2007MNRAS.380..911S,2008MNRAS.389..967M,2014ApJ...787..144L}. Observationally, important advances have come from spectroscopy of cluster members; \cite{2002A&A...386..816B} constrained the LOS velocities of the main cluster and of the subcluster using spectroscopic measurements of 78 cluster members. More recently, \cite{Benavides23} (henceforth \citetalias{Benavides23}) introduced a method of identifying and characterising the properties of group-sized substructures in galaxy clusters using positions and velocities of cluster members. When applied to the 231 cluster members, \citetalias{Benavides23} revealed the complex structure of the Bullet Cluster. It is composed of ten group-scale substructures, roughly following the two proposed axes of cluster assembly, 
both of which are supported by the detection of radio relic candidates to the east \citep[][]{2015MNRAS.449.1486S} and north-west of the main cluster \citep{2023MNRAS.518.4595S}, as well as by the recent weak lensing analysis by \cite{cho25}.

The most direct insight into the mass distribution in galaxy clusters is provided through weak lensing analyses, which enable low-resolution reconstructions on the cluster outskirts, and strong-lensing analyses, using the positions of multiple images to achieve precise reconstructions in the inner cluster regions. Strong lensing reconstructions rely on high-resolution, multi-band imaging, which is used to identify multiple image candidates. After their identification, it is crucial to obtain their spectroscopic redshifts, firstly to avoid multiple image misidentifications and secondly to provide lens model constraints. The inclusion of several spectroscopic redshifts in the lens model has been shown to substantially reduce systematic errors in multiple-image predictions, magnifications, and mass profiles \citep{2016ApJ...832...82J}. 

Over the past two decades, the Bullet Cluster has undergone several weak and strong lensing reconstructions. \cite{2001A&A...379...96M} provided the first strong lensing analysis using 16 multiple images (from six systems) from ESO-VLT-U1/FORS1 imaging and spectroscopy. This was followed by a weak lensing analysis by \cite{2004ApJ...604..596C} and the combined weak and strong lensing reconstruction by \cite{Bradac06} and \cite{2009ApJ...706.1201B} using 12 multiple image systems from the \textit{Hubble} Space Telescope (HST) observations (one system with spectroscopic redshift). \cite{Paraficz16} constrained the lens model with ESO-VLT-U1/FORS2 spectroscopy using the catalogue of 14 multiple image systems, 3 of which with confirmed spectroscopic redshifts. The \cite{Paraficz16} lens model was updated with additional spectroscopic redshift measurements with ESO-VLT-U4/MUSE (PID 094.A-0115, PI Richard) by \cite{Richard21} (henceforth \citetalias{Richard21}). With 15 systems and 5 spectroscopic redshifts, \citetalias{Richard21} is the most constrained lens model prior to the \textit{James Webb} Space Telescope (JWST) observations; hence, we use it as a reference for model comparison in this work. 

JWST has transformed strong-lensing studies, facilitating lensing reconstructions of galaxy clusters with an unprecedented number of multiple images and spectroscopic redshifts \citep[e.g.][]{2022A&A...666L...9C,2023ApJ...952...84B,2023ApJ...945...49M,2023MNRAS.523.4568F,2024ApJ...973...77G,Pearls24,2025A&A...703A.207D,rihtarsic25,2025arXiv250317498C}. JWST has observed the Bullet Cluster in early 2025 with NIRCam and NIRSpec (programme GO 4598, PIs: Brada\v{c}, Rihtar\v{s}i\v{c}, Sawicki). The NIRCam imaging facilitated the first JWST lensing reconstruction by \citetalias{cha25}, using a large catalogue of weak lensing constraints and a catalogue of 146 multiple image candidates. The lensing reconstruction has been extended to the cluster outskirts by \cite{cho25} through a weak lensing analysis using ground-based DECam observations, measuring the mass ratio between the main cluster and the subcluster to be $10^{+3}_{-2}$. However, all lensing analyses to date have been based on only a handful of multiple-image systems with spectroscopic redshifts; in particular, there has been only one spectroscopic system confirmed in the Bullet subcluster. In this work, we present an updated strong-lensing model of the Bullet Cluster, utilising the full capabilities of NIRCam imaging and NIRSpec spectroscopy from the JWST programme GO 4598. The lens model, presented in this work, is constrained using a catalogue of 135 secure multiple images from 27 distinct background galaxies, all with spectroscopic redshifts.  

The updated strong- and weak-lensing models based on JWST data are valuable for a wide range of applications, such as constraining dark matter self-interactions, as well as studies of high-redshift galaxies. By providing lensing magnifications, the Bullet Cluster has facilitated studies of several background sources, both with pre-JWST data \citep[e.g.][]{2001A&A...379...96M,2009ApJ...706.1201B,2009ApJ...703..348R,2010A&A...514A..77J,2010A&A...518L..13R,2012A&A...543A..62J,2018ApJ...863L..16M} and JWST observations. This includes the recent discovery of a highly magnified, extended arc at redshift 11 \citep[][]{Bradac25}, the study of a redshift 5.3 compact active galactic nucleus (i.e. little red dot), and its broad line region \citep[][]{2025ApJ...994L...6T}, and a spatially resolved, sub-kiloparsec analysis of the cold interstellar medium at $z\sim2.78$ with ALMA (Cornil-Ba\"{\i}otto et al. in prep.), all of which employ the lens model presented in this work.

This paper is organised as follows. In Section \ref{sec:data}, we present the imaging and spectroscopic data and in Section \ref{sec:multimcat} the derived multiple image catalogue used in this work. In Section \ref{sec:lensmodel}, we describe the lens model, its parametrisation, and mass components. In Section \ref{sec:results}, we discuss our lens model and compare it to earlier works, and we summarise our findings in Section \ref{sec:conclusion}. In Appendix \ref{app:multimcat} we provide the full multiple image catalogue with spectroscopic redshift measurements. In Appendix \ref{Appendix:NIRSpec spectra}, we give the details on individual NIRSpec spectra and redshifts and in Appendix \ref{app:modelparams}, we report the parameters of our fiducial model. The convergence $\kappa$, shown in this work, is defined for the ratio of angular diameter distances from the lens to the source and from the observer to the source $D_{\rm ls}/D_{\rm s }=1$. Throughout this work, we assume a flat lambda cold dark matter, $\Lambda$CDM, cosmology with $\Omega_\Lambda=0.7$, $\Omega_{\rm m}=0.3$, and $H_0=70~{\rm km\,s^{-1}\,Mpc^{-1}}$. At the redshift of the Bullet cluster $z_l=0.296$, a projected distance of $1''$ corresponds to a physical scale of $4.413$~kpc. Magnitudes are given in the AB system \citep{1983ApJ...266..713O}.

\section{Data}
\label{sec:data}
\subsection{Images and photometry}
The Bullet Cluster was observed with JWST Near Infrared Camera (NIRCam, \citealt{2023PASP..135b8001R}) as a part of the GO programme 4598  (PIs: Brada\v{c}, Rihtar\v{s}i\v{c}, Sawicki) in January 2025. The cluster was observed in eight filters (F090W, F115W, F150W, F200W, F277W, F356W, F410M and F444W) with $\sim 6.4$ ks exposure each. The RGB combination of NIRCam images is shown in Fig. \ref{fig:field}. The cluster was observed with the FULLBOX six-point dither pattern to cover the gap between the two NIRCam modules. The extent of the Bullet Cluster is well suited for NIRCam observations, with both cluster components (henceforth referred to as the main cluster and the subcluster on the left and the right side of Fig. \ref{fig:field}, respectively) falling within the full NIRCam FOV. Our JWST imaging was complemented by the archival HST/ACS images in F606W, F775W and F850LP filters (programmes 10200 and 10863, PI Jones \& Gonzalez) and F814W filter (programme 11099, PI Bradač). The images were processed following the Canadian NIRISS Unbiased Cluster Survey (CANUCS) data reduction pipeline, described in detail in \citetalias{Sarrouh25}. This includes the modified stage 1 and stage 2 STScI JWST pipeline, followed by redrizzling onto a 40 mas pixel grid using \texttt{grizli} code \citep{grizli}. The intracluster light (ICL) and bright cluster galaxies were modelled and subtracted from the images, following the procedure described in \cite{Martis24}. The bright-cluster-galaxy-subtracted images were point-spread-function (PSF) homogenised to match the PSF in the F444W filter.

\label{sec:images}

To obtain the photometry and photometric redshifts, we followed the procedures described in \citetalias{Sarrouh25}.
Photometric catalogues were produced with the \texttt{Photutils} package \citep{photutils} using $0.3''$ fixed apertures. Photometric redshifts $z_{\rm phot}$ were derived with \texttt{EAZY-py} \citep{2008ApJ...686.1503B} using  the latest standard templates (\texttt{tweak\_fsps\_QSF\_12\_v3}) and templates from \citet{2023ApJ...958..141L}. We also took into account the intergalactic medium (IGM) and circumgalactic medium (CGM)  attenuation correction following \cite{2025ApJ...983L...2A}, to avoid the overestimation of photometric redshifts above redshift 7. For the magnitudes of the cluster members used for scaling relations (Section \ref{sec:model-clustermembers}), we adopt the F277W fluxes in Kron apertures \citep{1980ApJS...43..305K} from \texttt{Photutils} catalogues. For the brightest cluster members, which were modelled and subtracted separately, we adopted the total modelled flux in the F277W filter. For testing the correlations between magnitudes and stellar masses (section \ref{sec:model-clustermembers}), we fit the spectral energy distribution with \texttt{BAGPIPES} \citep{2018MNRAS.480.4379C}, following the procedure used for CANUCS clusters, described in  \citetalias{Sarrouh25}.

\subsection{NIRSpec spectroscopy}
The Bullet Cluster was observed with the JWST Near Infrared Spectrograph (NIRSpec, \citealt{2022A&A...661A..80J}) in March 2025 as part of the GO programme 4598. It was observed in multi-object spectroscopy mode through the Micro-Shutter Assembly (MSA) with three MSA configurations with 3.5 ks exposure each. The observations used the PRISM/CLEAR disperser with nominal resolving power $\sim 100$ covering the wavelength range between 0.6  and 5.3 $\mu m$. The targets for the NIRSpec follow-up were selected based on various science goals, with an emphasis on obtaining redshifts for multiple images identified through NIRCam observations. The selection also included several photometrically selected cluster members without previous spectroscopic confirmation, as well as magnified background galaxies \citep[e.g.][]{2025ApJ...994L...6T,Bradac25}. In total, we observed spectra of 172 sources with NIRSpec.

The reduction of the spectra followed the procedure described in \citetalias{Sarrouh25} and \cite{2025A&A...693A..60H} using STScI \texttt{jwst} pipeline followed by \texttt{grizli} \citep{grizli} and \texttt{msaexp} \citep[][]{msaexp} packages. It includes custom snowball and 1/{\it f} noise correction, standard wavelength calibration with correction for the known intra-shutter offset, and photometric calibration. The spectral background is removed using the standard nodded background subtraction. One-dimensional spectra are extracted using a wavelength-dependent optimal extraction that accounts for the point-spread function variations with wavelength. In the present work, the spectroscopic redshifts, $z_{\rm NIRSpec}$, were obtained using the \texttt{msaexp} package or by fitting individual emission lines with Gaussian profiles. For several multiple-image systems, we derived the system redshift by combining spectra from individual images to increase the signal-to-noise ratio (S/N). We estimated the NIRSpec redshift uncertainty by comparing several $z_{\rm NIRSpec}$ measurements within individual multiple image systems and with VLT/MUSE redshifts published by \citetalias{Richard21}. We used the estimated uncertainty $\Delta z / (1+z) \sim 0.002$ for $z_{\rm NIRSpec}$ measurements throughout this work (unless specified otherwise in Appendix \ref{Appendix:NIRSpec spectra}). In Appendix \ref{Appendix:NIRSpec spectra}, we present details of the redshift derivations and give the individual NIRSpec spectra. Multiple image redshift measurements are listed in Table \ref{tab:multipleimagestable}. In addition, we provide spectroscopic redshifts for 34 cluster members, listed in Table \ref{tab:newnirspecclusterm}. We measure the mean cluster redshift of $\bar{z}=0.299\pm 0.002$, consistent with redshift measurements of \cite{2002A&A...386..816B}.

\section{Catalogue of multiple images}
\label{sec:multimcat}

\begin{figure*}
    \centering
    \includegraphics[width=\linewidth]{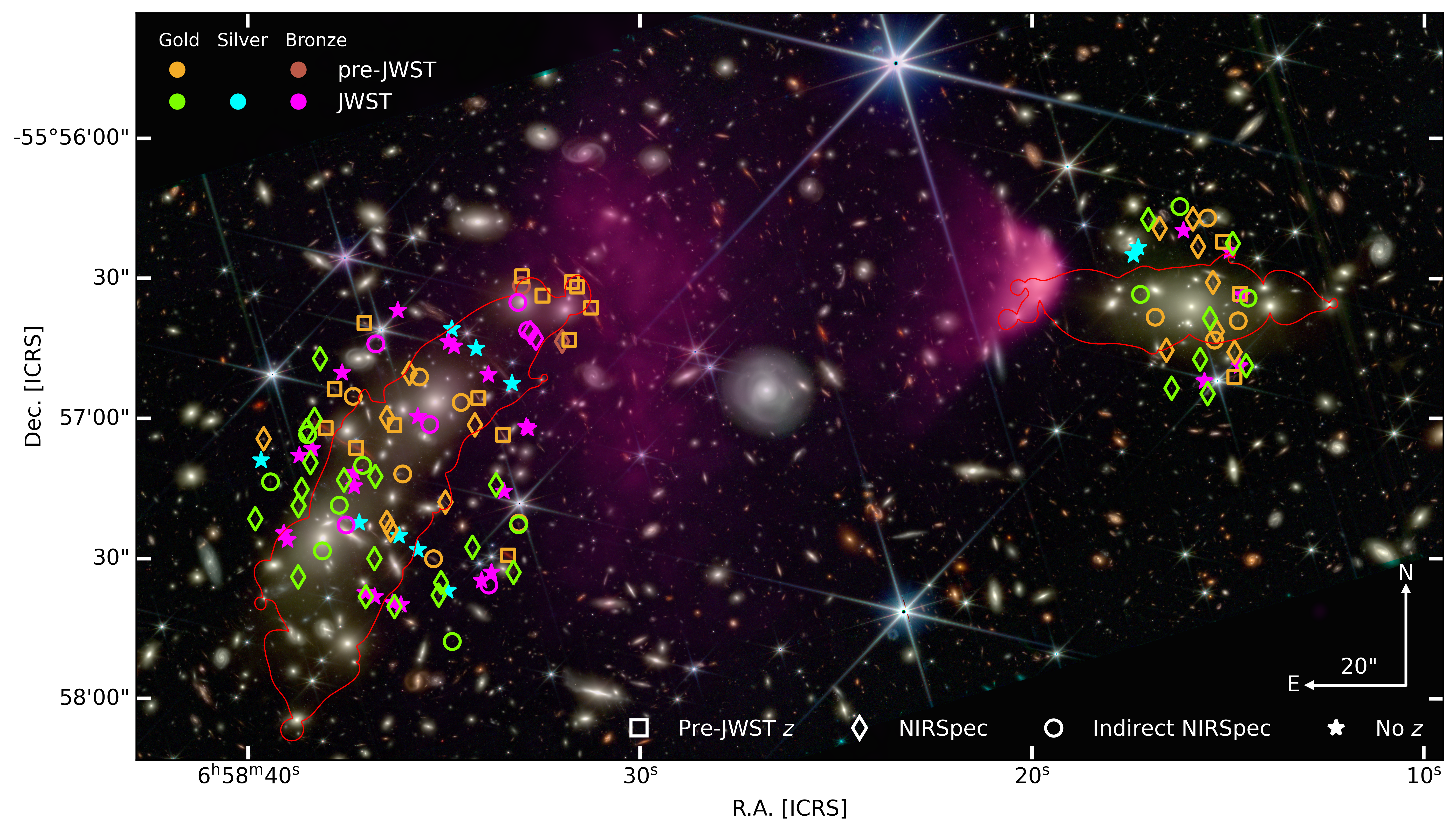}
    \caption{JWST/NIRCam image of the Bullet cluster with multiply imaged systems. The RGB image is composed of several NIRCam filters (F444W, F277W and F356W in red, F200W and F150W in green, and F115W and F090W in blue). A smoothed Chandra X-ray image, showing cluster gas, is shown in pink. The colour of multiple images represents their quality grades (gold, silver or bronze) and whether they were used in the pre-JWST models (\citetalias{Richard21}, \citealt{Paraficz16}) or were included after JWST observations (this work or \citetalias{cha25}). Squares represent systems which have known pre-JWST redshifts from the literature, diamonds represent images for which we obtained new spectroscopic redshifts. Circles indicate multiple images for which we did not measure spectroscopic redshifts directly, but have new $z_{\rm sys}$ from other images of the system. Stars represent newly identified multiple-image candidates without spectroscopic redshifts. For clarity, we show only one multiply lensed feature per galaxy.  The red solid line represents the longest tangential critical curves from our lensing model for redshift $z=9$.} 
    \label{fig:field}
\end{figure*}

We constructed our multiple-image catalogue by compiling the pre-JWST catalogues from \citetalias{Richard21}, \cite{Paraficz16} and \cite{Bradac06}, and by using the NIRCam imaging and NIRSpec spectroscopy from our programme. After acquiring the NIRCam data, we inspected the images to find multiply-imaged galaxies. We examined several RGB combinations of NIRCam and HST images, benefiting from the rich colour information spanning the wavelength range between 0.4 and 5 $\mu \mathrm{m}$ and the high resolution (20 mas) of the short-wavelength NIRCam channel images. Our visual inspection was aided by the multiple image predictions of the sources in our photometric catalogues using the \citetalias{Richard21} lens model. To predict the positions of each source, we primarily used the best-fit \texttt{EAZY} photometric redshifts, but have also computed predictions for a range of redshifts between 1 and 20 to mitigate the impact of potential redshift misestimates. We then visually inspected an extended region ($>10''$) around each prediction. New multiple image candidates were graded, based on their morphology, colours, photometric redshifts or lens-modelling arguments. We identified systems that required only a single spectroscopic redshift measurement and those that required spectroscopic confirmation of individual multiple images. This information was taken into account when preparing the MSA masks for the NIRSpec follow-up to maximise the utility of our observations. After deriving the fiducial lens model (see Section \ref{sec:lensmodel}), we recomputed counter-image predictions from our multiple image catalogue and verified that the missing images were either obscured by cluster members or stars, or were too faint to be reliably identified.

Our final catalogue is presented in Fig. \ref{fig:field} and Table \ref{tab:multipleimagestable}. It includes our quality grades (gold, silver, or bronze), following a scheme similar to that adopted for the lens models from CANUCS survey \citep{rihtarsic25,2024ApJ...973...77G}. The 'gold' catalogue of multiple images contains the most secure multiple images with known spectroscopic system redshift, $z_{\rm sys}$, which is used to constrain the lens model. The gold catalogue includes 135 multiple images from 49 lensed systems (individual multiply imaged components of background galaxies) from 27 distinct lensed galaxies. The 'silver' catalogue contains secure systems, based on their morphological features and lensing configuration, for which we were  not able to obtain spectroscopic redshifts (22 multiple images). In the 'bronze' catalogue (42 multiple images), we included systems and multiple images for which our data did not provide secure confirmation, as well as two systems with new spectroscopic redshifts (K17, N18) for which we could not securely identify the precise positions of lensed features. We did not use the silver and bronze multiple images to constrain the lens model. The complete catalogue of multiple images, provided in Table \ref{tab:multipleimagestable}, comprises 199 multiple images from 73 lensed systems (belonging to 43 distinct galaxies).

We indicate systems, known before JWST observations, with the letter K in the system IDs. From system K1 to system K15, the ID numbers follow the \citetalias{Richard21} catalogue. Systems K16 and K17, not included in \citetalias{Richard21}, correspond to systems D and C from \cite{Paraficz16}, respectively. All systems that were unknown prior to JWST observations are denoted by the letter "N". In 22 lensed galaxies, we identified several lensed features and included them in our catalogue of strong-lensing constraints, thus incorporating the local information on relative magnifications. We denote multiple image systems belonging to the same galaxy with lower-case letters from a to e, where the suffix after the dot specifies the multiple-image index. For example, K8b.3 represents the third multiple image of the lensed feature b in galaxy K8.

A similar set of multiple images, derived from NIRCam imaging (without new NIRSpec redshifts), was also published in \citetalias{cha25}. We cross-matched our catalogue with the \citetalias{cha25} catalogue, and we provide the \citetalias{cha25} IDs in Table \ref{tab:multipleimagestable}. We supplemented our bronze catalogue with eight candidate systems from \citetalias{cha25}. In turn, our catalogue contains several lensed galaxies not reported by \citetalias{cha25} (including 3 gold, 2 silver, and 3 bronze), as well as several new or alternative multiply-imaged clumps and new counter-images.

\begin{figure}
    \centering
    \includegraphics[width=\linewidth]{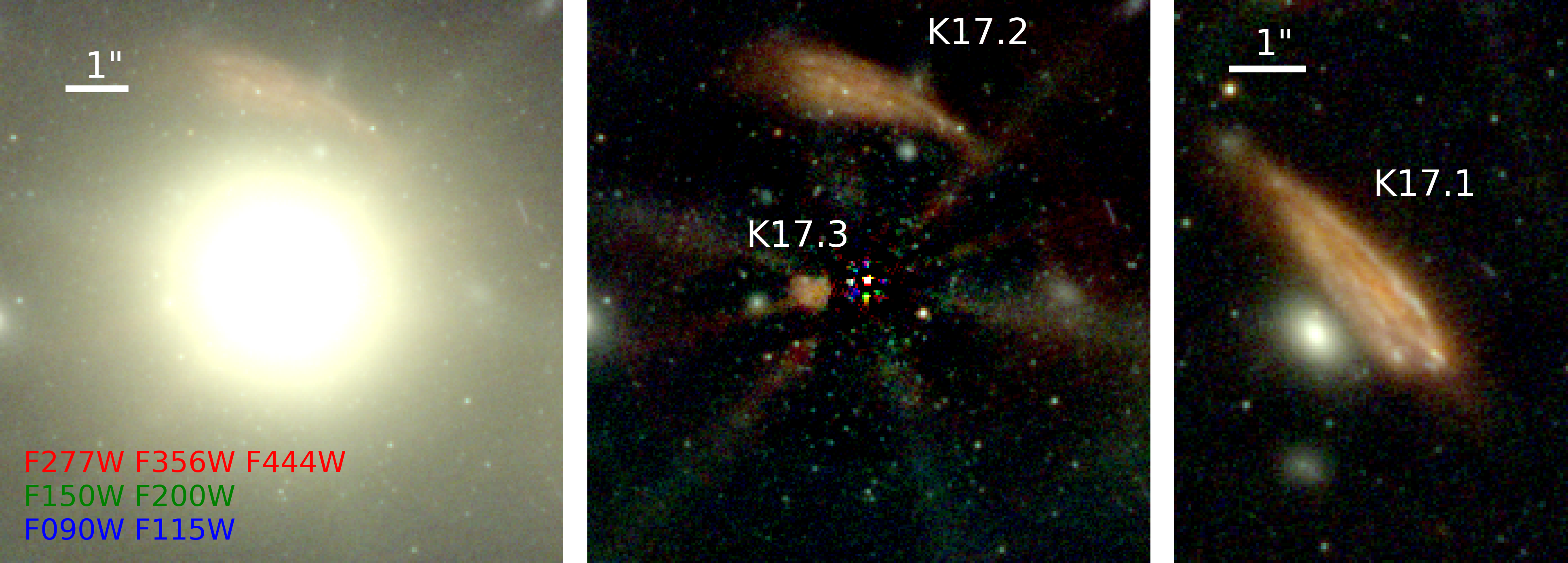}
    \caption{Three multiple images of system K17. \textit{Left}: NIRCam cutout centered on the cluster member G5. \textit{Middle}: bright-cluster-galaxy-subtracted cutout revealing multiple images K17.2 and K17.3. \textit{Right}: Cutout showing multiple image K17.1.} 
    \label{fig:K17cutout}
\end{figure}

For system K17 (system C from \citealt{Paraficz16}), we uncovered the third multiple image K17.3, hidden near the centre of the nearby cluster galaxy G5 (see Fig. \ref{fig:K17cutout}). K17.3 was missed in the original image, as it is embedded in the light of G5, but was revealed after the bright cluster galaxy subtraction, with its colours preserved, despite its proximity to the centre of the cluster member G5 ($\sim1''$). Although we do not use K17 to constrain the lens model, due to unclear positions of point-like lensed features we use in this work, we note its potential utility for constraining the mass and profile of G5.

\subsection{NIRSpec spectroscopic redshifts}
\begin{figure}
    \centering
    \includegraphics[width=\linewidth]{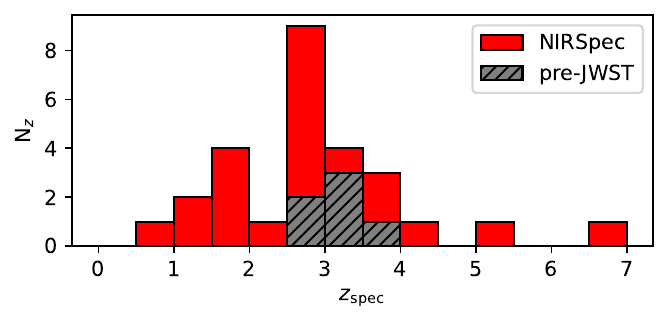}
    \caption{Spectroscopic redshift distribution of multiply lensed galaxies in the gold catalogue. Grey colour indicates the spectroscopic redshifts used in previous studies (\citetalias{cha25}). }
    \label{fig:speczs_histogram}
\end{figure}

51 of 271 spectra obtained with our NIRSpec follow-up observations were used to determine system redshifts $z_{\rm sys}$ and spectroscopically confirm multiple image systems. In total, we obtained new redshifts for 23 lensed galaxies, including all ten galaxies from pre-JWST catalogues (\citetalias{Richard21},\citealt{Paraficz16}) with previously unknown $z_{\rm sys}$. In addition, we confirmed the $z_{\rm sys}$ redshifts of 5 galaxies with previously known $z_{\rm sys}$. In total, we increased the number of lensed galaxies with spectroscopic redshifts from 6 (\citetalias{cha25}, \citetalias{Richard21}) to 30. In Fig. \ref{fig:speczs_histogram} we present the spectroscopic redshift distribution of multiply lensed galaxies, and in Fig. \ref{fig:field} their spatial distribution. With the addition of NIRSpec spectroscopic redshifts, our catalogue now spans a wide redshift range from 0.9 to 6.7.

In Appendix \ref{Appendix:NIRSpec spectra}, we provide the details of obtaining NIRSpec redshifts $z_{\rm NIRSpec}$ with uncertainties for all multiple images. Where no pre-JWST redshifts were available, we used $z_{\rm NIRSpec}$ as the system redshift, $z_{\rm sys}$, which is used as a lens model constraint. For systems with more than one $z_{\rm NIRSpec}$ measurement, we obtained  $z_{\rm sys} $ with a weighted mean, taking into account their uncertainties. For some systems, individual spectra did not yield secure redshift measurements; in these cases, we derive redshifts by combining the spectra to increase the S/N. All our $z_{\rm NIRSpec}$ measurements with uncertainties and the $z_{\rm sys}$ values are provided in Table \ref{tab:multipleimagestable}.

\subsection{Archival spectroscopic redshifts}
We included several systems with known spectroscopic redshifts from the literature. Systems K1, K3, and K15 all have MUSE spectroscopic redshifts $z_{\rm MUSE}$ and have been included in the \citetalias{Richard21} multiple image catalogue. We included those systems with their MUSE redshifts, as MUSE resolution outperforms the NIRSpec PRISM resolution. We did, however, observe all of them with NIRSpec and confirmed that their redshifts are consistent with $z_{\rm MUSE}$ (see Table \ref{tab:multipleimagestable}). System K8 was included in \citetalias{Richard21} lens model without spectroscopic redshift, although its multiple images have $z_{\rm MUSE}$ from \citetalias{Richard21} catalog of MUSE redshifts: K8.1 with $z_{\rm MUSE}=3.260 \pm	0.003$ and K8.2 with $z_{\rm MUSE}=3.260 \pm 0.002$, however, neither with the highest confidence (Confidence 2, see \citetalias{Richard21}). Our NIRSpec spectrum features a Balmer break and \Halpha emission line, which was used to measure \znirspecerrA{3.259}{0.007} (see Appendix \ref{Appendix:NIRSpec spectra}), thus confirming $z_{\rm MUSE}$ which is used as $z_{sys}$. 

System K10 (system H in \citealt{Paraficz16}) has a Lyman-$\alpha$ redshift measurement of 2.99,  reported by \cite{Paraficz16} using their FORS/VLT spectrum. Prior to our NIRSpec observations, this was the only system with a spectroscopic redshift available in the subcluster. We confirm the redshift using our K10.1 NIRSpec spectrum with several emission lines, and obtain \znirspecerrA{2.993}{0.006}, which we use for the $z_{sys}$.

The spectroscopic redshift of K4 was revised from the value reported in \cite{2018ApJ...863L..16M} to $2.7768 \pm 0.0002$, measured from the CO(3--2) line detected with ALMA Band 3 at 91.5 GHz (Cornil-Ba\"{\i}otto et al. in prep.).

\section{Lens model}
\label{sec:lensmodel}
For constraining the lens model, we used the parametric strong lens modelling code  \texttt{Lenstool}\footnote{\url{https://projets.lam.fr/projects/lenstool/wiki/}} \citep{Kneib96,Jullo07,Jullo09}, which models the mass distribution as a superposition of smooth elliptical mass profiles. For all our mass components (apart from cluster gas, see section \ref{sec:model-clustergas}), we adopt the
Pseudo Isothermal Elliptical Mass Distributions (PIEMD) \citep{2005MNRAS.356..309L,2007arXiv0710.5636E} with a 3D mass density profile
\begin{equation}
\rho(r)=\frac{\sigma_0^2}{2\pi G}\left(\frac{r_{\text {cut }}+r_{\text {core }}}{r_{\text {core }}^2r_{\text {cut }}}\right)\frac{1}{\left(1+r^2 / r_{\text {core }}^2\right)\left(1+r^2 / r_{\text {cut }}^2\right)},
\label{eq:piemdprofile}
\end{equation}
where $G$ represents the gravitational constant, $r$ the 3D distance from the centre, and the profile is defined with two characteristic radii (core radius $r_{\rm core}$ and cut radius $r_{\rm cut}$) and the central velocity dispersion, $\sigma_0$, related to the \texttt{Lenstool} fiducial velocity dispersion as $\sigma_{\rm lt}=\sqrt{2/3}\sigma_0$. The projected 2D elliptical profile is expressed as 
\begin{equation}
    \label{eq:projectedpiemd}
    \Sigma (R)=\frac{\sigma_0^2 r_{\text {cut }}}{2 G \left( r_{\text {cut }} - r_{\text {core }} \right)}
    \left(\frac{1}{\sqrt{r_{\text {core }}^2+R^2} }- \frac{1}{\sqrt{r_{\text {cut}}^2+R^2} }\right),
\end{equation}
where $R$ is expressed with ellipticity, $e$, and the 2D coordinates along the semi-major ($x$) and semi-minor ($y$) axes, relative to the center $(x_0,y_0)$ via


\begin{equation}
    R^2=
    \left(\frac{1+q}{2}\right)^2(x-x_0)^2
    +
    \left(\frac{1+q}{2q}\right)^2(y-y_0)^2 .
\end{equation}

where $q^2=(1-e)/(1+e)$. Therefore, the PIEMD halos can be characterised with eight parameters: coordinates of the centre ($x_0$, $y_0$), redshift $z$, which we set to 0.296 for all halos,  ellipticity, $e$, position angle, $\theta$ (counterclockwise from the west), characteristic radii, $r_{\rm core}$ and $r_{\rm cut}$, and the \texttt{Lenstool} fiducial velocity dispersion, \sigmalt. Due to the analytical projected potential and the flexibility, with an additional free parameter compared to the NFW profile, they have been successfully used to model both cluster-scale \citep[e.g.][]{Kneib96,2007ApJ...662..781R,2022A&A...657A..83C} and early-type galaxy halos \citep[e.g.][]{1998ApJ...499..600N,2007A&A...461..881L}. The total mass of the PIEMD mass profile is computed as

\begin{equation}
M_{\text {tot }}=\frac{\pi \sigma_0^2 r_{\text {cut }}}{G} .
\label{eq:piemdmass}
\end{equation}
We modelled the Bullet Cluster with several mass components, as described in the following subsections.

\begin{figure*}
    \centering
    \includegraphics[width=\linewidth]{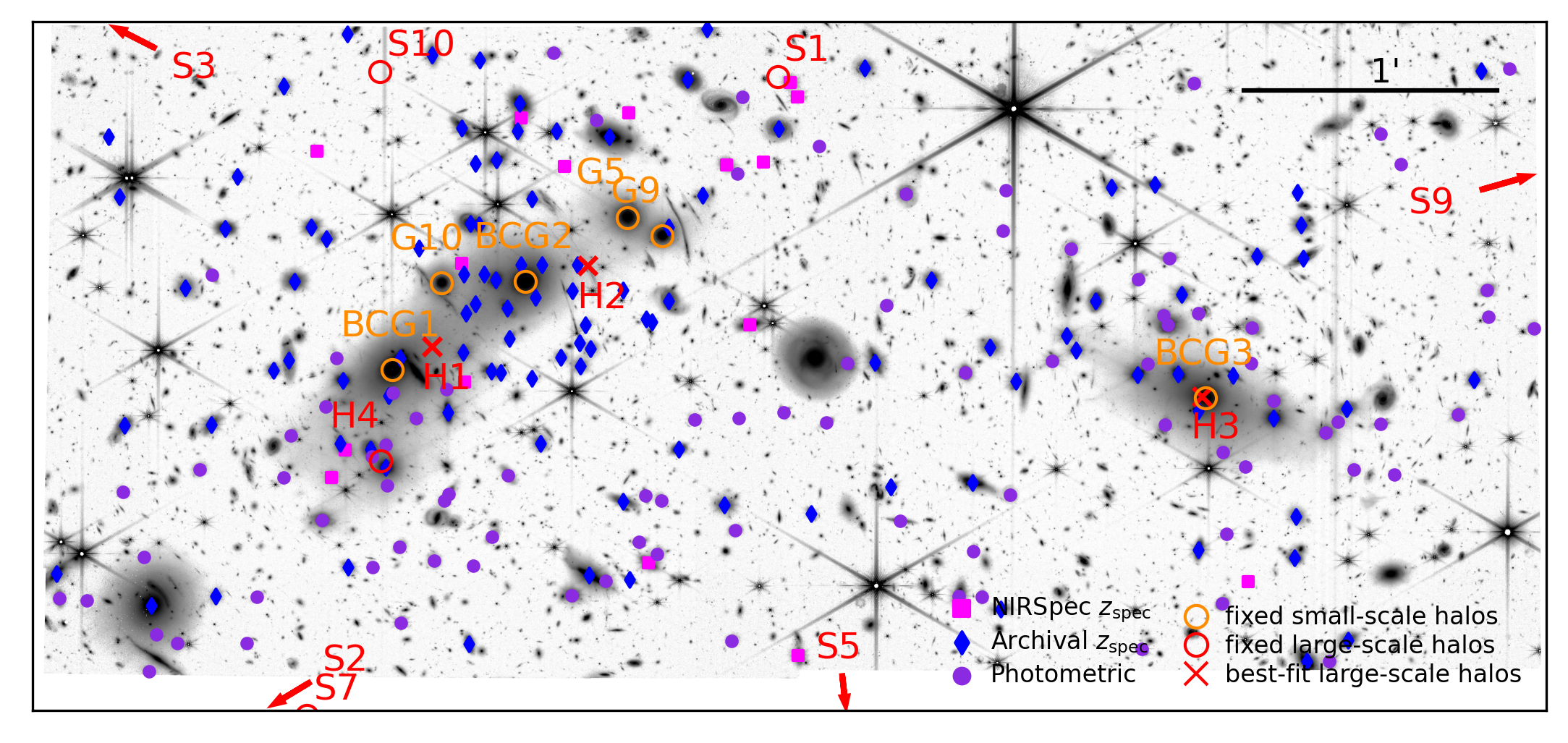}
    \caption{Inverted greyscale image (F277W) of the Bullet cluster covering the full NIRCam FOV, with indicated photometric and spectroscopic cluster members and other halos included in the lens model. Cluster members selected using the NIRSpec $z_{\rm spec}$ are shown in magenta, those confirmed with literature $z_{\rm spec}$ in blue, and photometrically selected candidates in violet. The positions of other lens model components (halos) are indicated with empty circles if they have a fixed position or with X if their position was left as a free parameter (in which case we mark the best-fit position). Galaxy-scale halos modelled outside the cluster member scaling relations and large-scale halos are shown in orange and red, respectively. For halos that are outside the NIRCam FOV, we indicate the direction of their positions with arrows.} 
    \label{fig:model_clustermembers}
\end{figure*}

\subsection{Large-scale halos}
\label{sec:model-largescalehalos}
We modelled the large-scale DM distribution with four PIEMD halos. We fit all PIEMD profile parameters with the uniform prior ranges listed in Table \ref{tab:parameterpriors}.  We show the best-fit parameter values with uncertainties in Table \ref{tab:bestfitparam}. We set the cut radii $r_{\rm cut}$ to 2 Mpc (similar to $r_{200}$\footnote{$r_{200}$ is the radius of a sphere enclosing a mass $M_{200}$ within which the mean enclosed density is equal to 200 times the critical density of the Universe.} of the cluster, estimated from weak lensing, \citealt{cho25}, and galaxy kinematics, \citetalias{Benavides23}). The cut radii extend far beyond the distance between the two subclusters and the radius containing multiple images; hence, their exact values cannot be constrained with our data. Furthermore, they are extended beyond the NIRCam field of view and, thus, their exact values have little impact on our analysis. The ellipticity, $e$, was restricted to be less than 0.7 to avoid unrealistically elongated halos. Halos H1, H2, and H3 are loosely associated with the brightest cluster galaxies (BCGs). Specifically, BCG1, BCG2, and BCG3, respectively, are each free to move over a prior range $\gtrsim20''$ (see Figure \ref{fig:model_clustermembers}). BCG1, situated in the southeast of the main cluster, is the brightest galaxy in the cluster. BCG2 is situated to the north-west, tracing the second component of the main cluster, and BCG3 is the brightest galaxy in the subcluster. We opt for the wide position prior ranges instead of the close association of the mass with light (e.g. \citealt{2022A&A...664A..90L, 2025A&A...703A..10L}) to capture the complex cluster morphology using a parametric lens modelling technique (see Section \ref{sec:discuss_parametrisation}). The wide ranges were required to ensure that the positions converged well within the prior limits. Testing an alternative model with a restricted prior range resulted in the best-fit positions near the imposed limits. Apart from halos H1, H2, and H3, which were all included in the \citetalias{Richard21} and \cite{Paraficz16} lens models, we added an additional halo H4 to the south of the BCG1, motivated by the southern extension of the ICL (Fig. \ref{fig:model_clustermembers}). As the halo is situated outside the region with multiple images, we were not able to constrain its position, and instead fix it to a peak in the smoothed F277W image, using a $4''$ Gaussian kernel, in the region surrounding several cluster members to the south of the BCG1  (Fig. \ref{fig:model_clustermembers}). 

\subsection{Cluster gas}
\label{sec:model-clustergas}
To accurately study the distribution of DM in the Bullet cluster, it is important to disentangle the contribution of DM and cluster gas, which lags behind the main cluster (see Fig. \ref{fig:field}). In this work, we included the cluster gas density model derived from Chandra observations for the analyses by \cite{Bradac06}, \cite{Randall08} and \cite{clowe06}. The cluster gas was modelled using a simple three-dimensional mass model composed of several analytic components, optimised by fitting to the Chandra X-ray image. For the fit, it was assumed that the X-ray surface brightness scales with the line-of-sight integral of the squared gas density. The gas temperature and metallicity, used to normalize the gas density, were measured from the X-ray spectra. The obtained cluster gas mass maps are robust with $10\%$ uncertainty \citep{clowe06}. In our fiducial model, we included gas as a fixed mass map and briefly investigate the effect of excluding the gas in Section \ref{sec:discuss_parametrisation}. We used the gas map to derive the convergence ($\kappa$) map and deflection angle maps by performing a discrete Fourier transform. The fixed gas maps are added to our \texttt{Lenstool} models.

\subsection{Cluster members}
\label{sec:model-clustermembers}
Cluster galaxy halos have been included in our lens model as circular PIEMD halos with small $r_{\rm core}$ values and with individual $r_{\rm cut,i}$ and $\sigma_{\rm lt,i}$ scaling with brightness, $L_i$, of each galaxy, according to scaling relations \citep[e.g.][]{1996MNRAS.280..167J, 1997MNRAS.287..833N} expressed as

\begin{equation}
  \begin{split}
\sigma_{\mathrm{lt}, i}=\sigma_{\mathrm{lt}}^{\mathrm{ref}}\left(\frac{L_i}{L_{\mathrm{ref}}}\right)^\alpha,
  \end{split}
\quad
  \begin{split}
r_{\rm cut, \it i}=r_{\mathrm{cut}}^{\mathrm{ref}}\left(\frac{L_i}{L_{\mathrm{ref}}}\right)^{\beta_{\rm cut}}.
  \end{split}
  \label{eq:scalingrelations}
\end{equation}

In this work, we set $\alpha=0.25$ and $\beta_{\rm cut}=0.5$ so that the mass would scale proportionally to luminosity,  leaving the normalization $\sigma_{\mathrm{lt}}^{\mathrm{ref}}$ and $r_{\mathrm{cut}}^{\mathrm{ref}}$ as free parameters. We verified that using circular halos instead of elliptical halos in the scaling relations speeds up the computations with negligible impact on multiple image positions. For the choice of the photometric filter to use for the scaling relation, we compared the relations between magnitudes of cluster members in individual filters and stellar mass $M_*$ derived with \texttt{BAGPIPES} using all available filters. We found that the relation between magnitude and $M_*$ is the tightest, with the smallest scatter around the $\log M_*-m$ relation, in the F277W filter (with $\Delta \log M_*/M_\odot$ standard deviation of 0.12). This is also the filter in which the cluster members are measured with the highest S/N and, hence, we used $m_{\rm F277W}$ magnitudes for the scaling relations. As a reference magnitude, we chose $m_{\rm F277W}=15.77$ of the BCG1. 

The quality of lens models relies heavily on the accuracy of the cluster member catalogue, which in turn depends on spectroscopic confirmation of cluster membership. \citep[e.g.][]{2015ApJ...800...38G}. Through our JWST programme, we obtained NIRSpec spectroscopic confirmation for 34 cluster members, 16 of which are included in the cluster member catalogue (See Table \ref{tab:newnirspecclusterm} in Appendix \ref{Appendix:NIRSpec spectra}). Furthermore, we compiled spectroscopic redshifts of the cluster members from several sources in the literature \citep{2002A&A...386..816B,2009A&A...499..357G,2017A&A...606A.122F} using the NASA/IPAC Extragalactic Database (NED\footnote{\texttt{https://ned.ipac.caltech.edu/}}) with additional spectroscopic redshifts obtained with Inamori Magellan Areal Camera
and Spectrograph (IMACS), used in \cite{2010ApJ...725.1536C}, and the VLT/MUSE redshift catalog from \cite{Richard21}, which covers a part of the main cluster. We selected  galaxies in the NIRCam FOV in the redshift range of $0.28<z_{\rm spec}<0.36$ to incorporate the Bullet Cluster and a small background group at a nearby redshift of $\sim0.35$ \citep[][]{2010A&A...518L..14R,2010ApJ...725.1536C}{}. A handful of confirmed foreground galaxies (with $z_{\rm spec}<0.28$) lie far from the closest multiple image in projection (>20$''$) and their influence was expected to be small. Since these objects are not expected to follow the cluster member scaling relations and could not be properly treated within our single-plane lens modelling framework, we excluded them from the lens model. 

We complemented the catalogue of spectroscopic cluster members with photometric redshift selection using our \texttt{EAZY} $z_{\rm phot}$ catalogue. We selected galaxies with \texttt{EAZY} $z_{\rm phot}$ consistent with cluster redshift (i.e. the interval between $0.28 <z< 0.32$ overlaps with the $68\%$ confidence interval of $z_{\rm phot}$). We also visually inspected and refined the catalogue (for instance, by correcting incorrect centroid positions or spurious matches between our and archival $z_{\rm spec}$ catalogues). Finally, we applied a brightness cut using $m_{\rm F277W}<21.3$, which limits the number of all cluster members to 219. The brightness cut was chosen to include faint cluster members with a small influence on multiple images (for instance, \citetalias{Richard21} included only the 100 cluster members), while keeping the computational cost manageable. Out of 219 cluster members, 129 were selected with spectroscopic redshifts and 16 with NIRSpec redshifts provided by our programme (see Table \ref{tab:newnirspecclusterm}). The catalogue is displayed in Fig. \ref{fig:model_clustermembers}. 

A few of the cluster members were modelled separately and were not included in the scaling relations. This includes the brightest galaxies in the main cluster (BCG1 and BCG2) and in the subcluster (BCG3), since they do not necessarily follow the scaling relations \citep[e.g.][]{2013ApJ...765...25N,2013ApJ...765...24N}, as well as three fainter cluster members close to lensed systems, with many constraints. These fainter cluster members include galaxy G10, located near arc K4, as well as G5 and G9, situated near K1. Galaxies were modelled with free \sigmalt, their positions, ellipticities, and orientations were fixed based on photometry.  The cut radii, $r_{\rm cut}$, were fixed using scaling relations, after verifying that they remained uncostrained when left as free parameters in the lens model.

\subsection{Group-scale substructures}
\label{sec:model-substructures}
Identifying and constraining the properties of group-scale substructure halos outside the region with multiple images is beyond the power of strong-lensing analyses alone. It requires a weak lensing analysis using high-resolution space-based imaging on the cluster outskirts \citep[e.g.,][]{2023MNRAS.524.2883N}, which is currently unavailable for the Bullet Cluster. Alternatively, such substructure candidates can be identified through the kinematics of cluster members using spectroscopy. \cite{2002A&A...386..816B} used the ESO NTT spectroscopy of 78 cluster members to identify the Bullet subcluster and estimate its mass. More recently, \citetalias{Benavides23} presented a new method of identifying substructures and applied it to an extended sample of the 231 archival spectroscopic redshifts of the Bullet cluster. They identified ten group-scale substructures (see Table 2 of \citetalias{Benavides23}), including the Bullet subcluster and two substructures in the main cluster traced by BCG1 and BCG2. Beyond the main cluster and the subcluster, they identify seven other substructures at distances between 1.4 and 6.2 arcminutes from BCG1. They roughly follow two proposed axes of cluster assembly: one along the axis of the Bullet merger with the main cluster, and the other along the axis of the two components in the main cluster. The two merger axes, traced by the substructures, are thus roughly aligned with the recent Bullet merger axis and the proposed merger between the two main cluster components \citep[e.g.][]{cho25}. We will henceforth refer to those groups as substructures (as opposed to the main cluster and subcluster) and denote them as S1, S2, S3, S5, S7, S9, S10. Their positions, given by \citetalias{Benavides23}, are shown in Figures \ref{fig:model_clustermembers} and \ref{fig:kappamaps}. \citetalias{Benavides23} also provide the velocity dispersion $\sigma_g$ with uncertainty for each substructure. The substructures S4, S6 and S8 from \citetalias{Benavides23} correspond to the inner components traced by BCG2, BCG3, and BCG1, respectively, which are modelled with large-scale halos, well constrained by multiple images. We therefore do not adopt their \citetalias{Benavides23} substructure properties.

Since all modelled substructures are at least $35''$ away from the closest multiple image, we were not able to recover all their properties with strong-lensing analysis. Instead, we implemented them with priors or fixed parameters set to \citetalias{Benavides23} estimates. In this way, we attempted to incorporate a physically motivated influence of the cluster surroundings (rather than adding a constant shear applied to both the main cluster and the subcluster), while limiting the possible degeneracies that would arise from having flexible mass components outside the region with strong-lensing constraints.

We estimated the mass of each substructure using the method described in \citetalias{Benavides23}. We first corrected $\sigma_g$ by a factor of $\sim1.8$ to account for the overestimation of measured $\sigma_g$ relative to the true group velocity dispersion (Table 1 from \citetalias{Benavides23}). This factor is derived from simulations and carries a large relative uncertainty ($>100\%$, see \citetalias{Benavides23}), which adds to the uncertainties in the measured velocity dispersions (Table 2 of \citetalias{Benavides23}). We computed the mean $M_{200}$ by using two relations between $M_{200}$ and velocity dispersion from \cite{2013MNRAS.430.2638M} (assuming NFW profile and from "AGN gal" simulations, see Table 1 from \citealt{2013MNRAS.430.2638M}) to derive masses $M_{200}^{\sigma}$ (shown in Table \ref{tab:substrmass}). We furthermore calculate a second mass estimate, $M^r_{200}$, using the richness of each group. We assume that the number of galaxies belonging to each substructure (Table 2 of \citetalias{Benavides23}) scales proportional to its mass and that the same ratio between the number of galaxies and mass holds for the cluster ($1.0^{+1.1}_{-0.2}\times10^{14} M_{\odot}$ with 141 galaxies, see \citetalias{Benavides23}). The two mass estimates (Table \ref{tab:substrmass}), with large individual uncertainties, differ by up to an order of magnitude. This is unsurprising, as inferring the properties of groups after their infall from their member galaxies is difficult; tidal effects lower group densities, enlarge their sizes and velocity dispersions \citep[][]{2020MNRAS.498.3852B}, and strip their member galaxies following pericentric passage \citep[][]{2023MNRAS.518.1316H}. Furthermore, the identified groups are affected by contamination from interlopers associated with the cluster (\citetalias{Benavides23}). While the uncertainties were estimated by \citetalias{Benavides23} to be large when compared against simulations, they are difficult to quantify precisely. Hence we select the average of the two estimates as our best $M_{200}$ estimate. The radius $r_{200}$ of each substructure is derived from $M_{200}$.

We implemented the substructures in our lens model as circular halos, fixed to measured positions from \citetalias{Benavides23}. We assumed PIEMD profiles for consistency with other lens model components and fix their $r_{\rm core}$ to 50 kpc. We note that due to their being far-removed from the multiple images, the shape of the profile and the core radius will have little impact on multiple image positions. In this work, we set the $r_{\rm cut}$ to our $r_{200}$ estimates and fix the velocity dispersion \sigmalt of most substructures so that the mass enclosed in $r_{\rm cut}$ 2D aperture is equal to our average mass estimate $M_{200}$. We verified that the $r_{200}$ radii of the resulting PIEMD profiles matched the $r_{\rm cut}$ within $10\%$ and that using a different normalisation (i.e. normalising to 3D radius $r_{\rm cut}$) did not change the performance of the fiducial model.  Due to large mass uncertainties, we add the normalisation \sigmalt of substructures S1 and S7 as a free parameter so that their mass could vary between 0 and the upper limit of $M^{\sigma}_{200}$ from Table \ref{tab:substrmass}. The two substructures were selected, based on their proximity to the inner cluster regions (Fig. \ref{fig:kappamaps}), and their high estimated mass $M_{200}$, compared to S10. They are therefore expected to have a significant influence on multiple image deflection angles. Adding them with fixed parameters based on uncertain $M_{200}$ estimates may significantly bias the lens model.

Our resulting fiducial model includes seven substructure halos with a total of two free parameters, which equals the number of free parameters when adding the constant shear component with free strength and orientation. All parameter values of substructure PIEMD halos, including the prior ranges of S1 and S7 \sigmalt are given in Table \ref{tab:parameterpriors}. The substructures in our fiducial model are discussed in Section \ref{sec:discuss-substructures}.

\begin{table}[]
    \centering
    \renewcommand{\arraystretch}{1.1}
    \caption{Mass estimates using the spectroscopic velocity dispersions ($M_{200}^\sigma$) and richness ($M_{200}^r$) of the substructures from \citetalias{Benavides23} with their average $M_{200}$.   }
    \begin{tabular}{c c c c}
    ID & $M_{200}^\sigma$ & $M_{200}^r$ & $M_{200}$ \\
       
        & ($10^{13}M_{\odot}$) & ($10^{13}M_{\odot}$) & ($10^{13}M_{\odot}$) \\
\hline

 S1 & $56_{-33}^{+78}$ & 6 & 31\\
S2 & $7_{-4}^{+9}$ & 6 & 7 \\
S3 & $39_{-22}^{+48}$ & 6 & 23 \\
S5 & $15_{-9}^{+21}$ & 6 & 10 \\
S7 & $32_{-19}^{+44}$ & 6 & 19 \\
S9 & $23_{-11}^{+20}$ & 10 & 17 \\
S10 & $6_{-3}^{+6}$ & 8 & 7 \\
    \end{tabular}
    \tablefoot{The uncertainty of  $M_{200}^\sigma$ reflect only the uncertainty of measured $\sigma_{g}$ from \citetalias{Benavides23}, not taking into account any uncertainties in the $\sigma_g$ correction factor. The relative uncertainties in the estimated $M_{200}^r$ values are large (of the order of $\sim200\%$, based on the purity, completeness and total mass uncertainty, reported by \citetalias{Benavides23}).}
    \label{tab:substrmass}
\end{table}

\subsection{Model selection and optimisation}
The parameters of the mass distribution in \texttt{Lenstool} are optimised using the \texttt{BayeSys} \citep{skilling2004} Monte Carlo sampler. The $\chi^2$ used to compute the likelihood is defined as
\begin{equation}
  \chi^2= \sum_{i=1}^{N_{\rm im}} \chi_i^2=\sum_{i=1}^{N_{\rm im}} \frac{\lVert \pmb{r}_{i}^o - \pmb{r}_{i}^m \rVert^2}{\sigma_p^2},
\end{equation}
where the sum goes over all $N_{\rm im}$ multiple images and their individual contributions $\chi_i^2$.  $\pmb{r}_{i}^o $ represents the observed multiple image positions, and $\pmb{r}_{i}^m$ the model-predicted positions, which are derived by computing the barycenter of the source-plane positions of multiple images of each system, weighted by their magnifications \citep[e.g.][]{2021MNRAS.506.2002B}, and then lensing it back to the image plane. For the position uncertainty, $\sigma_p$, we use $0.5''$. The adopted uncertainty is significantly higher than the true position uncertainty derived from JWST images ($\sim20~\mathrm{mas}$). However, such precision is not achievable by the current lens modelling methods, due to the complexity of the cluster mass distribution and the contribution from the structures distributed along the line-of-sight \citep[e.g.][]{2010Sci...329..924J}. Using the true position uncertainty would result in a very low statistical uncertainty of the mass distribution, not accounting for systematic biases due to insufficient model complexity.  Our increased $\sigma_p$ does not influence the best-fit mass model, however, it provides more conservative parameter uncertainties. Our adopted $0.5''$ uncertainty results in best-fit $\chi^2$ value lower, but of the same order of magnitude, as the number of degrees of freedom (DoF) of our fiducial model:
\begin{equation}
    \mathrm{DoF}=2N_{\rm im}-2N_{\rm sys}-N_{\rm par},
\end{equation}
where the number of multiple images $N_{\rm im}=135$, the number of lensed systems $N_{\rm sys}=49$ and the number of free parameters in the fiducial model $N_{\rm par}=32$. In this work, we used two modes of optimisation. For most models that we tested, we used the standard \texttt{BayeSys} sampling, as described in \cite{Jullo07}, to fully sample the parameter space, derive the parameter uncertainties, and calculate the Bayesian evidence. We sampled the parameter space with 40\,000 samples,  following at least 16\,000 burn-in samples - with \texttt{Lenstool} rate parameter, governing the length of the burn-in phase \citep[][]{Jullo07} set to 0.02 or lower to ensure that the models reached a stable $\chi^2$ distribution before transitioning to the sampling phase. For our best-fit model, we also carried out optimisation by allowing the cooling factor in the selective annealing technique (see \citealt{Jullo07}) to increase above 1. This mode is intended for finding the true minimum of $\chi^2$ and is used to identify the best-fit model and its $\chi^2$ and root-mean-square (rms) separation between the observed and predicted lensed image positions, $\Delta_{\rm rms}$ (see equation \ref{eq:deltarms}). Our fiducial model parametrisation was selected by comparing several parametrisations across different metrics, including the Bayesian evidence derived by \texttt{Lenstool} \citep{Jullo07}. We note that the evidence between different runs of the same model parametrisation (e.g. when doubling the Lenstool rate parameter) showed deviations of $\Delta\log E\sim5$. We also quote the approximation of the Bayesian evidence, which depends only on the $\chi^2$ minimum - Bayesian Information Criterion (BIC) \citep{1978AnSta...6..461S}:
\begin{equation}
    {\rm BIC}= \chi^2 + N_{\rm par} \ln N_{\rm im}.
\end{equation}
The accuracy of the lens model is evaluated by computing the rms distance from the observed and best-fit model predicted positions of multiple
images:

\begin{equation}
    \Delta_{\rm rms}=\sqrt{\frac{1}{N_{\rm im}} \sum_{i=1}^{N_{\rm im}} \lVert \pmb{r}_{i}^o - \pmb{r}_{i}^m  \rVert^2}.
    \label{eq:deltarms}
\end{equation}
We tested several models, varying the number of main cluster halos, their prior ranges, and the number of galaxies we model outside of the scaling relations. Our selected fiducial model, described above, includes 32 free parameters and achieves the image position accuracy $\Delta_{\rm rms}=0.41''$ with $\chi^2/\mathrm{DoF=0.7}$. While above the positional uncertainty, this level of precision is expected for \texttt{Lenstool} models and is higher but comparable to that achieved by the CANUCS lensing models ($0.41''$, \citealt{rihtarsic25}, $0.56''$, \citealt{2024ApJ...973...77G}). Other model metrics are provided in Table \ref{tab:modelompare}, where they are compared to a few selected parametrisations, discussed in Section \ref{sec:results}. We provide the fixed parameter values and the prior ranges in Table \ref{tab:parameterpriors} and the best-fit values with uncertainties in Table \ref{tab:bestfitparam}.

\section{Results and discussion }
\label{sec:results}

\subsection{Main cluster and subcluster}

\label{sec:discuss_parametrisation}

\begin{table}[]
 \centering
\caption{Comparison of the performance of the various model parametrisations discussed in this work.}
\begin{tabular}{c | c c c c}
Model & $\chi^2$&$\Delta_{\rm rms}('')$ & BIC & $\log E$ \\
\hline
fiducial &  102 (92) & 0.43 (0.41) & 267 (257) & -205  \\
\hline

no H4 & 125 & 0.48 & 269 & -214 \\
no substructures & 137 & 0.50 & 292 & -213 \\

only S1 \& S7 & 120 & 0.47 & 285 & -215 \\
constant shear & 86 & 0.40 & 250 & -207 \\

no X-ray & 100 & 0.43 & 265 & -209 \\

\end{tabular}
\tablefoot{For all models, the values are derived using the full Bayesian sampling of parameter space with 40 000 samples for a straightforward comparison. For the fiducial model, we also provide the results from the optimisation, intended to find the precise minimum of $\chi^2$, in the parentheses.}
\label{tab:modelompare}
\end{table}

\begin{figure*}
    \centering
    \includegraphics[width=\linewidth]{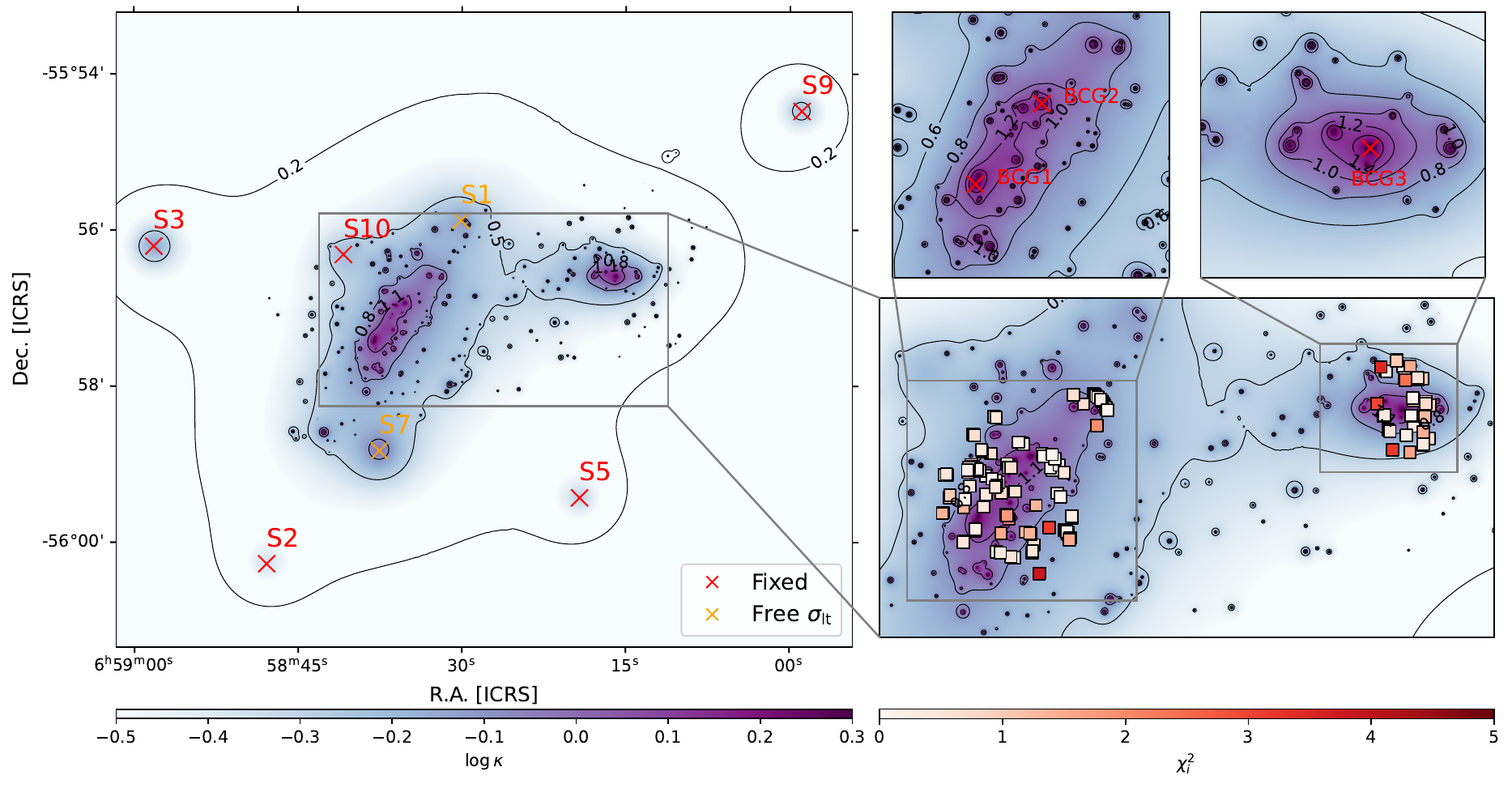}
    \caption{\textit{Left} panel: Convergence (total density) from our best-fit fiducial model, covering a large FOV with indicated positions of substructure halos with fixed (red) or free (orange) normalisation \sigmalt. The contour values indicate the values of $\kappa$ while the colourmap shows $\log \kappa$. \textit{Lower right} panel: Convergence map of the inner regions with multiple images. The colour of multiple images represents their contribution to the goodness of fit $\chi^2$. \textit{Upper right} panel: Convergence map of the main cluster (left) and the subcluster (right) with indicated BCG positions.} 
    \label{fig:kappamaps}
\end{figure*}

In Fig. \ref{fig:kappamaps}, we show the mass distribution derived from our best-fit mass model together with the $\chi^2_i$ contributions of individual multiple images. The best-fit positions of the mass components are also indicated in Fig. \ref{fig:model_clustermembers} together with other mass components, overlaid over the F277W NIRCam image. 

The main cluster displays a complex mass distribution, with halos H1 and H2  loosely associated with the corresponding BCGs: the center of H1 is $49_{-13}^{+1}$  kpc away from BCG1, and H2 $66_{-0}^{+46}$ kpc away from BCG2, where the quoted values correspond to the distance derived from the best-fit model, and the uncertainties represent the $68\%$ confidence interval from the model samples. The distribution of halos in the main cluster, displaced from the BCGs, suggests a complex,  asymmetric mass distribution. Due to the degeneracy between the masses of the main halos and the BCGs, we caution against overinterpreting the displacement in the main cluster and note that the mass distribution should be considered in total. The mass map in Fig. \ref{fig:kappamaps} shows two clear mass peaks at the BCGs with an elevated density between. The H1 halo, whose best-fit position lies between the BCGs (Fig.~\ref{fig:model_clustermembers}) and which displays high ellipticity (Table \ref{tab:bestfitparam}), acts as a bridge between the two components. Halo H2, situated to the northeast between the BCG2 and galaxies G5 and G9, serves as a lower-density tail of the main cluster, oriented towards the northeastern direction of the hypothesised merger axis traced by the two BCGs and substructures. We further add the H4 cluster halo at a fixed position centred on the galaxies to the south of the cluster (Fig. \ref{fig:model_clustermembers}). The halo serves as an elongation of the mass distribution to the south of the BCG1 and is motivated by the extension of the ICL (see Fig. \ref{fig:model_clustermembers}), and an improvement in the BIC and $\Delta_{\rm rms}$ (Table \ref{tab:modelompare}). 
Overall, despite the limitations of the parametric lens modelling technique, our fiducial model achieved a good overall description of multiple image positions (see $\chi^2_i$ in Fig. \ref{fig:kappamaps}). Our model of the main cluster exhibits a highly elongated, curved mass distribution with a downturn in the south-east (e.g. see Fig. \ref{fig:kappamaps} and the critical curve in Fig. \ref{fig:field}), roughly following the ICL, with two mass peaks at the positions of the two BCGs. 

We note that H1 and H4 approach the ellipticity limit of 0.7 (Table \ref{tab:bestfitparam}), which was imposed to avoid highly elongated individual cluster components. This limit is sometimes adopted in cluster lens modelling studies \cite[e.g.][]{2022A&A...664A..90L}, and is motivated by the shapes of relaxed, isolated halos in simulations \citep{2017MNRAS.466..181D}, different from the Bullet Cluster. Therefore, we briefly investigated the impact of the adopted prior range by rerunning our fiducial model with an increased ellipticity upper limit of 0.9. The resulting lens model (run in the standard Bayesian sampling mode), provided a small increase in $\chi^2$ (96), and resulted in a $\kappa$ map, similar to that of the fiducial model, within the model uncertainty over most of the multiple-image region. In particular, the best-fit ellipticity of the main halo H1 remains close to our fiducial limit (0.74), however, halo H4 becomes very elongated, with its best-fit ellipticity at the new prior limit (0.89). We conclude that the 0.7 ellipticity limit in our fiducial model does not have a drastic impact on our results and note that the total ellipticity of the main cluster, consisting of three large-scale halos, can exceed the value of individual components.

In contrast to the approach taken for the main cluster, we went on to model the subcluster with a simpler mass distribution, consisting of the large-scale cluster halo H3 and the subcluster BCG (BCG3). All model parameters of the subcluster are well within the prior limits, including ellipticity, and the position of the H3 halo coincides well with the BCG3, with a BCG3-H3 distance of only $4_{-2}^{+3}$ kpc. When projecting the position of H3 along the axis connecting the subcluster gas peak and BCG3, we obtain an offset of $-4\pm3$ kpc with a $95\%$ confidence interval of ($-9$, $1$) kpc, and $-0.1_{-0.8}^{+0.7}$ kpc in the perpendicular direction. Halo H3 thus lies on the merger axis, with its best-fit position lagging behind BCG3 by a few kpc, but remaining consistent with it within $2\sigma$. For a direct comparison of the improvement in uncertainty brought on by the newly discovered systems and redshifts, we re-optimised our fiducial model with pre-JWST multiple image systems only (obtaining the displacement of $9_{-1}^{+14}$ kpc), and by including the new systems while leaving their redshifts as free parameters (obtaining the displacement of $7.01_{-5}^{+0.01}$ kpc). Increasing the number of systems led to a threefold improvement in precision and a systematic shift to lower displacement values. Although increasing the number of spectroscopic redshifts alone did not improve precision, it lowered the best-fit value from 7 to 4 kpc. We note that the decrease of the halo position uncertainty upon the addition of JWST systems has potentially important implications for constraints on the DM self-interaction cross-section, which scales with the expected displacement between DM and galaxies \citep[e.g.][]{Robertson17}. However, we caution against overinterpretation of the measured displacement. First, we do not break the degeneracy between the masses of halo H3 and BCG3, both of which are modelled with several free parameters. This degeneracy could be lifted in future work by spectroscopically constraining the velocity dispersion of the BCG \citep[e.g.][]{Bergamini19,2024MNRAS.527.3246B}. Second, \cite{Robertson17} demonstrated that the inferred galaxy–DM displacement depends on the aperture used to define the DM centre and is largely driven by asymmetries in the mass distribution, while the DM peaks themselves are expected to remain coincident even in SIDM scenarios. Our simplified parametrisation, which assumes a symmetric large-scale halo, lacks the flexibility to capture self-interaction–induced asymmetries. A robust interpretation of our displacement will therefore require reproducing our strong-lensing analysis on the simulations of the Bullet Cluster.

\begin{table}[]
    \centering
    \renewcommand{\arraystretch}{1.05}
    \caption{Gas fractions in circular (250 kpc) apertures with different centres and in the whole NIRCam FOV.}
    \begin{tabular}{c c}
    Aperture & Gas fraction \\
    \hline
    BCG1 (250 kpc) & 0.10 \\
    BCG2 (250 kpc) & 0.12 \\
    BCG3 (250 kpc) & 0.10 \\
    Main cluster gas (250 kpc) & 0.16\\
    Subcluster gas (250 kpc) & 0.15\\
    \hline
    NIRCam FOV & 0.13 \\
    
    \end{tabular}
    \tablefoot{Centre of apertures, centred on the two gas components, were obtained from Table 2 in \cite{clowe06}. The NIRCam FOV represents the rectangular aperture ($5.9\times2.6$ arcmin) and is shown in Fig. \ref{fig:model_clustermembers}. The uncertainty on the fraction is dominated by the gas mass uncertainty ($\sim10\%$), which we do not consider in the lens model.
    }
    \label{tab:gasfraction}
\end{table}

In Table \ref{tab:gasfraction}, we show the gas mass fraction in 250 kpc apertures centred in different regions of the Bullet Cluster. The fraction ranges between $10\%$ and $16\%$ depending on the region, as expected due to the displacement between dark matter and gas. The gas fraction in the BCG1-centred aperture is consistent with the one reported in \cite{Paraficz16}. The fraction within the full NIRCam FOV, encompassing both cluster components and their respective gas peaks, is $13\%$. The baryon fraction contained in the intracluster medium is lower than the universal baryon fraction (15.6\%, \citealt{2020A&A...641A...6P}), within the range of expected values in the inner regions of galaxy clusters \citep[e.g.][]{2006ApJ...640..691V,2007ApJ...666..147G}. To assess its impact on our lens model, we tested a model without the gas component (see model "no X-ray" in Table \ref{tab:modelompare}). In contrast to previous studies \citep{Paraficz16}, the addition of the gas component does not improve the fit and results in a similar $\Delta_{\rm rms}$. The gas furthermore does not have a drastic influence on the halo displacement. Removing the gas completely resulted in an H3–BCG3 offset of $10.6_{-5.6}^{+0.2}$ kpc, which is larger but consistent with our fiducial model.

\subsection{Substructures and constant shear}
\label{sec:discuss-substructures}

\begin{figure}
    \centering
    \includegraphics[width=\linewidth]{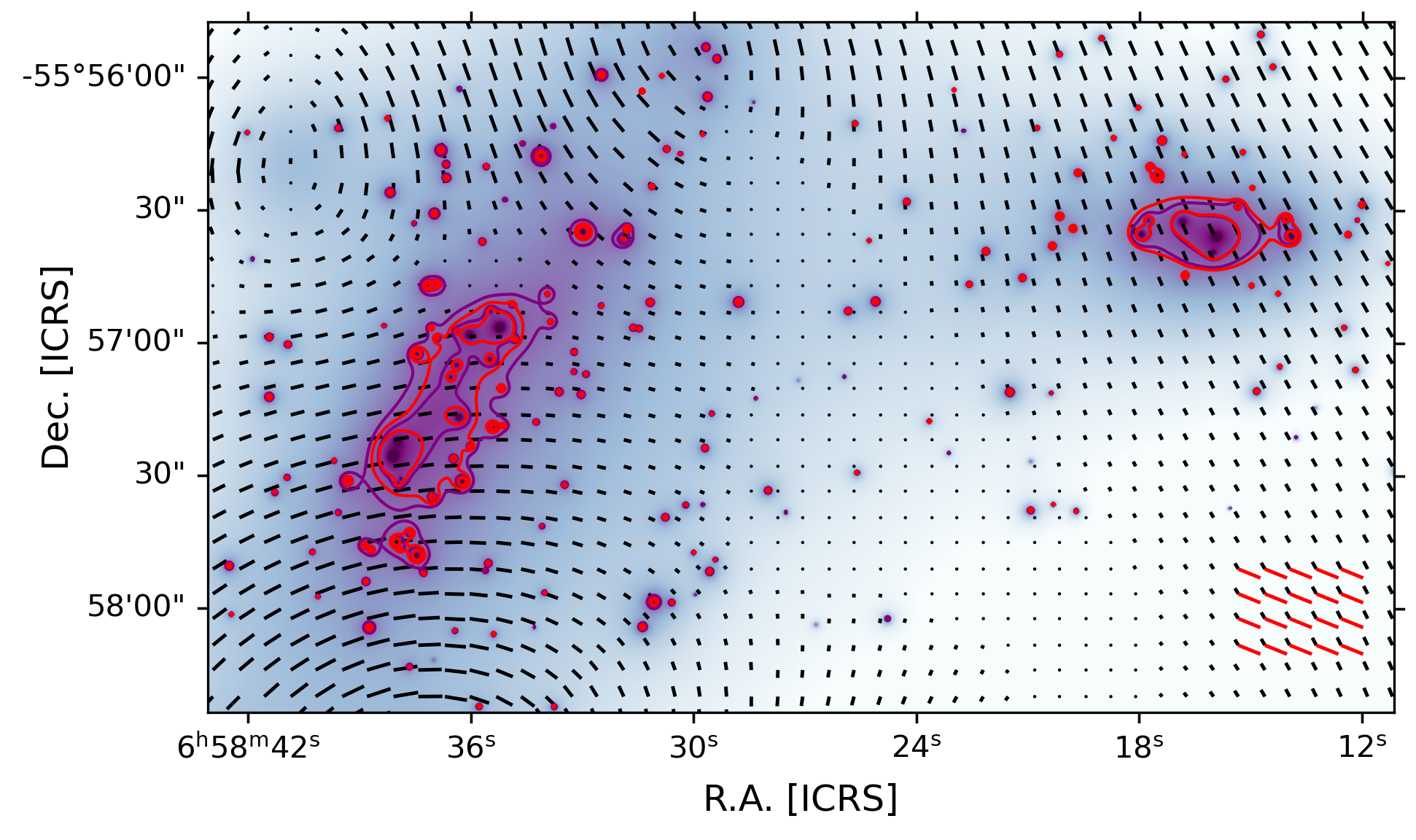}
    \caption{Density, $\kappa$, overplotted with the line field indicating the lensing shear, contributed by the group-scale substructures on the cluster outskirts. The lines indicate the shear orientation, and their length is proportional to the strength of the shear in each grid point. In the bottom right corner, we also indicate the strength (using the same $D_{\rm ls}/D_{\rm s}$ normalization), and direction of the constant shear model, where the substructures are replaced by a constant shear component with free strength and orientation. The contours represent the $\kappa$ values of 1 and 1.2 from the fiducial (purple) and the constant shear models (red).   } 
    \label{fig:shear}
\end{figure}

Our fiducial lens model includes several group-scale substructures surrounding the inner cluster regions, with their parameters based on velocity dispersion measurements from \citetalias{Benavides23} (see section \ref{sec:model-substructures}). Although most halos (apart from the two with free normalisation) are included with fixed parameters based on highly uncertain values (see Table \ref{tab:substrmass}) derived from spectroscopic measurements, the substructures serve as an important element in our model, decreasing the $\Delta_{\rm rms}$ of multiple images in the inner regions from $0.5''$ to $0.43''$ (see model labelled 'no substructures' in Table \ref{tab:modelompare}). In our fiducial model, we include two substructures, S1 and S7 (see Fig. \ref{fig:kappamaps}), which were selected on the basis of their mass estimates and distances, with free normalisation \sigmalt. To determine whether the improvement in the model is driven solely by the two substructures, we ran a model excluding all the other substructures (model "only S1 \& S7"). Excluding the fixed substructure halos marginally degraded the model quality across all metrics. Their addition with fixed parameters, despite their uncertainty, provides an improvement in the $\Delta_{\rm rms}$ of multiple images without introducing additional free parameters.

In addition, we compared our model with one in which the substructure component was replaced with constant lensing shear with free strength and orientation. On galaxy scales, constant shear is frequently invoked to account for the unknown mass components in the vicinity of the lens and along the line of sight, as well as for complexity in the lens itself \citep[e.g.][]{2024MNRAS.531.3684E}. It has also been successfully applied to account for the large-scale environment on cluster scales \citep[e.g.][]{2018MNRAS.473..663M}. Our "constant shear" model, with strength $\gamma=0.12\pm0.01$ at $D_{\rm ls}/D_{\rm s}=1$, and orientation $\theta=158_{-1}^{+6}$ counterclockwise from the west, provides a better fit to our data, decreasing the $\Delta_{\rm rms}$ from 0.43 to 0.40 arcseconds (see Table \ref{tab:modelompare}), although providing no increase in evidence. In Fig. \ref{fig:shear}, we show the strength and orientation of the shear due to the substructures, compared to the constant shear model. Because their distances from the inner cluster regions are comparable to the overall extent of the system, the substructures produce a spatially varying shear with differing strengths and orientations across the multiply imaged region. While the constant shear model yields a representative orientation (see the red arrows in Fig. \ref{fig:shear}), it does not capture those variations.  Comparing the mass distributions of our fiducial and constant shear models (contours in Fig. \ref{fig:shear}), we find that the constant shear model underpredicts mass in the main cluster, which is expected, since the shear component is not associated with any mass. The substructures, on the other hand, can contribute with their density profile even in the inner cluster regions. A similar mass difference was also reported by \citet{2018MNRAS.473..663M} when comparing substructure and constant shear models in Abell 2744. Because the substructures have a clearer physical interpretation and are associated with mass components, we favour them over constant shear in our fiducial model. However, we note that the higher amplitude of constant shear, compared to the overall strength, contributed by the substructures, as well as the better $\Delta_{\rm rms}$, suggests that the substructure parameters should be refined in future studies. We again note the high uncertainty of the substructure parameters, which are not considered in our lens model (section \ref{sec:model-substructures}). The shear could be better constrained, for example, through weak-lensing analyses using JWST data targeting substructure candidates, or with Euclid or Roman observations at larger radii. In addition, constant shear could be compensating for contributions not captured by our model parametrisation, such as the line-of-sight structure, the large-scale environment, or the complexity of the Bullet Cluster mass distribution itself.

Although the introduced substructures improve the model quality, we note that, as they lie beyond the regions with multiple images, they should be regarded primarily as tools for refining the mass distribution within the inner regions — well inside the NIRCam FOV — rather than as secure confirmation of their individual properties. Nevertheless, our analysis highlights the impact of accounting for the mass distributed along the two merger axes in the Bullet Cluster, which are supported by cluster-member kinematics (\citetalias{Benavides23}), weak-lensing constraints \citep[][]{cho25}, and radio relic candidates \citep[][]{2015MNRAS.449.1486S,2023MNRAS.518.4595S}. The importance of including the substructures on the outskirts, which influence multiple image positions through external shear, has also been reported in other observed \citep[e.g.][]{2018MNRAS.473..663M} and simulated \citep[][]{2017MNRAS.470.1809A} clusters.

\subsection{Lens model comparison}

\label{sec:modelcomparisson}

\begin{figure}
    \centering
    \includegraphics[width=\linewidth]{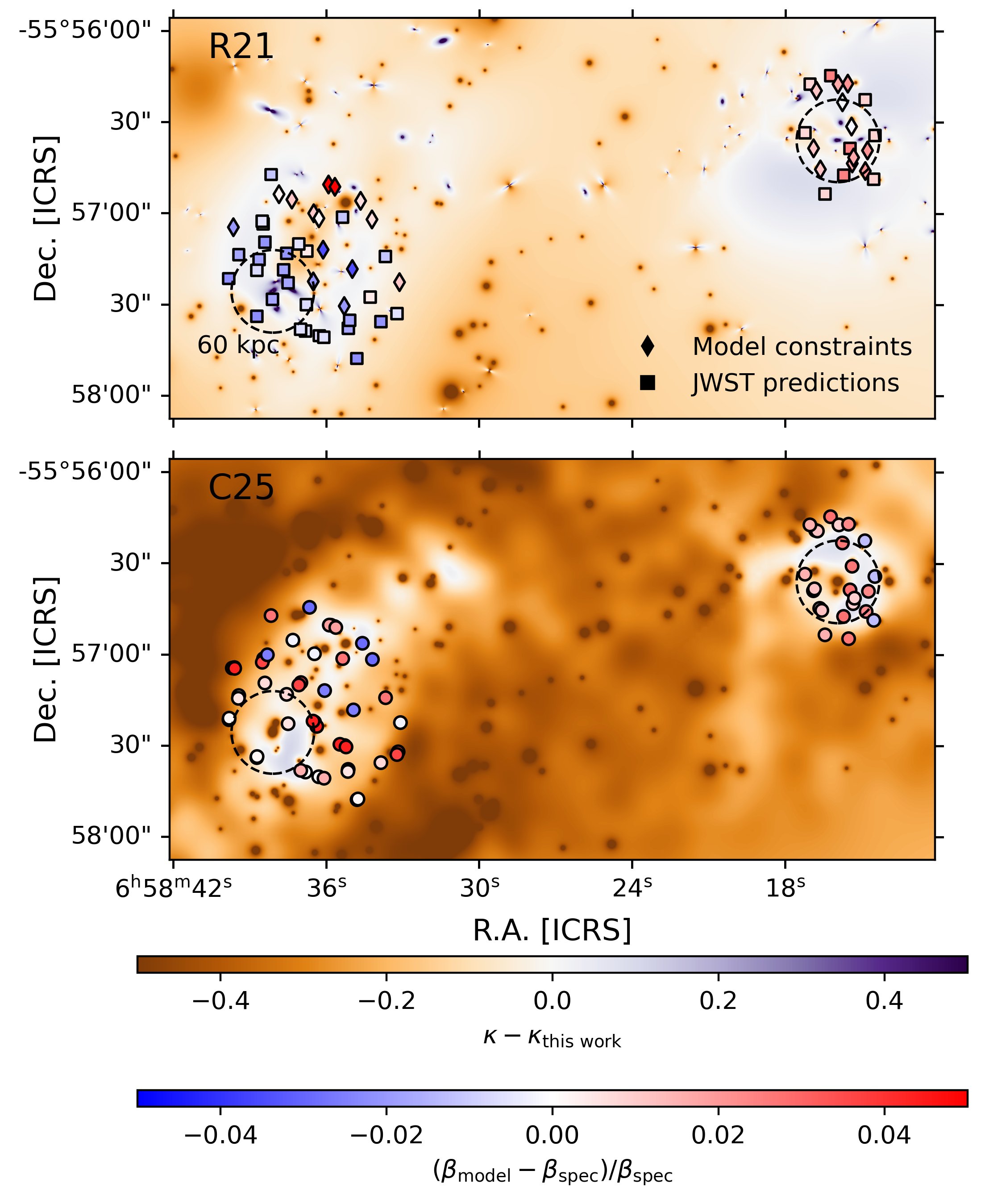}
    \caption{$\kappa$ map differences between our model and the lens models of \citetalias{Richard21} (upper panel) and \citetalias{cha25} (lower panel), overplotted with multiple images with newly obtained NIRSpec redshifts. The colours of multiple images indicate the relative difference between the model-predicted and spectroscopic lensing efficiencies $(\beta_{\rm model}-\beta_{\rm spec})/\beta_{\rm spec}$, which scale with redshift offsets (see also relative $\beta$ differences in Fig. \ref{fig:redshiftcompare}). In the first panel, we plot multiple images used in the \citetalias{Richard21} model without $z_{\rm spec}$ (diamonds) and \citetalias{Richard21} best-fit model predictions for JWST-discovered systems (squares), showing a single clump per galaxy. In the second panel, we show the \citetalias{cha25} systems without $z_{\rm spec}$. The dashed 60 kpc radius circle marks the region where the enclosed masses show good agreement between models (Figure \ref{fig:cumulativemass}).   } 
    \label{fig:redshift_spacialdiffs}
\end{figure}

\begin{figure}
    \centering
    \includegraphics[width=\linewidth]{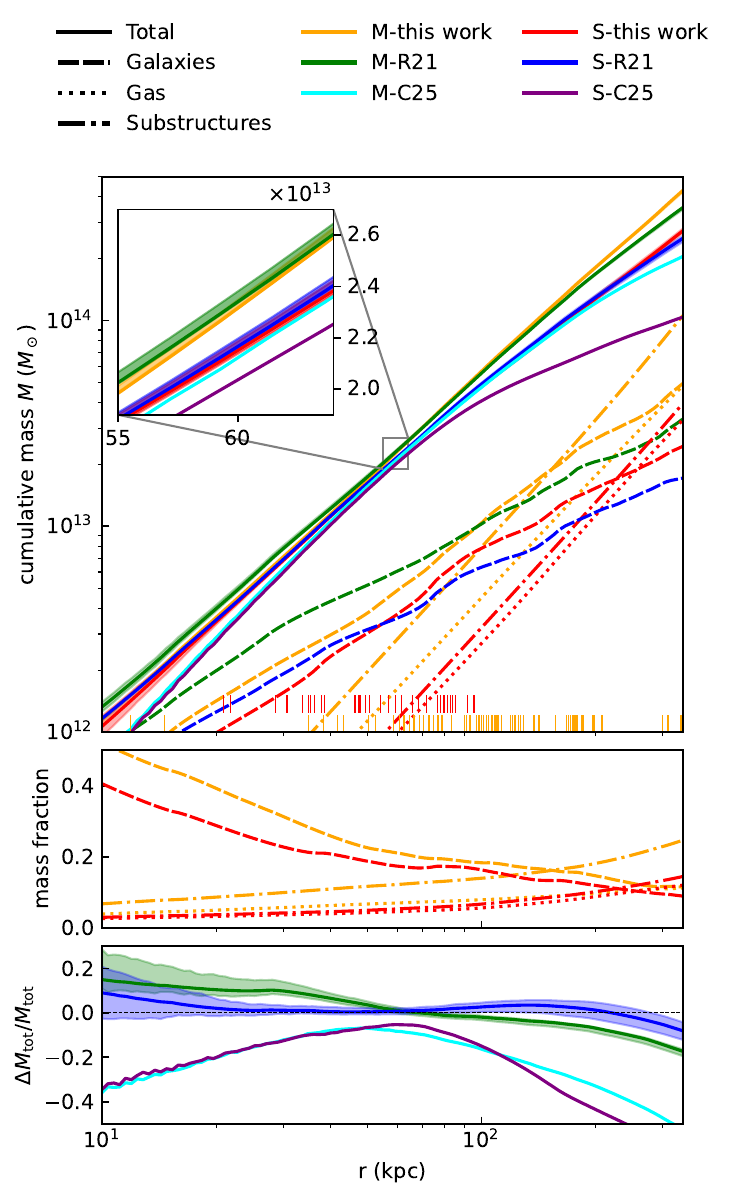}
    \caption{\textit{Top}: Cumulative mass in the 2D aperture $r$ as a function of the distance from BCG1 in the main cluster (labelled as M) and the BCG3 in the subcluster (labelled as S) in kpc. The plot shows the mass distributions in \citetalias{Richard21}, \citetalias{cha25} and our fiducial model. We show the total mass profiles of the cluster and the subcluster with uncertainties (where available), as well as the best-fit profiles of individual mass components (cluster gas, galaxies and group-scale substructures). \textit{Middle}: Fraction of each mass component relative to the total mass of the fiducial model as a function of radius. \textit{Bottom}: Relative mass difference, $(M-M_{\rm this\ work})/M_{\rm this\ work}$, between the models of \citetalias{Richard21} and \citetalias{cha25} and our fiducial lens model.} 
    \label{fig:cumulativemass}
\end{figure}

In this section, we compare our fiducial model with the lens models presented in \citetalias{Richard21} and \citetalias{cha25}. The lens models differ in the datasets, as well as in the lens modelling codes and strategies. \citetalias{Richard21} used a catalogue containing 40 multiple images from 15 systems, 5 of which had spectroscopic redshifts. The lens-modelling strategy in \citetalias{Richard21} is broadly similar to ours: both analyses employ the parametric lens-modelling code \texttt{Lenstool} and adopt comparable parametrisations, assigning PIEMD large-scale halos to the three BCGs. The \citetalias{cha25} model, on the other hand, utilises a larger dataset of 146 multiple images, derived from NIRCam imaging, although only six systems have spectroscopic redshifts. The \citetalias{cha25} lens model is constrained using a free-form reconstruction code MARS \citep[][]{2022ApJ...931..127C,2024ApJ...961..186C}. Unlike \citetalias{Richard21} and our strong-lensing only models, \citetalias{cha25} also incorporates the weak lensing information from NIRCam imaging data. Compared to \citetalias{cha25}, our catalogue of lens model constraints contains more than four times more spectroscopic system redshifts (27) than \citetalias{cha25}. On the other hand, we have used a comparable total number of multiple images, as we included only the most secure spectroscopic systems. The general features of the mass distribution between the three models share several similarities. For instance, we found the main cluster to be composed of two mass peaks, with elevated density in between, accounted for by halo H1 (Fig. \ref{fig:kappamaps}). This behaviour is consistent with the findings of \citetalias{cha25}, which report a higher density between the two BCGs, than expected from a superposition of two BCG-centred cluster halos. We further find that the mass distribution extends towards the north-east from BCG2, in agreement with the \citetalias{cha25}. We also note that our measured H3-BCG3 displacement in the subcluster ($4_{-2}^{+3}$ kpc) is consistent with the offset found by \citetalias{cha25} using the first moment of convergence ($4.09 \pm  0.63$ kpc using strong-lensing-only maps), although the two measurements are not strictly equivalent, and the displacement in \citetalias{cha25} increases when including weak lensing ($17.78 \pm 0.66$ kpc).  

Nevertheless, our model differs from previous models in several respects, which are compared in this section. In Fig. \ref{fig:redshift_spacialdiffs}, we show the difference in convergence $\kappa$ maps between our model and the models of \citetalias{Richard21} and \citetalias{cha25}. In Fig. \ref{fig:cumulativemass}, we plot the mass enclosed within the aperture with radius $r$, centred on BCG1 and BCG3. We plot the total mass profiles and their relative differences, along with the profiles of the individual mass components. At small scales, the most prominent differences among the three lens models arise from their different treatment of cluster member galaxies. The free-form model by \citetalias{cha25} does not allow for the small-scale variations and does not include cluster member subhalos. This is reflected in the large mass differences observed in the vicinity of the BCGs (Fig. \ref{fig:cumulativemass}). While both \citetalias{Richard21} and our model include cluster member galaxies, our model includes approximately twice as many across the NIRCam field of view, with the selected catalogue differing slightly between the two models. \citetalias{Richard21} furthermore models cluster members with elliptical halos, whereas we adopt circular halos, resulting in quadrupole residuals in Fig. \ref{fig:redshift_spacialdiffs}.

As expected, the large-scale differences between the three best-fit lens models are the smallest in the region well constrained by multiple images, with the aperture mass showing the closest agreement at $\sim 60\ \mathrm{kpc}$ from BCG1 and BCG3 (see Fig. \ref{fig:cumulativemass} and the indicated aperture in Fig. \ref{fig:redshift_spacialdiffs}). The 60 kpc aperture mass around BCG1 (BCG3) in the \citetalias{cha25} model is 8\% (4\%) lower than in our best-fit fiducial model. In comparison, the differences relative to \citetalias{Richard21} are smaller, with the mass enclosed within 60 kpc only 1.1\% (0.7\%) higher than in our fiducial model around BCG1 (BCG3). In the subcluster in particular, the \citetalias{Richard21} model is consistent with our fiducial model over a wide range of radii, despite being constrained by only a single previously known spectroscopic redshift. We attribute the overall agreement between the \citetalias{Richard21} and our lens models to the use of the same lens-modelling code and a similar parametrisation. Conversely, the \citetalias{cha25} model contains a smaller aperture mass than our fiducial model at all radii. 

Despite the relatively good agreement of the aperture mass profiles at radii constrained by multiple images, our model redistributes mass in the inner regions compared to previous models. (see Fig. \ref{fig:cumulativemass}). Compared to \citetalias{Richard21}, our model assigns more mass to the northern region of the main cluster and less mass to the southern part, where very few pre-JWST multiple images were available (see Fig. \ref{fig:field}). In the subcluster, we find a more elongated mass distribution along the east–west direction than in the \citetalias{Richard21} model, which instead assigns more mass to the northern and southern sides of the inner region containing multiple images. Compared to the \citetalias{cha25} model, which does not include cluster member galaxies, our model places more mass in the BCGs and cluster member halos rather than distributing it smoothly between them.

On the outskirts, beyond the radius that contains multiple images, the disagreements between \citetalias{cha25} and the two \texttt{Lenstool} models increase. This can first be attributed to the different modelling strategies. Whereas \citetalias{Richard21} and our model both impose the $r^{-2}$ profile between the $r_{\rm core}$ and $r_{\rm cut}$ radii (from PIEMD profile, see equation \ref{eq:piemdprofile}), \citetalias{cha25} employs a more flexible grid-based model where the profile can fall off much more rapidly on the outskirts. The second reason is constraints. While \citetalias{cha25} model includes weak lensing constraints to model the region outside the area containing multiple images, the \texttt{Lenstool} models rely only on multiple image positions. Beyond the inner region, locally constrained by multiple images in the subcluster or the main cluster, the constraints are provided only by multiple images in the other cluster component. Thus, extrapolation of the PIEMD profiles in \texttt{Lenstool} strong-lensing-only models beyond the radii containing multiple images should not be considered reliable, and any meaningful comparison of the three lens models should primarily be restricted to the inner regions, where the models show better agreement.

\begin{figure}
    \centering
    \includegraphics[width=\linewidth]{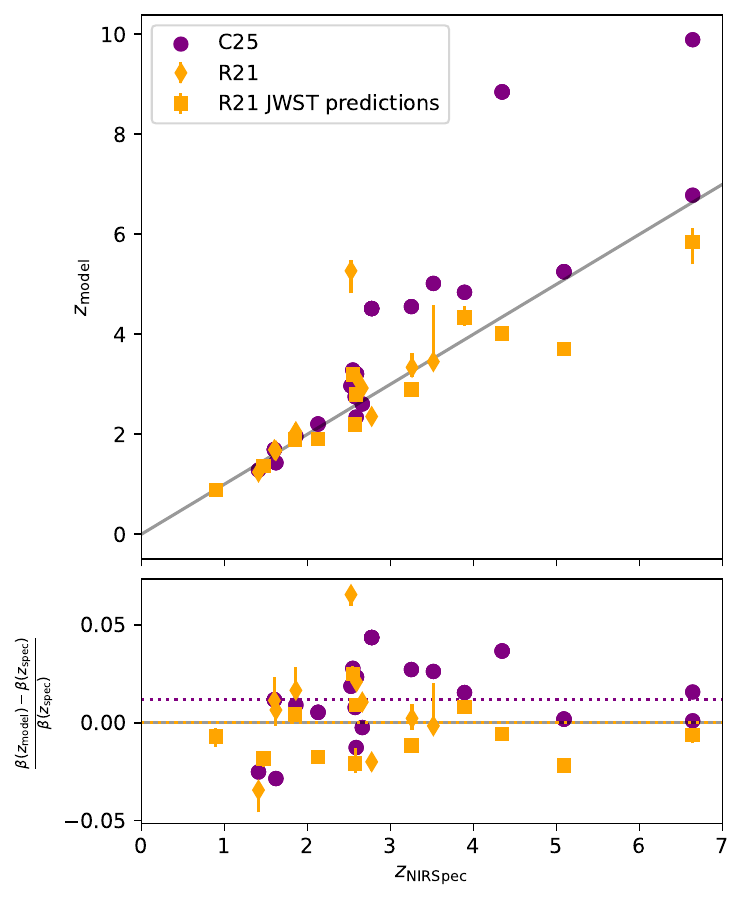}
    \caption{\textit{Top}: Comparison between the NIRSpec spectroscopic redshifts and the lens-model predicted redshifts from \citetalias{Richard21} and \citetalias{cha25} as a function of spectroscopic redshifts. For \citetalias{Richard21}, we plot the pre-JWST systems used for model optimisation (diamonds) with statistical uncertainties, as well as the best-fit model predictions of JWST-discovered systems that were not used to optimise the model (squares). The solid grey line indicates where $z_{\rm model}=z_{\rm NIRSpec}$. \textit{Bottom}: Redshift differences, plotted as a relative difference in lensing efficiencies $\beta$. For each model, we indicate the median relative difference with a dotted line.  } 
    \label{fig:redshiftcompare}
\end{figure}

Both \citetalias{Richard21} and \citetalias{cha25} lens models included all but a few multiple image systems, which had pre-JWST spectroscopic redshift measurements, as free parameters. In Fig. \ref{fig:redshiftcompare}, we show the comparison between their model-predicted and the spectroscopic redshifts presented in this work. In addition to redshifts, we also examine the relative differences in the corresponding lensing efficiencies,
\begin{equation}
    \beta={D_{\rm ls}/D_{\rm s}},
    \label{eq:betadef}
\end{equation}
 where $D_{\rm ls}$ and $D_{\rm s}$ denote the angular diameter distances between the lens and the source, and the observer and the source, respectively. Lensing efficiencies are directly proportional to deflection angles, which govern multiple image positions; thus, they provide insight into how redshift mismatches translate into model disagreements. For example, for an image with a deflection angle of $30''$ (a typical deflection angle of multiple images in our lens model), the relative $\beta$ mismatch of 0.02 (which is a typical mismatch observed in Fig. \ref{fig:redshiftcompare}) translates into $0.6''$ difference in the image positions (higher than our adopted position uncertainty), given the same mass distribution. The redshift deviations in Fig.~\ref{fig:redshiftcompare} increase at high redshift, as also reported in previous studies (e.g. \citealt{2016A&A...587A..80C}); however, this behaviour arises from the highly non-linear scaling of $\beta$ with redshift. When expressed in terms of $\beta$ differences, the deviations remain relatively uniform. For both the \citetalias{Richard21} and \citetalias{cha25} models, the predicted $\beta$ deviations from the new spectroscopic redshifts in Fig. \ref{fig:redshiftcompare} are small. In particular, we highlight the predictive power of the pre-JWST \citetalias{Richard21} model in reproducing the redshifts of JWST systems not included as lensing constraints, but which we reoptimised using the best-fit \citetalias{Richard21} model. Nonetheless, \citetalias{Richard21} systems exhibit greater disagreement than lens-model redshift uncertainties, and the overall deviations of both models exceed both the assumed $\sigma_p$ and the $\Delta_{\rm rms}$ of our model, highlighting the utility of the newly acquired spectroscopic redshift measurements from NIRSpec.

Figure \ref{fig:redshift_spacialdiffs} furthermore shows that the relative $\beta$ differences exhibit spatially coherent offsets. In the southern part of the main cluster, the redshifts of new JWST systems are clearly underpredicted when refitted using \citetalias{Richard21} model. By contrast, they are overpredicted by the \citetalias{cha25} model, which includes them as constraints. In the subcluster, where both studies previously relied on only a single spectroscopic redshift, both \citetalias{cha25} and \citetalias{Richard21} show a slight tendency to overpredict redshifts. On average, the $\beta$ differences in \citetalias{Richard21} model show no systematic shift, while  \citetalias{cha25} shows a small systematic offset toward higher predicted redshifts (Fig. \ref{fig:redshiftcompare}). 

We note that both $\beta$ and the mass distribution influence the deflection angles; systematic offsets in the predicted $\beta$ may therefore reflect inaccuracies in the mass model in those regions. Previous lensing studies have reported the tendency to overpredict redshifts, when few spectroscopic redshifts are available \cite[e.g.][]{2016ApJ...832...82J,2016A&A...587A..80C,2021ApJ...910..146R}. Such trends are usually associated with an underestimated cluster mass due to the mass-redshift degeneracy. In the ideal case of identical relative density distributions, the mass normalisation is inversely proportional to $\beta$ of the multiple-image systems. We observe similar anticorrelation in parts of the cluster. For instance, in the southern region of the main cluster, the $1-2\%$ mass overestimation by \citetalias{Richard21} (Fig. \ref{fig:cumulativemass}) is matched by a similar relative underestimation of $\beta$ (Fig. \ref{fig:redshift_spacialdiffs}). However, in general, the relation between $\beta$ and density deviations between the three models is not straightforward, given their substantially different relative density distributions. In the subcluster, both \citetalias{Richard21} and \citetalias{cha25} overestimate $\beta$ by a small amount ($\sim 2\%$), while showing vastly different levels of agreement with our model. While \citetalias{cha25} underestimates the mass by at least $4\%$, the \citetalias{Richard21} mass shows much better agreement with ours and is even slightly higher at some radii (Fig. \ref{fig:cumulativemass}). We again note that the three lens models differ in several respects, including lens-modelling approaches and parametrisations. We particularly note the relative redistribution of mass between models (Fig. \ref{fig:redshift_spacialdiffs}) and the influence of the environment beyond individual multiply-imaged regions, including two cluster components, cluster gas, and group-scale substructures, which can affect multiple image positions through external shear.

The importance of including spectroscopic redshifts has been thoroughly investigated by \cite{2016ApJ...832...82J} using the simulated cluster ARES \citep[][]{2017MNRAS.472.3177M}, who found that the addition of spectroscopic redshifts significantly reduces systematic errors in multiple-image predictions, magnifications, and mass profiles. In addition, they emphasise the importance of an even spatial distribution of spectroscopic systems, as well as a wide and uniform distribution of system redshifts, which helps to establish the mass slope and lift the mass-sheet degeneracy \citep[e.g.][]{1995A&A...294..411S}. With our gold catalogue of 135 multiple images, all with spectroscopic redshifts, we achieved a more uniform spatial distribution of spectroscopic redshifts, with 19 spectroscopic lensed galaxies in the main cluster and 8 in the subcluster. We note that most of the spectroscopic systems used in previous works (\citetalias{cha25}, \citetalias{Richard21}) were situated in the northern half of the main cluster, with only one system in the subcluster  (Fig. \ref{fig:field}). We also significantly expanded the spectroscopic redshift range, from  $z=2.8-3.5$ to $z=0.9-6.7$ (Fig. \ref{fig:speczs_histogram}). In the ARES cluster, $\gtrsim25$ systems with spectroscopic redshifts were required to achieve the agreement between $\Delta_{\rm rms}$ from multiple images and the true model precision \citep{2016ApJ...832...82J}. Our catalogue, containing  27  spectroscopic redshifts, is thus expected to substantially reduce systematic uncertainties relative to previous studies.

\section{Summary and conclusion}
\label{sec:conclusion}
In this work, we present an updated strong-lensing model of the Bullet Cluster, using our JWST NIRCam imaging and NIRSpec spectroscopic observations. Our lens model has been constrained using the catalogue of 135  secure multiple images from 49 multiply lensed systems (galaxy components) belonging to 27 distinct galaxies, all with spectroscopic redshifts. With our NIRSpec observations, the number of lensed galaxies with spectroscopic redshifts increases by a factor of four - from 5 to 19 in the main cluster and from 1 to 8 in the subcluster - compared to previous works (\citetalias{cha25}). This includes obtaining spectroscopic redshifts for all previously known systems and several systems discovered in the NIRCam imaging. Our catalogue achieves a wider, more uniform spatial distribution of spectroscopic systems, and it includes eight spectroscopic redshifts in the subcluster, which had previously contained only a single spectroscopic system. We also significantly expanded the redshift range of multiple images from $z=2.8-3.5$ to $z=0.9-6.7$. In addition to our most secure catalogue of constraints, we provide a full catalogue of 199 multiple image candidates, split into gold, silver, and bronze catalogues based on their quality. We also provide 34 NIRSpec spectroscopic redshifts of cluster members. 

We modelled the Bullet Cluster with the parametric lens modelling code \texttt{Lenstool}. We parametrised the mass distribution using several large-scale dark-matter halos, cluster members following luminosity-based scaling relations, and a small number of galaxies modelled individually. In the outskirts of the cluster, we included several group-scale substructures identified by \citetalias{Benavides23} using the kinematics of cluster galaxies. We included most substructures with fixed parameters, based on approximate masses derived from their velocity dispersion. In addition, two substructures were included with free normalisation. We also included cluster gas as a fixed mass map derived from X-ray observations. 

In the main cluster, our model features a complex mass distribution with two peaks at the two BCGs. The three large-scale halos are displaced from the two BCGs and altogether they form a highly elliptical density distribution with elongation to the north-east, elevated density between the BCGs, and an extension in the south. The mass distribution in the subcluster is simpler, described by a single large-scale halo that closely coincides  with the subcluster BCG. With a measured displacement of $4_{-2}^{+3}$ kpc, we achieved a three-fold improvement in precision and a systematic shift towards lower displacements, compared to a model with the same parametrisation, but without the inclusion of JWST systems. The halo position is consistent with the subcluster BCG within $2\sigma$. Although we caution against overinterpreting the displacement given our idealised model description and degeneracy with the mass of the BCG, this highlights the potential utility of our JWST observations for studies of dark matter via galaxy-DM displacement. 

A notable component in our model are the group-scale substructures surrounding the inner cluster regions. Despite the high uncertainty of the substructure properties derived from spectroscopy, their addition improves the model performance in the inner regions. Despite the fact that an even better fit can be achieved when replacing them with a constant shear component, which warrants further refinement of their properties in future work, the substructures provide a physically motivated source of shear tied to mass components. They furthermore highlight the importance of considering the mass, extending beyond the regions covered by multiple images, along the two proposed merger axes in the Bullet Cluster.

We compared our lens model with the pre-JWST model by \citetalias{Richard21} and the free-form reconstruction by \citetalias{cha25} (which includes new NIRCam candidates without $z_{\rm spec}$). The most striking differences between the three models are due to different lens modelling approaches (e.g. treatment of cluster members, assumed DM profiles, and inclusion of group-scale substructures). The three models are most consistent in the inner regions, covered by multiple images, with the aperture mass profiles showing the closest agreement at $\sim 60 \mathrm{\ kpc}$. We find that the \citetalias{cha25} lens model contains less mass than our fiducial model at all aperture radii, whereas the \citetalias{Richard21} model is in closer agreement with our results, particularly in the subcluster. We also investigated the agreement between our spectroscopic redshifts and redshifts, predicted by previous models. Although the overall agreement is reasonable for both models, the remaining differences are significant and spatially coherent. In the subcluster, which had previously been constrained by only a single spectroscopic redshift, both earlier models show a slight tendency to overpredict redshifts, consistent with reports that redshifts are systematically overestimated when spectroscopic constraints are sparse (e.g. \citealt{2016ApJ...832...82J}).

The lens model presented in this work relies only on systems with spectroscopic redshifts, increasing their number from 5 to 19 in the main cluster and from 1 to 8 in the subcluster compared to previous studies. Our dataset provides improvements in both statistical and systematic errors, which is essential for applications requiring precise mass reconstruction in the inner cluster regions, such as studies of dark matter self-interactions.

\section{Data availability}
Our fiducial lens model, including \texttt{Lenstool} parameter and output files, multiple image and cluster member catalogues, best-fit and sample lensing maps, are available on the website of the CAnadian NIRISS Unbiased Cluster Survey (CANUCS)\footnote{\url{https://niriss.github.io/lensing.html}}, as well as in the Strong Lensing Cluster Atlas Data Base, hosted at Laboratoire d’Astrophysique de Marseille\footnote{\url{https://data.lam.fr/sl-cluster-atlas/home}}. The associated data products (BCG-subtracted images and photometric catalogues) will also be released on the CANUCS website. Table A.1 and the 1D spectra, presented in this work, are also available in electronic form at the CDS via anonymous ftp to cdsarc.u-strasbg.fr (130.79.128.5) or via \url{http://cdsweb.u-strasbg.fr/cgi-bin/qcat?J/A+A/}.

\begin{acknowledgements}
We thank the anonymous referee and Marceau Limousin for useful comments that greatly enhanced the quality of this paper. We thank Gabriel Brammer for support with archival HST images and NIRSpec spectra reductions, Johan Richard for sharing the MUSE data cubes from \citetalias{Richard21}, Sangjun Cha for sharing the convergence map from \citetalias{cha25}, and Benjamin Beauchesne for a useful discussion on \texttt{Lenstool} capabilities. GR, MB, AH, NM, JJ and VM acknowledge support from the ESA PRODEX Experiment Arrangement No. 4000146646, ERC Grant FIRSTLIGHT, and Slovenian National Research Agency ARIS, through grants N1-0238 and P1-0188. AH acknowledges support by the Science and Technology Facilities Council (STFC), by the ERC through Advanced Grant 695671 “QUENCH”, and by the UKRI Frontier Research grant RISEandFALL. This research was also enabled by grants 18JWST-GTO1 and 24JWGO3A12 from the Canadian Space Agency and funding from the Natural Sciences and Engineering Research Council of Canada. ML acknowledges support from the National Recovery and Resilience Plan (NRRP), funded by the European Union – NextGenerationEU; Project title "GRAVITY", project code PNRR\_BAC24MLOMB\_01, CUP C53C22000350006. Support for SWR was provided by the Chandra X-ray Center through NASA contract NAS8-03060, the Smithsonian Institution, and by the Chandra X-ray Observatory grant GO3-14134X. This research used the Canadian Advanced Network For Astronomy Research (CANFAR) operated in partnership by the Canadian Astronomy Data Centre and The Digital Research Alliance of Canada, with support from The National Research Council of Canada, the Canadian Space Agency, CANARIE and the Canadian Foundation for Innovation. This work is based on observations made with the NASA/ESA/CSA James Webb Space Telescope (associated with programme \#4598), and NASA/ESA Hubble Space Telescope (associated with programs \#10200, \#10863, and \#11099), which are operated by the Space Telescope Science Institute (STScI), which is operated by the Association of Universities for Research in Astronomy, Inc., under NASA contracts NAS 5–26555 and NAS 5-03127.  Support for programme \#4598 was provided by NASA through a grant from STScI under NASA contract NAS5-03127. The data were obtained from the Mikulski Archive for Space Telescopes at the STScI. We furthermore acknowledge support from programs JWST-GO-03362 and JWST-GO-05890, provided through a grant from the STScI under NASA contract NAS5-03127. This work has made use of resources provided by the Strong Lensing Cluster Atlas Database hosted at the Laboratoire d'Astrophysique de Marseille (LAM).

\end{acknowledgements}

\bibliographystyle{aa} 
\bibliography{references}{}

\appendix

\onecolumn
\section{Multiple Image catalogue}
\label{app:multimcat}

\begin{table}[h]
    \centering
    \caption{Multiple image catalogue with coordinates, new NIRSpec redshift measurements ($z_{\rm NIRSpec}$), pre-JWST spectroscopic redshifts $z_{\rm old}$ (\citetalias{Richard21}, \citealt{Paraficz16}), system redshifts used for lens modelling ($z_{\rm sys}$), their class and the corresponding ids from \citetalias{cha25} lens model. }
    \begin{tabular}{c c c c c c c c c}
      \hline
      \hline
        ID & RA (ICRS) & DEC (ICRS) &$z_{\mathrm{NIRSpec}}$&$z_{\mathrm{old}}$&$z_{\mathrm{sys}}$&Class &ID$_{\mathrm{Cha25}}$ \\
      
\hline
K1b.1 & 104.630175 & $-$55.943430 & $3.249\pm0.006$ & 3.24 & 3.24 & \gold &   \\
K1b.2 & 104.631670 & $-$55.942180 &  &   &   &   & 28.1b \\
K1b.3 & 104.632213 & $-$55.941898 & $3.256\pm0.007$ &   &   &   & 28.1a \\
K1c.1 & 104.630308 & $-$55.943397 &  &   &   &   &   \\
K1c.2 & 104.631449 & $-$55.942382 &  &   &   &   &   \\
K1c.3 & 104.632728 & $-$55.941691 &  &   &   &   &   \\
K1d.1 & 104.630501 & $-$55.943126 &  &   &   &   &   \\
K1d.2 & 104.631044 & $-$55.942631 &  &   &   &   &   \\
K1d.3 & 104.632873 & $-$55.941582 &  &   &   &   &   \\
K1e.1 & 104.630771 & $-$55.942980 &  &   &   &   & 28.2b \\
K1e.2 & 104.630911 & $-$55.942859 &  &   &   &   &   \\
K1e.3 & 104.633185 & $-$55.941501 & $3.244\pm0.006$ &   &   &   & 28.2a \\
K1a.2 & 104.631955 & $-$55.942078 &  &   &   & \bronze &   \\
K1a.3 & 104.632155 & $-$55.941981 &  &   &   & \bronze &   \\
\hline
K2a.1 & 104.651904 & $-$55.956198 & $2.772\pm0.006$ &   & 2.774 & \gold & 12.1b \\
K2a.2 & 104.646940 & $-$55.958370 &  &   &   &   & 12.1c \\
K2a.3 & 104.664995 & $-$55.951223 & $2.772\pm0.006$ &   &   &   & 12.1a \\
K2b.1 & 104.651493 & $-$55.956653 & $2.780\pm0.006$ &   &   &   & 12.2b \\
K2b.2 & 104.647864 & $-$55.958230 &  &   &   &   & 12.2c \\
K2b.3 & 104.665380 & $-$55.951282 &  &   &   &   & 12.2a \\
K2c.1 & 104.652138 & $-$55.956124 &  &   &   &   & 12.3b \\
K2c.2 & 104.646726 & $-$55.958483 &  &   &   &   & 12.3c \\
K2c.3 & 104.664938 & $-$55.951295 &  &   &   &   & 12.3a \\
\hline
K3a.1 & 104.642163 & $-$55.948816 & $3.249\pm0.006$ & 3.2541 & 3.2541 & \gold & 19.1b \\
K3a.2 & 104.639550 & $-$55.950996 &  &   &   &   & 19.1c \\
K3a.3 & 104.654274 & $-$55.944329 &  &   &   &   & 19.1a \\
K3b.1 & 104.641568 & $-$55.949328 & $3.254\pm0.007$ &   &   &   & 19.2b \\
K3b.2 & 104.640619 & $-$55.950095 & $3.260\pm0.007$ &   &   &   & 19.2c \\
K3b.3 & 104.654678 & $-$55.944245 &  &   &   &   &   \\
\hline
K4a.1 & 104.658414 & $-$55.950599 &  & 2.7768 & 2.7768 & \gold & 11.3a \\
K4a.2 & 104.655161 & $-$55.951769 &  &   &   &   & 11.3b \\
K4a.3 & 104.638998 & $-$55.958171 &  &   &   &   & 11.3c \\
K4b.1 & 104.657857 & $-$55.950815 &  &   &   &   & 11.4a \\
K4b.2 & 104.656568 & $-$55.951427 &  &   &   &   & 11.4b \\
K4b.3 & 104.638707 & $-$55.958270 &  &   &   &   & 11.4c \\
K4c.1 & 104.658805 & $-$55.950458 &  &   &   &   & 11.2a \\
K4c.2 & 104.654332 & $-$55.951960 &  &   &   &   & 11.2b \\
K4c.3 & 104.639207 & $-$55.958089 &  &   &   &   & 11.2c \\
K4d.1 & 104.659243 & $-$55.950332 &  &   &   &   & 11.1a \\
K4d.2 & 104.653743 & $-$55.952223 &  &   &   &   & 11.1b \\
K4d.3 & 104.639536 & $-$55.957985 &  &   &   &   & 11.1c \\
\hline
K5.1 & 104.637870 & $-$55.956249 &  &   & 2.661 & \gold & 14c \\
K5.2 & 104.651893 & $-$55.949967 & $2.661\pm0.005$ &   &   &   & 14b \\
K5.3 & 104.655474 & $-$55.948702 &  &   &   &   & 14a \\
\hline
K6.1 & 104.650212 & $-$55.953322 &  &   & 1.411 & \gold & 10b \\
K6.2 & 104.645675 & $-$55.954998 & $1.425\pm0.009$ &   &   &   & 10c \\
K6.3 & 104.659589 & $-$55.950068 & $1.410\pm0.003$ &   &   &   & 10a \\
    \end{tabular}
    \tablefoot{The class of each multiple image is equal to the first one in the system unless indicated otherwise. Asterisk ($^*$) indicates NIRSpec redshift obtained by stacking the spectra of several multiple images of the system. (\dag) indicates spectroscopic confirmation of the multiple image without redshift measurement (see Appendix \ref{Appendix:NIRSpec spectra}).}
        \label{tab:multipleimagestable}
        \end{table}

\begin{table}
                  \ContinuedFloat
        \centering
        \caption{}
        \begin{tabular}{c c c c c c c c c}
      \hline
      \hline
         ID & RA (ICRS) & DEC (ICRS) &$z_{\mathrm{NIRSpec}}$&$z_{\mathrm{old}}$&$z_{\mathrm{sys}}$&Class &ID$_{\mathrm{Cha25}}$ \\
\hline
K7a.1 & 104.643999 & $-$55.949069 &  &   & 1.621 & \gold & 17b \\
K7a.2 & 104.642537 & $-$55.950405 & $1.621\pm0.006$ &   &   &   & 17c \\
K7a.3 & 104.653066 & $-$55.945583 &  &   &   & \bronze & 17a \\
K7b.1 & 104.644317 & $-$55.948777 &  &   &   &   &   \\
K7b.2 & 104.642084 & $-$55.950774 &  &   &   &   &   \\      
\hline
K8a.1 & 104.657447 & $-$55.948266 & $3.259\pm0.007$ & 3.26 & 3.26 & \gold & 13a \\
K8a.2 & 104.651086 & $-$55.950426 &  &   &   &   & 13b \\
K8a.3 & 104.637908 & $-$55.956349 &  &   &   &   & 13c \\
K8b.1 & 104.657247 & $-$55.948512 &  &   &   &   &   \\
K8b.2 & 104.651516 & $-$55.950387 &  &   &   &   &   \\
K8b.3 & 104.637822 & $-$55.956558 &  &   &   &   &   \\
\hline
K9a.1 & 104.649493 & $-$55.947333 & $2.525^{+0.028}_{-0.007}$ &   & 2.525 & \gold & 18.2a \\
K9a.2 & 104.648427 & $-$55.947554 &  &   &   &   & 18.2b \\
K9b.1 & 104.649222 & $-$55.947342 &  &   &   &   & 18.1a \\
K9b.2 & 104.648627 & $-$55.947472 &  &   &   &   & 18.1b \\
\hline
K10a.1 & 104.563058 & $-$55.939497 & $2.993\pm0.006$ & 2.99 & 2.993 & \gold & 24.1c \\
K10a.2 & 104.561217 & $-$55.942591 &  &   &   &   & 24.1b \\
K10a.3 & 104.561812 & $-$55.947543 &  &   &   &   & 24.1a \\
K10b.1 & 104.562924 & $-$55.939715 &  &   &   &   & 24.2c \\
K10b.2 & 104.561317 & $-$55.942373 &  &   &   &   & 24.2b \\
K10b.3 & 104.561879 & $-$55.947675 &  &   &   &   & 24.2a \\
\hline
K11.1 & 104.561803 & $-$55.946063 & $2.592^{+0.016}_{-0.005}$ &   & 2.592 & \gold & 25a \\
K11.2 & 104.561422 & $-$55.944206 &  &   &   &   & 25b \\
K11.3 & 104.564676 & $-$55.938087 & \dag &   &   &   & 25c \\
\hline
K12a.1 & 104.570238 & $-$55.943990 &  &   & 1.602 & \gold & 20.2b \\
K12a.2 & 104.569039 & $-$55.945957 & $^*$$1.602\pm0.003$ &   &   &   & 20.2a \\
K12a.3 & 104.569778 & $-$55.938706 & $^*$$1.602\pm0.003$ &   &   &   & 20.2c \\
K12b.1 & 104.570421 & $-$55.944162 &  &   &   &   & 20.1b \\
K12b.2 & 104.569378 & $-$55.945774 &  &   &   &   & 20.1a \\
K12b.3 & 104.570051 & $-$55.938655 &  &   &   &   & 20.1c \\
\hline
K13a.1 & 104.565692 & $-$55.939794 & $3.506\pm0.007$ &   & 3.517 & \gold & 23c \\
K13a.2 & 104.564106 & $-$55.941917 & $3.524\pm0.007$ &   &   &   & 23b \\
K13a.3 & 104.564662 & $-$55.948543 & $3.522\pm0.007$ &   &   &   & 23a \\
K13b.1 & 104.565896 & $-$55.939489 &  &   &   &   &   \\
K13b.2 & 104.563896 & $-$55.942234 &  &   &   &   &   \\
\hline
K14a.1 & 104.563931 & $-$55.945349 &  &   & 1.861 & \gold & 22a \\
K14a.2 & 104.563704 & $-$55.944842 & $^*$$1.861\pm0.004$ &   &   &   & 22b \\
K14a.3 & 104.566233 & $-$55.938144 & $^*$$1.861\pm0.004$ &   &   &   & 22c \\
K14b.1 & 104.563886 & $-$55.945159 &  &   &   &   &   \\
K14b.2 & 104.563817 & $-$55.945017 &  &   &   &   &   \\
K14b.3 & 104.566494 & $-$55.938037 &  &   &   & \bronze &   \\
\hline
K15.1 & 104.637513 & $-$55.941569 &  & 3.537 & 3.537 & \gold & 29a \\
K15.2 & 104.635354 & $-$55.942714 & $3.535\pm0.007$ &   &   &   & 29b \\
K15.3 & 104.632501 & $-$55.945342 &  &   &   &   & 29c \\
\hline
K16a.1 & 104.645000 & $-$55.944683 &  &  &  & \silver & 40a \\
K16a.2 & 104.642406 & $-$55.945848 &  &   &   &   & 40b \\
K16b.1 & 104.645261 & $-$55.944535 &  &   &   & \bronze &   \\
K16b.2 & 104.641821 & $-$55.946045 &  &   &   & \bronze &   \\
K16c.2 & 104.640457 & $-$55.946739 &  &   &   &   &   \\
K16c.3 & 104.638609 & $-$55.947930 &  &   &   &   &   \\
K16d.2 & 104.640031 & $-$55.946915 &  &   &   &   &33.3a   \\
K16d.3 & 104.638990 & $-$55.947590 &  &   &   &   &33.3b   \\
\hline
K17.1 & 104.633265 & $-$55.945437 & $2.078\pm0.004$ &   & 2.078 & \bronze &   \\
K17.2 & 104.637577 & $-$55.942113 &  &   &   &   &   \\
K17.3 & 104.637970 & $-$55.943117 &  &   &   &   &   \\
\hline
N1.1 & 104.653123 & $-$55.953472 & $1.853\pm0.004$ &   & 1.852 & \gold &   \\
N1.2 & 104.660313 & $-$55.950983 &  &   &   &   &   \\
N1.3 & 104.642801 & $-$55.957689 & $1.852\pm0.004$ &   &   &   &   \\
    \end{tabular}
        \label{tab:multipleimagestable}
        \end{table}

\begin{table}
                  \ContinuedFloat
        \centering
        \caption{}
        \begin{tabular}{c c c c c c c c c}
      \hline
      \hline
         ID & RA (ICRS) & DEC (ICRS) &$z_{\mathrm{NIRSpec}}$&$z_{\mathrm{old}}$&$z_{\mathrm{sys}}$&Class &ID$_{\mathrm{Cha25}}$ \\
        
\hline
N2a.1 & 104.664268 & $-$55.953788 &  &   & 2.128 & \gold & 6.1a \\
N2a.2 & 104.646395 & $-$55.960536 & $2.128^{+0.013}_{-0.005}$ &   &   &   & 6.1c \\
N2a.3 & 104.658760 & $-$55.957898 &  &   &   &   &   \\
N2a.4 &	104.656240 & $-$55.956356 &  &   &   & \bronze   & 6.1b  \\
N2b.1 & 104.664359 & $-$55.954042 &  &   &   &   & 6.2a \\
N2b.2 & 104.646493 & $-$55.960728 &  &   &   &   & 6.2c \\
N2b.3 & 104.658847 & $-$55.958070 &  &   &   &   &   \\
\hline
N3a.1 & 104.565462 & $-$55.946499 & $2.550\pm0.005$ &   & 2.547 & \gold & 21a \\
N3a.2 & 104.567622 & $-$55.937412 &  &   &   &   & 21c \\
N3a.3 & 104.564431 & $-$55.944059 & $2.545\pm0.005$ &   &   &   & 21b \\
N3b.1 & 104.565408 & $-$55.946416 &  &   &   &   &   \\
N3b.2 & 104.567628 & $-$55.937350 &  &   &   &   &   \\
N3b.3 & 104.564410 & $-$55.944148 &  &   &   &   &   \\
\hline
N4.1 & 104.562003 & $-$55.939597 & $^*$$2.587\pm0.005$ &   & 2.587 & \gold & 26a \\
N4.2 & 104.560374 & $-$55.942851 &  &   &   &   & 26b \\
N4.3 & 104.560568 & $-$55.946877 & $^*$$2.587\pm0.005$ &   &   &   & 26a \\
\hline
N5a.1 & 104.640293 & $-$55.953977 & $3.249\pm0.006$ &   & 3.252 & \gold & 15c \\
N5a.2 & 104.647337 & $-$55.950375 &  &   &   & \bronze & 15b \\
N5a.3 & 104.658992 & $-$55.946470 & $3.26\pm0.01$ &   &   &   & 15a \\
N5b.1 & 104.641085 & $-$55.953524 &  &   &   &   &   \\
N5b.2 & 104.647103 & $-$55.950477 &  &   &   & \bronze &   \\
N5b.3 & 104.659328 & $-$55.946389 &  &   &   &   &   \\
\hline
N6.1 & 104.570983 & $-$55.938178 & $3.894\pm0.008$ &   & 3.892 & \gold & 31c \\
N6.2 & 104.571803 & $-$55.942637 &  &   &   &   & 31b \\
N6.3 & 104.568491 & $-$55.948215 & $3.890\pm0.008$ &   &   &   & 31a \\
\hline
N7a.1 & 104.660439 & $-$55.950736 & $^*$$4.345^{+0.166}_{-0.008}$ &   & 4.345 & \gold & 9.2a \\
N7a.2 & 104.654465 & $-$55.952811 &  &   &   &   & 9.2b \\
N7a.3 & 104.638440 & $-$55.959176 & $^*$$4.345^{+0.166}_{-0.008}$ &   &   &   & 9.2c \\
N7b.1 & 104.660211 & $-$55.950402 &  &   &   &   & 9.1a \\
N7b.2 & 104.654132 & $-$55.952591 &  &   &   &   & 9.1b \\
N7b.3 & 104.638227 & $-$55.958927 &  &   &   &   & 9.1c \\
\hline
N8a.1 & 104.661335 & $-$55.959443 & $5.10\pm0.01$ &   & 5.09 & \gold & 5.1b \\
N8a.2 & 104.665888 & $-$55.955981 & $5.09\pm0.01$ &   &   &   & 5.1a \\
N8a.3 & 104.644957 & $-$55.963292 &  &   &   &   & 5.1c \\
N8b.1 & 104.661280 & $-$55.959358 &  &   &   &   & 5.2b \\
N8b.2 & 104.665814 & $-$55.955881 &  &   &   &   & 5.2a \\
N8b.3 & 104.644815 & $-$55.963229 &  &   &   &   & 5.2c \\
\hline
N9a.1 & 104.653383 & $-$55.960771 &  &   & 6.64 & \gold & 1a \\
N9a.2 & 104.651091 & $-$55.961211 & $6.64\pm0.01$ &   &   &   & 1b \\
N9b.1 & 104.654145 & $-$55.960626 & $6.63\pm0.01$ &   &   &   & 2a \\
N9b.2 & 104.650364 & $-$55.961337 &  &   &   &   & 2b \\
\hline
N10.1 & 104.656963 & $-$55.955183 &  &   & 1.471 & \gold &   \\
N10.2 & 104.660961 & $-$55.954234 & $^*$$1.471\pm0.003$ &   &   &   &   \\
N10.3 & 104.646140 & $-$55.959789 & $^*$$1.471\pm0.003$ &   &   &   &   \\
\hline
N11.1 & 104.661292 & $-$55.955211 & $^*$$0.897^{+0.018}_{-0.008}$ &   & 0.897 & \gold &   \\
N11.2 & 104.653229 & $-$55.958355 & $^*$$0.897^{+0.018}_{-0.008}$ &   &   &   &   \\
\hline
N12.1 & 104.656461 & $-$55.953679 & $2.577\pm0.005$ &   & 2.575 & \gold & 8b \\
N12.2 & 104.660021 & $-$55.952650 & $2.572\pm0.005$ &   &   &   & 8a \\
N12.3 & 104.641073 & $-$55.959933 &  &   &   & \bronze & 8c \\
\hline
N13a.1 & 104.665281 & $-$55.952492 &  &   &   & \silver & 7.1a \\
N13a.2 & 104.654858 & $-$55.956210 &  &   &   &   & 7.1b \\
N13a.3 & 104.645459 & $-$55.960262 &  &   &   &   & 7.1c \\
N13b.1 & 104.665157 & $-$55.952402 &  &   &   &   & 7.2a \\
N13b.2 & 104.654687 & $-$55.956114 &  &   &   &   & 7.2b \\
N13b.3 & 104.645264 & $-$55.960177 &  &   &   &   & 7.2c \\
N13c.1 & 104.665134 & $-$55.952326 &  &   &   &   & 7.3a \\
N13c.2 & 104.654527 & $-$55.956089 &  &   &   &   & 7.3b \\
N13c.3 & 104.645253 & $-$55.960074 &  &   &   &   & 7.3c \\
    \end{tabular}
        \label{tab:multipleimagestable}
        \end{table}

\begin{table}
                  \ContinuedFloat
        \centering
        \caption{}
        \begin{tabular}{c c c c c c c c c}
      \hline
      \hline
         ID & RA (ICRS) & DEC (ICRS) &$z_{\mathrm{NIRSpec}}$&$z_{\mathrm{old}}$&$z_{\mathrm{sys}}$&Class &ID$_{\mathrm{Cha25}}$ \\
        
\hline
N14.1 & 104.650601 & $-$55.956994 &  &   &   & \silver &   \\
N14.2 & 104.648622 & $-$55.957819 &  &   &   &   &   \\
\hline
N15a.1 & 104.572579 & $-$55.940279 &  &   &   & \silver &   \\
N15a.2 & 104.572042 & $-$55.939833 &  &   &   &   &   \\
N15b.1 & 104.572422 & $-$55.940191 &  &   &   &   &   \\
N15b.2 & 104.572111 & $-$55.939934 &  &   &   &   &   \\
\hline
N16.1 & 104.662889 & $-$55.956836 &  &   &   & \bronze & 35a \\
N16.2 & 104.662466 & $-$55.957232 &  &   &   &   & 35b \\
\hline
N17.1 & 104.645362 & $-$55.945477 &  &   &   & \bronze &   \\
N17.2 & 104.644755 & $-$55.945728 &  &   &   &   &   \\
N17.3 & 104.637082 & $-$55.950513 &  &   &   &   &   \\
\hline
N18a.1 & 104.636916 & $-$55.944786 &  &   & 5.27 & \bronze &   \\
N18a.2 & 104.636659 & $-$55.944908 & $5.27\pm0.01$ &   &   &   &   \\
N18b.1 & 104.637248 & $-$55.944654 &  &   &   &   &   \\
N18b.2 & 104.636051 & $-$55.945260 & $5.28\pm0.01$ &   &   &   &   \\
\hline
N19.1 & 104.653206 & $-$55.960616 &  &   &   & \bronze & 3a \\
N19.2 & 104.651297 & $-$55.960986 &  &   &   &   & 3b \\
\hline
N20.1 & 104.654212 & $-$55.960388 &  &   &   & \bronze & 4a \\
N20.2 & 104.650418 & $-$55.961094 &  &   &   &   & 4b \\
\hline
N21.1 & 104.656664 & $-$55.947287 &  &   &   & \bronze & 16a \\
N21.2 & 104.648631 & $-$55.949908 &  &   &   &   & 16b \\
N21.3 & 104.639477 & $-$55.954359 &  &   &   &   & 16c \\
\hline
N22.1 & 104.561294 & $-$55.946890 &  &   &   & \bronze & 27a \\
N22.2 & 104.562365 & $-$55.940048 &  &   &   &   & 27c \\
N22.3 & 104.561059 & $-$55.942663 &  &   &   &   &  \\
\hline
N23.1 & 104.641107 & $-$55.947424 &  &   &   & \bronze & 32a \\
N23.2 & 104.636902 & $-$55.950610 &  &   &   &   & 32b \\
N23.3 & 104.650747 & $-$55.943593 &  &   &   &   &  \\
\hline
N24.1 & 104.659853 & $-$55.951808 &  &   &   & \bronze & 36a \\
N24.2 & 104.655412 & $-$55.953225 &  &   &   &   & 36b \\
N24.3 & 104.640786 & $-$55.959162 &  &   &   &   & 36c \\
\hline
N25.1 & 104.661221 & $-$55.952219 &  &   &   & \bronze & 37a \\
N25.2 & 104.655368 & $-$55.954040 &  &   &   &   & 37b \\
N25.3 & 104.641831 & $-$55.959661 &  &   &   &   & 37c \\
\hline
N26.1 & 104.565024 & $-$55.947790 &  &   &   & \bronze & 41a \\
N26.2 & 104.567245 & $-$55.938829 &  &   &   &   & 41b \\
    \end{tabular}
    \label{tab:multipleimagestable}
\end{table}

\clearpage

\onecolumn
\section{NIRSpec spectra}

\label{Appendix:NIRSpec spectra}
In this section, we present the NIRSpec spectra of multiple images used to measure their redshifts. Most redshift fits are obtained using \texttt{msaexp} with a redshift prior range $z=0.3-9$. The continuum of the spectra, which contained several clear emission lines, was modelled with splines. The continuum of some spectra is instead modelled using \texttt{EAZY} templates to achieve the correct redshift solution, consistent with significant emission lines and continuum features (K5.2, K6.2, K6.3, K7a.2, K13a.1, K13a.2, K17.2, N2a.3, N3a.1, N3a.3, N5a.1, N5a.3, N9b.1) or in the absence of any emission lines, where the redshift relies on spectral break (K9a.1, K11.1, K12, N4, N7, N10). In some cases (if the redshift solution was clear from spectral features or the spectra of counterimages), we restricted the redshift range to achieve the correct solution (e.g. K6.3, R8a.2, R18a.2) or obtained the redshift by fitting a Gaussian profile to \Halpha emission line (K6.2, K8a.1, K13a.3). The NIRSpec spectra of multiple images, together with RGB cutouts are shown in Figure \ref{fig:nirspec_spectra}, where we also indicate some prominent spectral features. In spectra with a clear spectral break, we indicate the wavelength of the Balmer break (marked as B.B.) regardless of whether the \texttt{msaexp} fit relied on the Balmer or the 4000 \AA\  break. In the multiple image system K11, we did not measure the redshift of K11.3. The spectrum shows a clear spectral break, consistent with K11.1, providing confirmation of the image. However, the source is close to a nearby star (see the cutout in Fig. \ref{fig:nirspec_spectra}) and the spectrum is relatively contaminated. We therefore rely on K11.1 to obtain the system redshift. For system K17 we could obtain a spectroscopic redshift from the spectrum of K17.1. Despite the missing spectral coverage between 1 and 3.6 $\mu$m, the \texttt{msaexp} fit on the full redshift range gives a clear redshift solution, consistent with the photometric redshift ($z_{\rm phot}=2.06_{-0.22}^{+ 0.06}$), and driven by the Pa-$\beta$ emission line. In some multiple-image systems, where spectra of several secure (based on colours and/or morphology) multiple images were available, we found that stacking them to increase the S/N significantly helped reliably determine the redshift when individual redshift fits were insufficient. Stacking increased the S/N of the emission lines (K12) or helped to better resolve the shape of the spectral break, used for the redshift fit with \texttt{msaexp} and \texttt{EAZY} templates (K14, R4, R7, R10). The individual spectra of such systems and the combined spectra used for redshift fitting are shown in Fig. \ref{fig:stackedspectra}. System K7 required special treatment even after stacking the spectra. To obtain a reliable redshift, we excluded the low-S/N part of the spectrum below $1 \mu$m. The redshift solution is consistent with the Balmer break and the possible \Halpha detection.

To estimate the redshift errors, including random and systematic \citep[e.g.][]{2024A&A...690A.288B} uncertainty, we compared the redshifts between different (secure) multiple images of a single system and between NIRSpec and MUSE redshifts from \citetalias{Richard21} where available. Based on the redshift scatter among 11 systems with several NIRSpec redshift measurements, we estimate the redshift uncertainty $\Delta z / (1+z) \sim 0.002$, which we use for most NIRSpec redshifts in this work. We note that this is the same uncertainty as the one found for the NIRSpec prism spectra of multiple images in MACS J0416.1-2403 \citep[][]{rihtarsic25}. We furthermore compared secure NIRSpec redshifts of 7 sources with their existing $z_{\rm MUSE}$ from \citetalias{Richard21} MUSE catalog (most of them cluster members) and found that the maximal difference $|z_{\rm NIRSpec}-z_{\rm MUSE}| / (1+z)$ is 0.0035 (and the mean difference is 0.0016), which further corroborates our uncertainty estimate. In a few spectra, we found that the statistical uncertainty obtained with \texttt{msaexp} is larger than estimated above, especially in sources where redshifts rely on spectral breaks. In those cases, we obtained the uncertainty by convolving the \texttt{msaexp} redshift probability density with a Gaussian uncertainty, estimated above. We obtain \onlyvalueA{1.425}{0.009} for K6.2, \onlyvalueA{1.621}{0.006} for K7a.2, \onlyvalueB{2.525}{0.007}{0.028} for K9a.1, 
\onlyvalueB{2.592}{0.005}{0.016} for K11.1, 
\onlyvalueB{2.128}{0.005}{0.013} for N2a.2,
\onlyvalueA{3.26}{0.01} for N5a.3,
\onlyvalueB{4.345}{0.008}{0.166} for N7 and 
\onlyvalueB{0.897}{0.008}{0.018} for N11.

\begin{figure}[h]
\centering
\includegraphics[width=\linewidth]{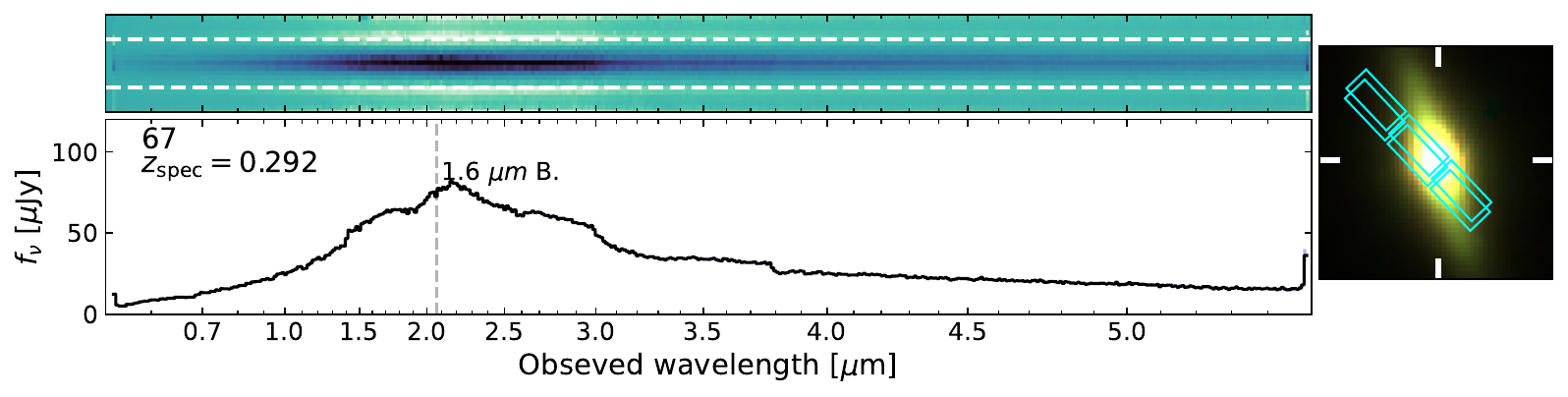}
\includegraphics[width=\linewidth]{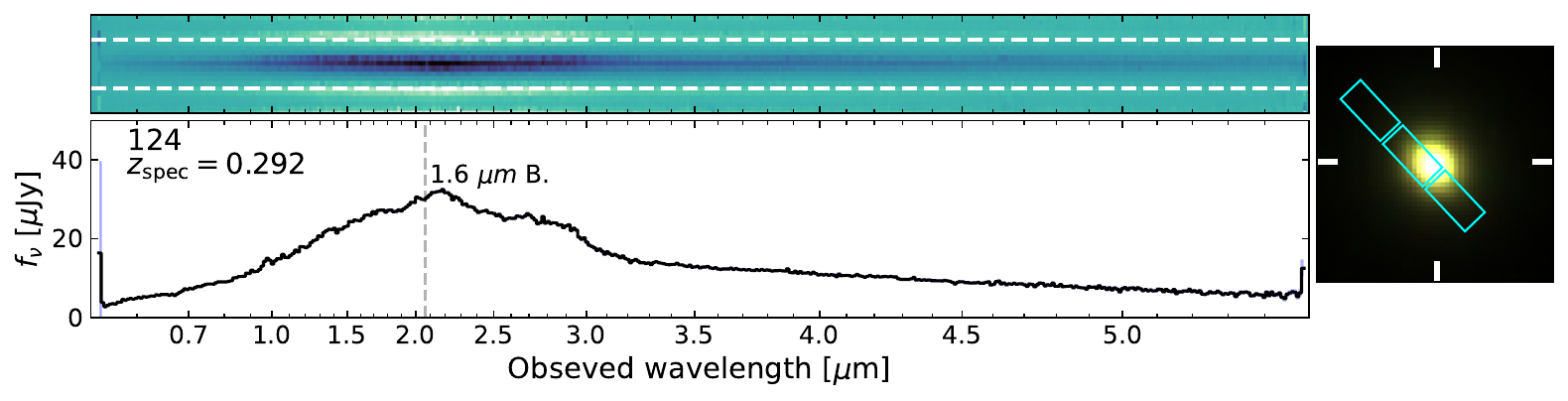}
\includegraphics[width=\linewidth]{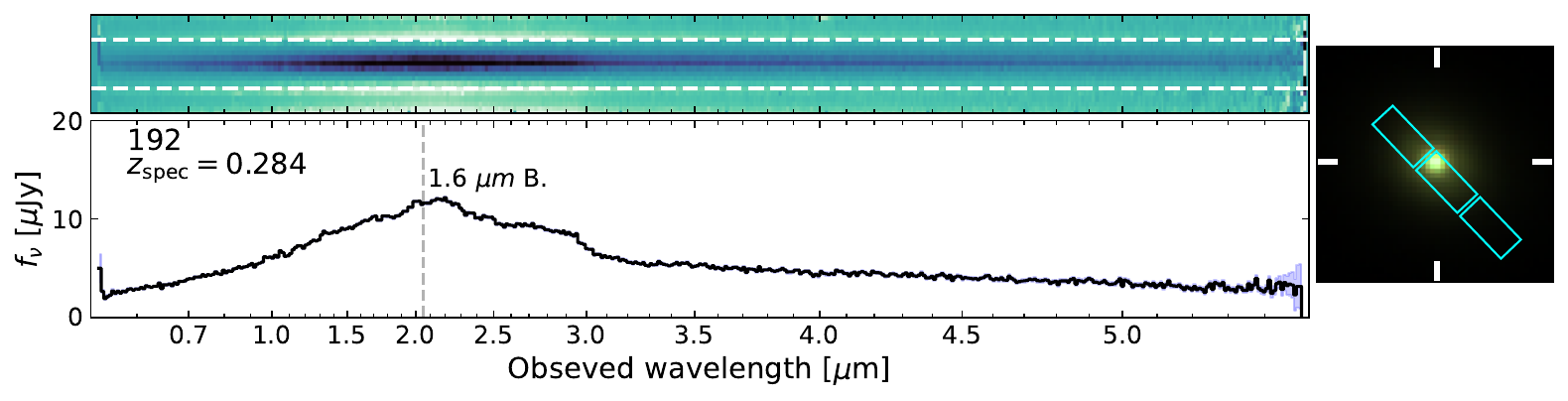}
\caption{Three example NIRSpec prism spectra of cluster members with indicated 1.6 $\mu\mathrm{m}$ stellar bump.}
\label{fig:clustermembersspectra}
\end{figure}

In addition to multiple images, we obtained NIRSpec spectroscopic confirmation of 34 cluster members, 16 of which are included in the lens model. The list of cluster members with our spectroscopic redshifts and their $m_{\rm F277W}$ magnitudes is given in Table \ref{tab:newnirspecclusterm}. In Fig. \ref{fig:clustermembersspectra}, we show the example NIRSpec spectra of three cluster members.

\twocolumn

\begin{figure}[h]
\centering
\includegraphics[width=\linewidth]{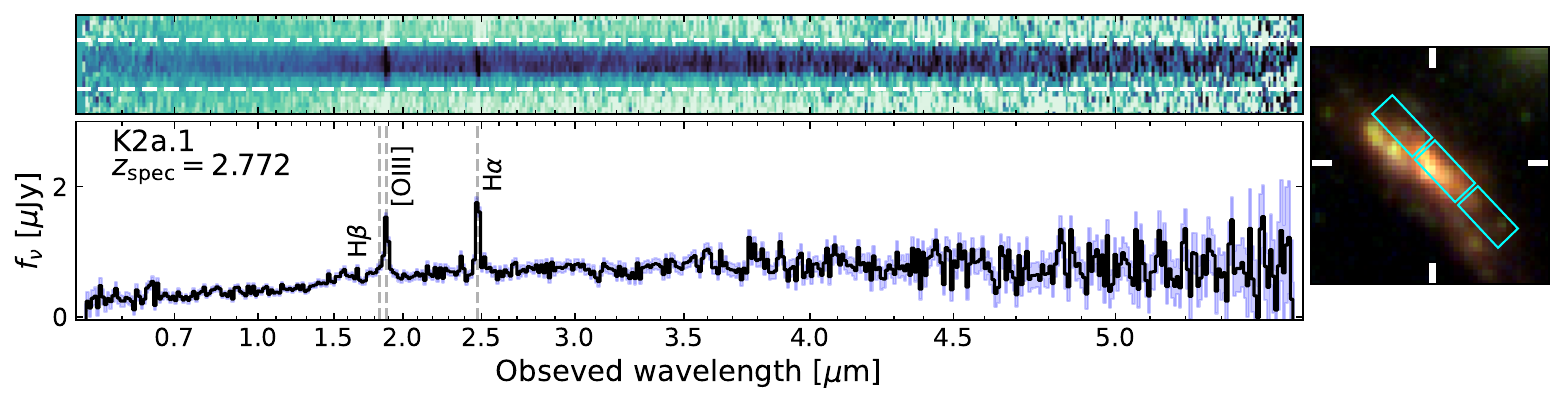}\\ 
\includegraphics[width=\linewidth]{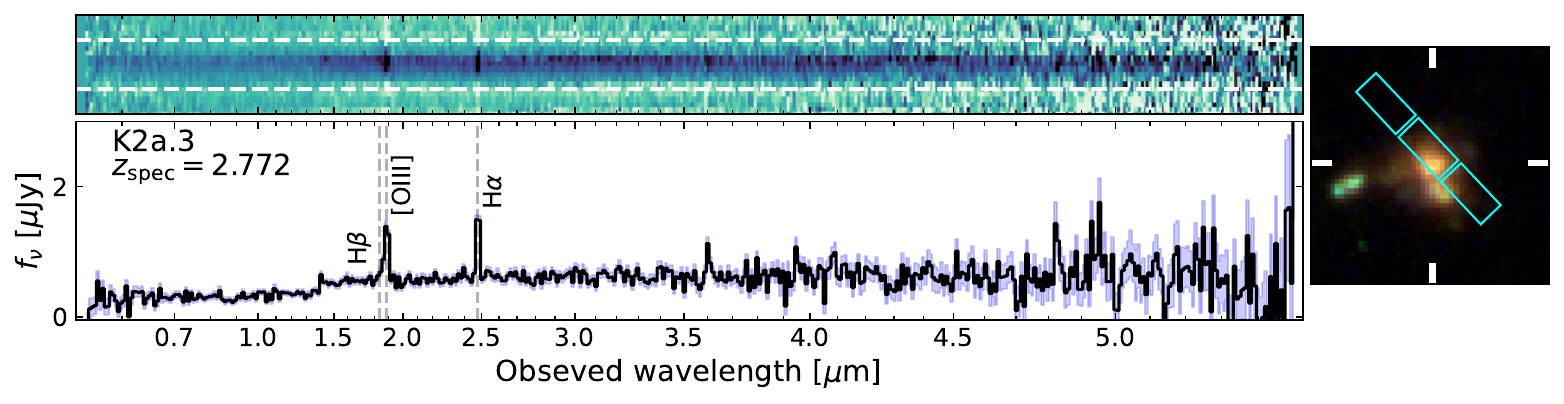}\\ 
\includegraphics[width=\linewidth]{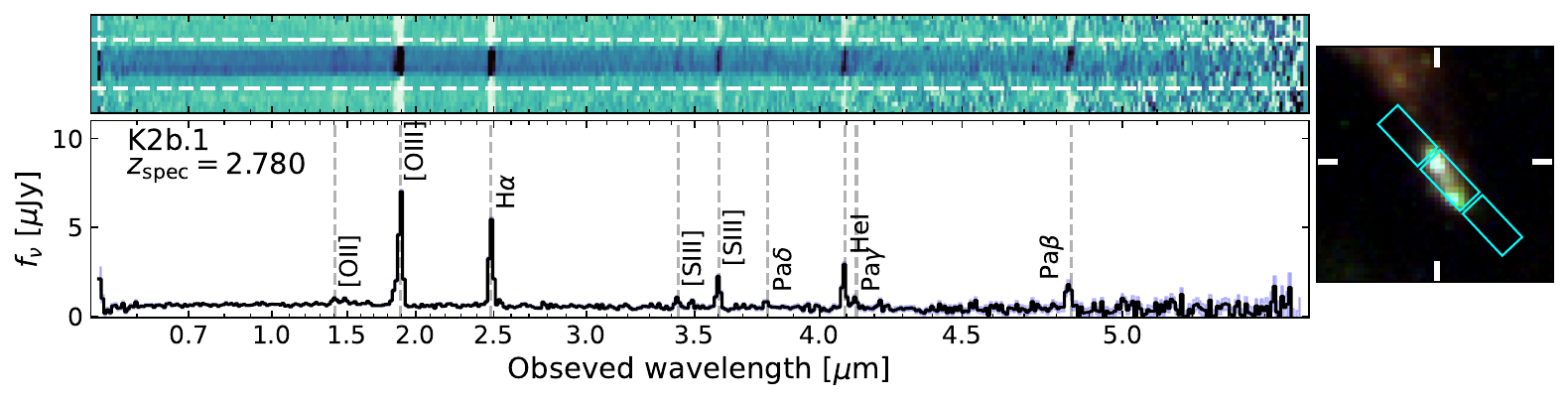}\\ 
\includegraphics[width=\linewidth]{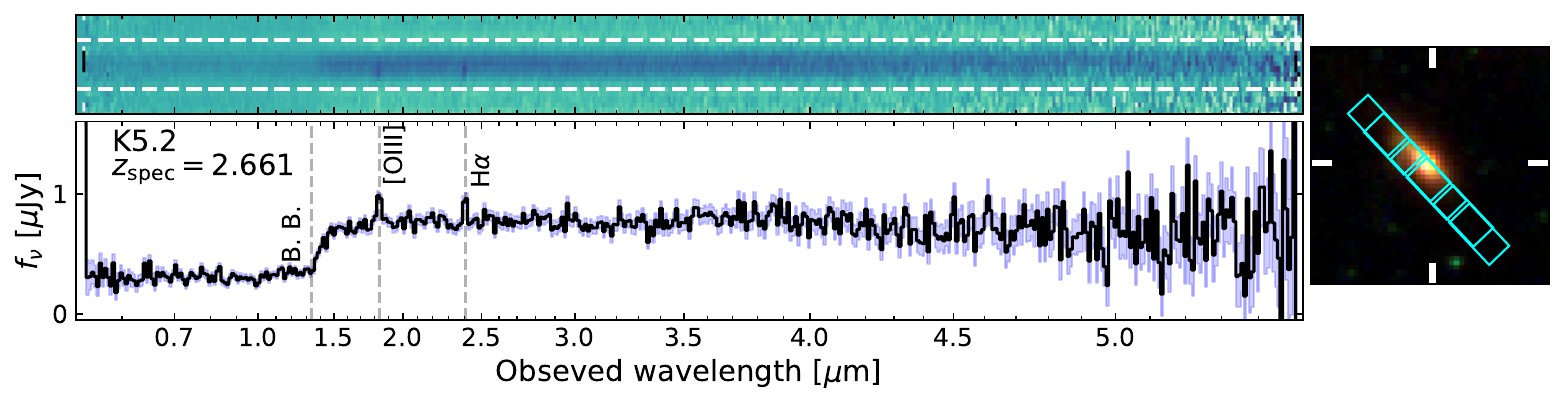}\\
\includegraphics[width=\linewidth]{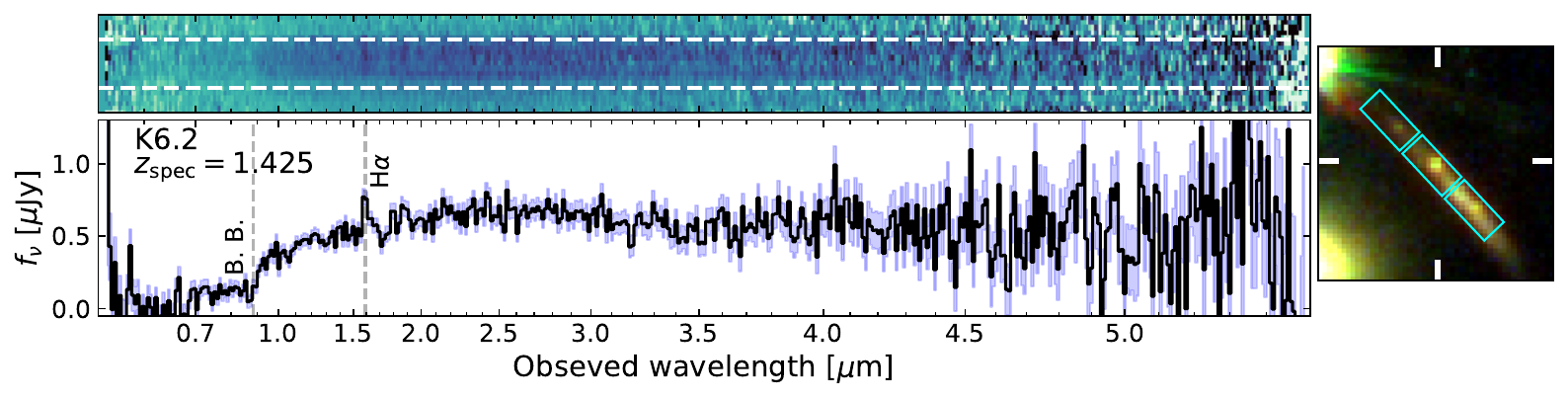}\\
\includegraphics[width=\linewidth]{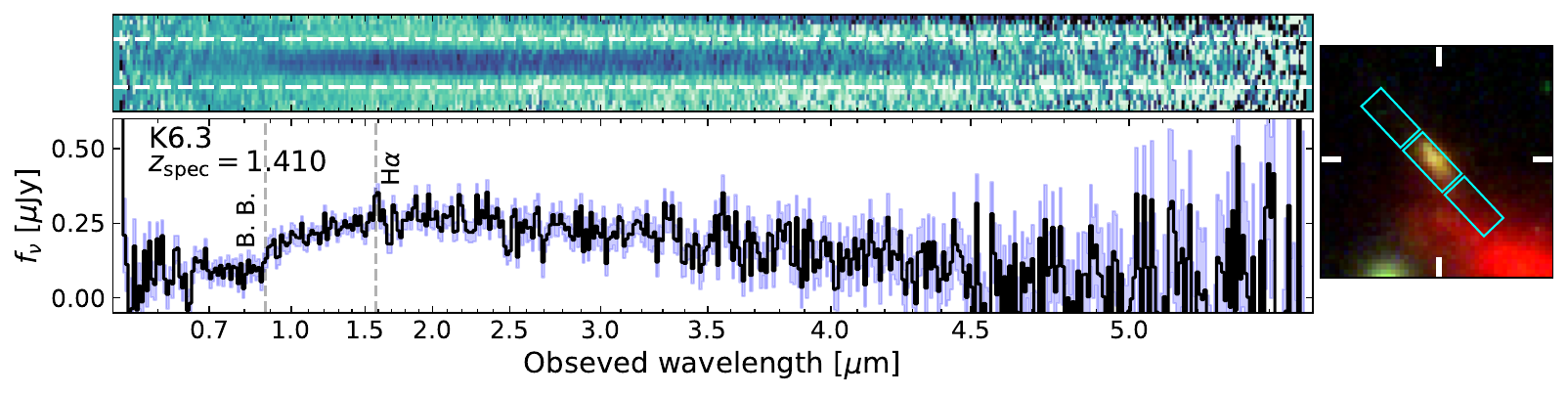}\\
\includegraphics[width=\linewidth]{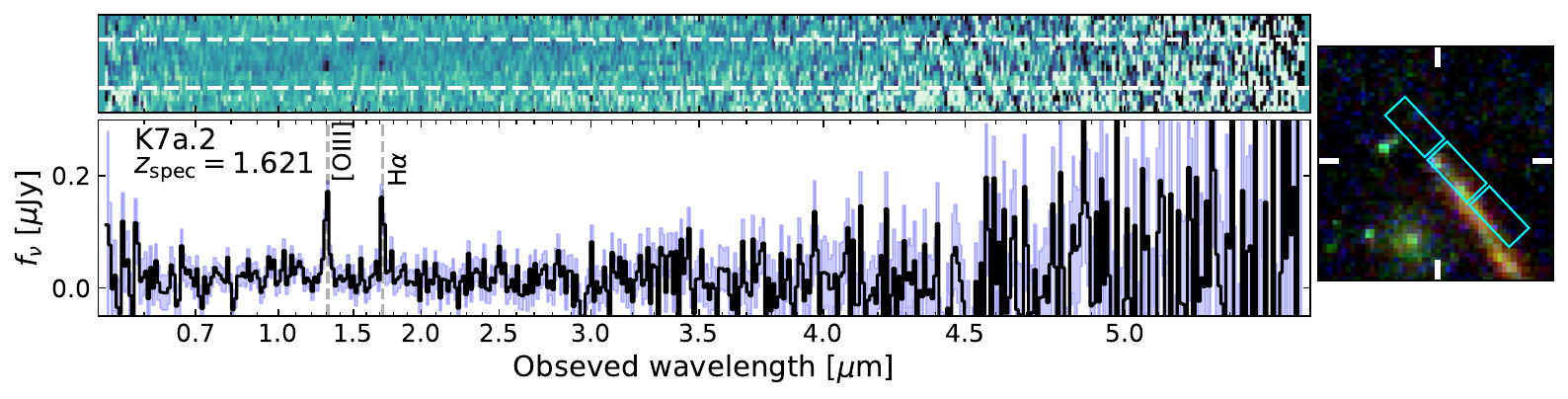}\\
\includegraphics[width=\linewidth]{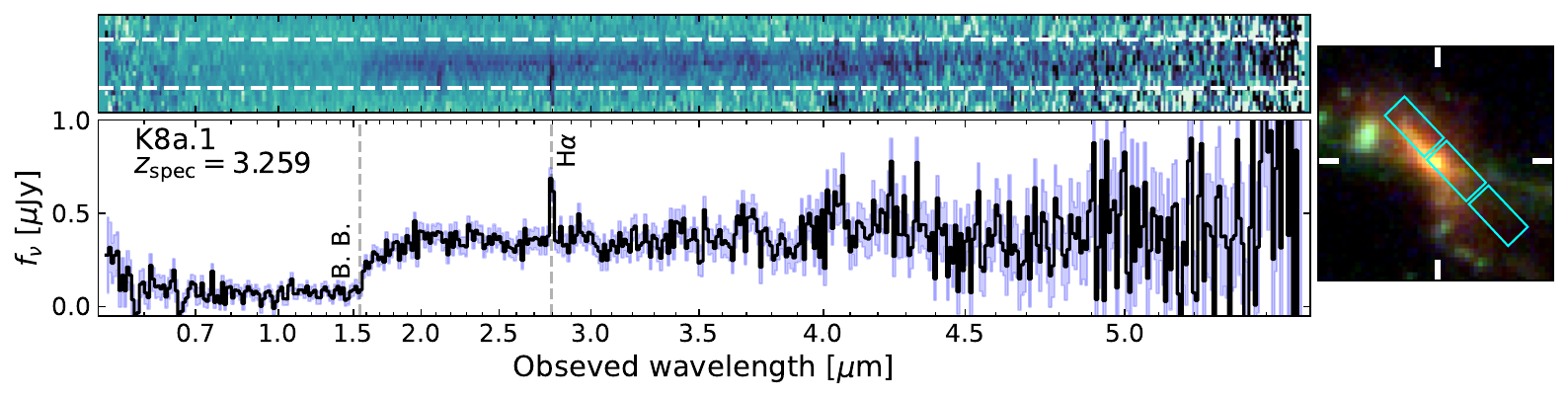}\\
\includegraphics[width=\linewidth]{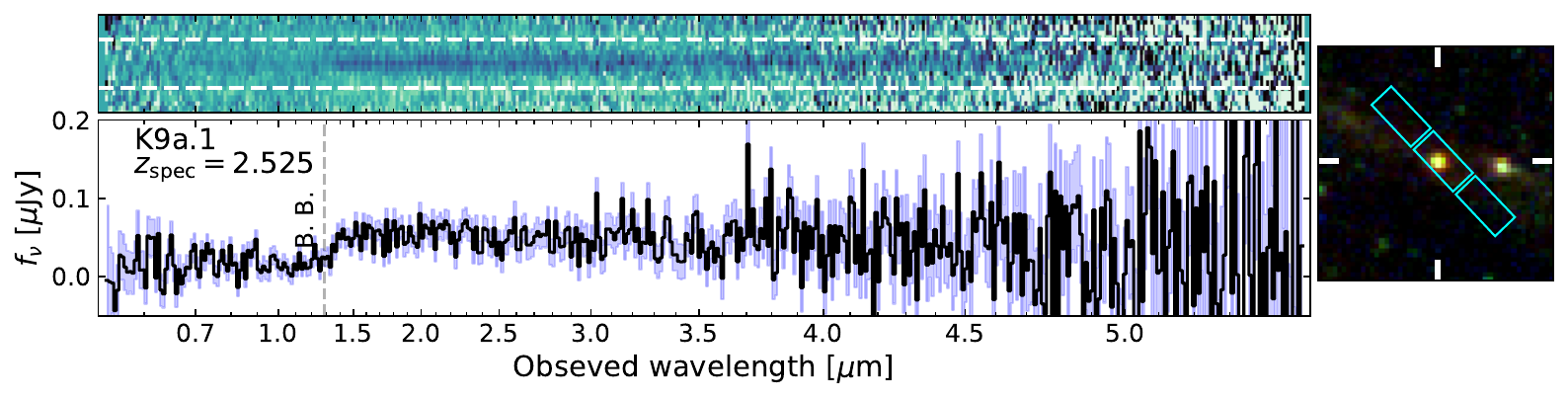}\\
\caption{2D and 1D NIRSpec prism spectra of multiple images, used in this work.  Vertical dashed lines mark some of the visible spectral lines and breaks. The RGB cutouts span $2''$ and are composed of F277W, F356W, F410M, F444W filters shown in red, F115W, F150W and F200W in green, and 
F814W, F606W and F090W in blue. Cyan rectangles mark the MSA slit positions.}
\label{fig:nirspec_spectra}
\end{figure}

\begin{figure}[h]
\ContinuedFloat
\centering
\includegraphics[width=\linewidth]{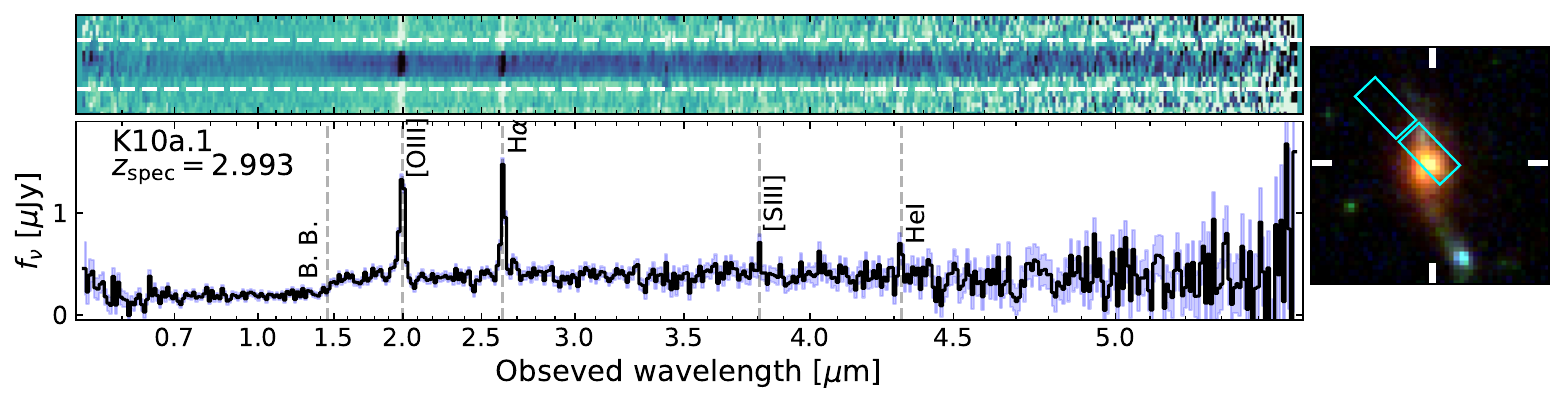}\\ 
\includegraphics[width=\linewidth]{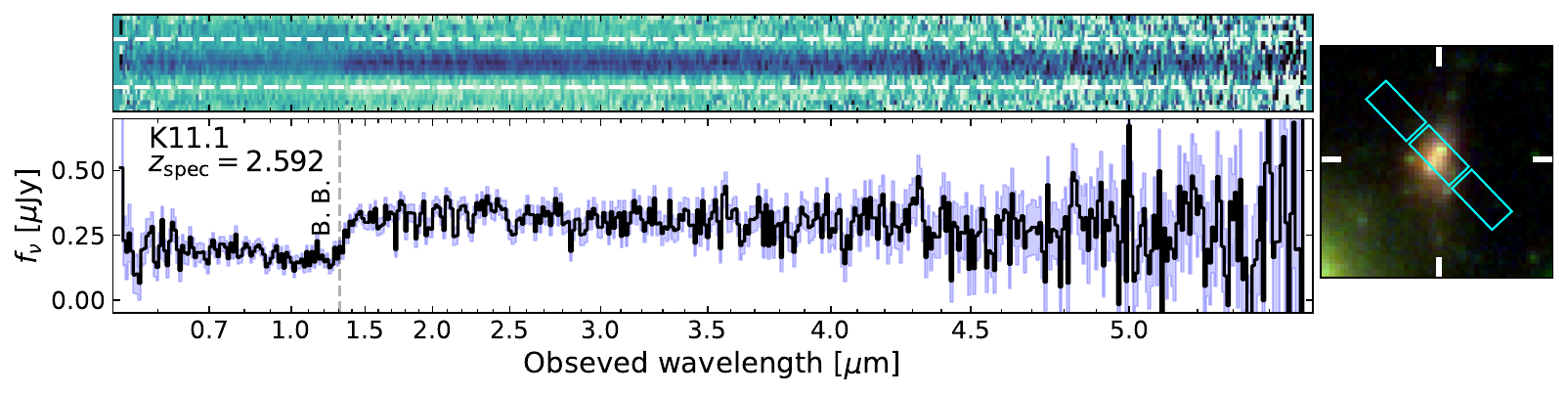}\\ 
\includegraphics[width=\linewidth]{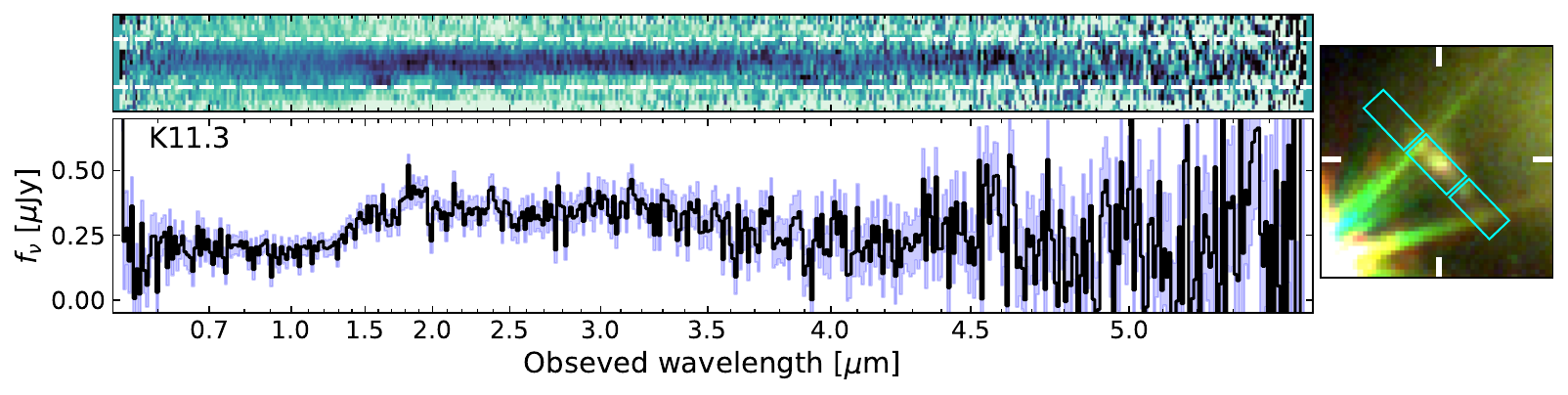}\\ 
\includegraphics[width=\linewidth]{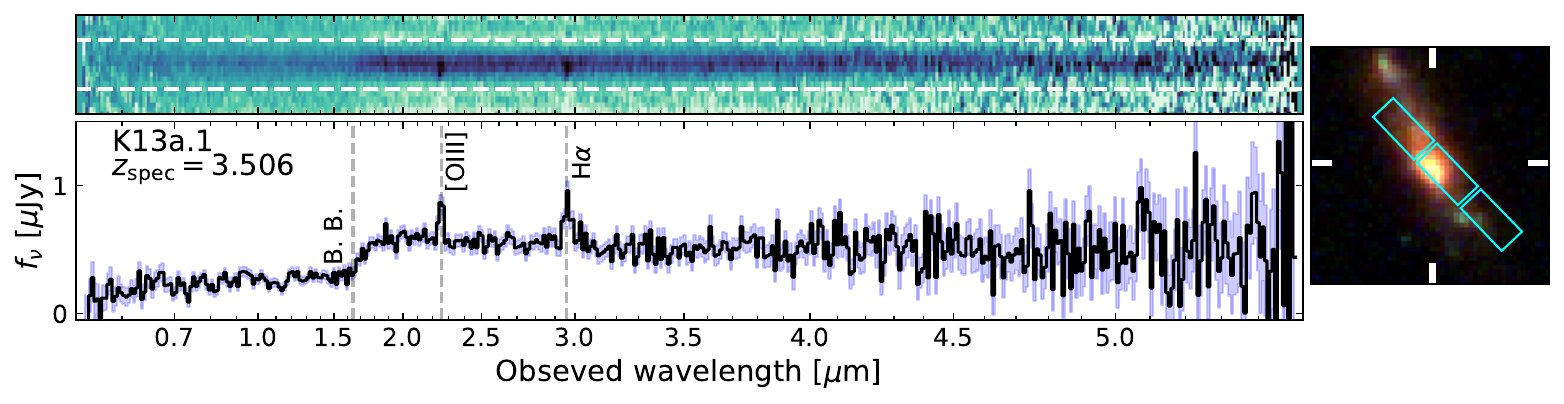}\\ 
\includegraphics[width=\linewidth]{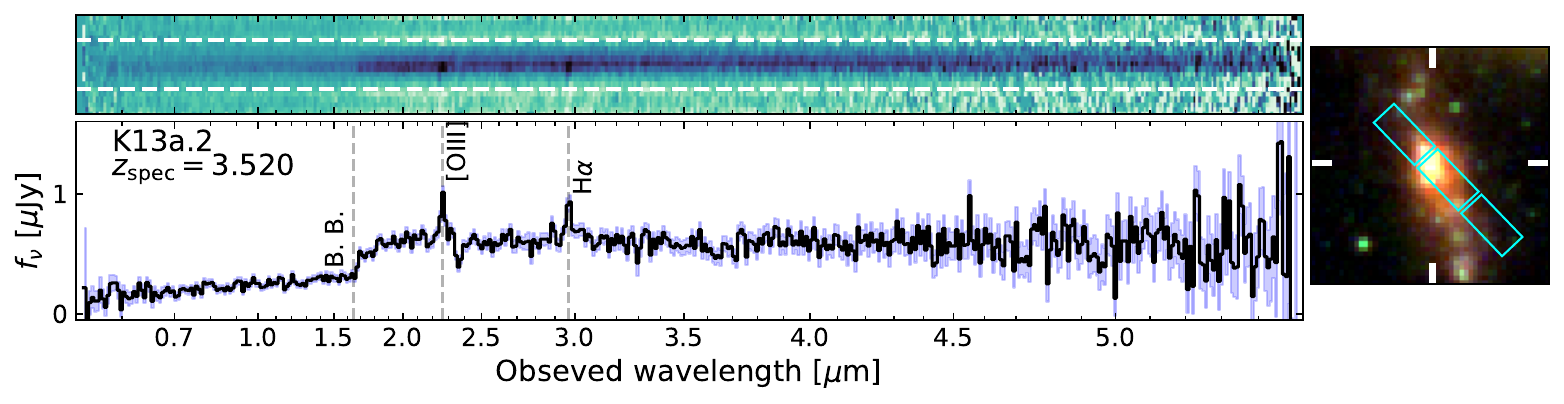}\\ 
\includegraphics[width=\linewidth]{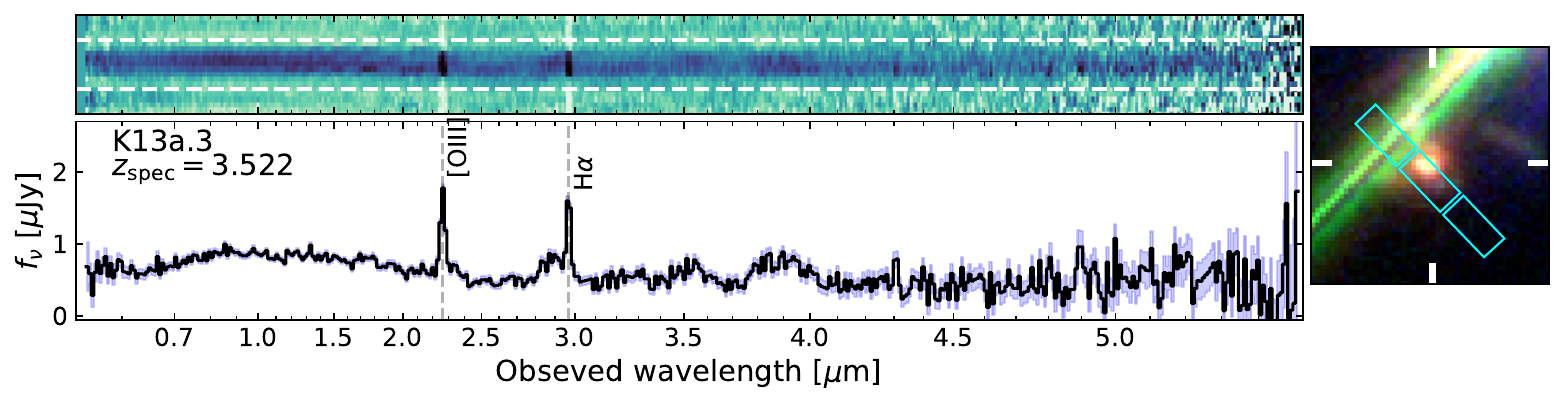}\\ 
\includegraphics[width=\linewidth]{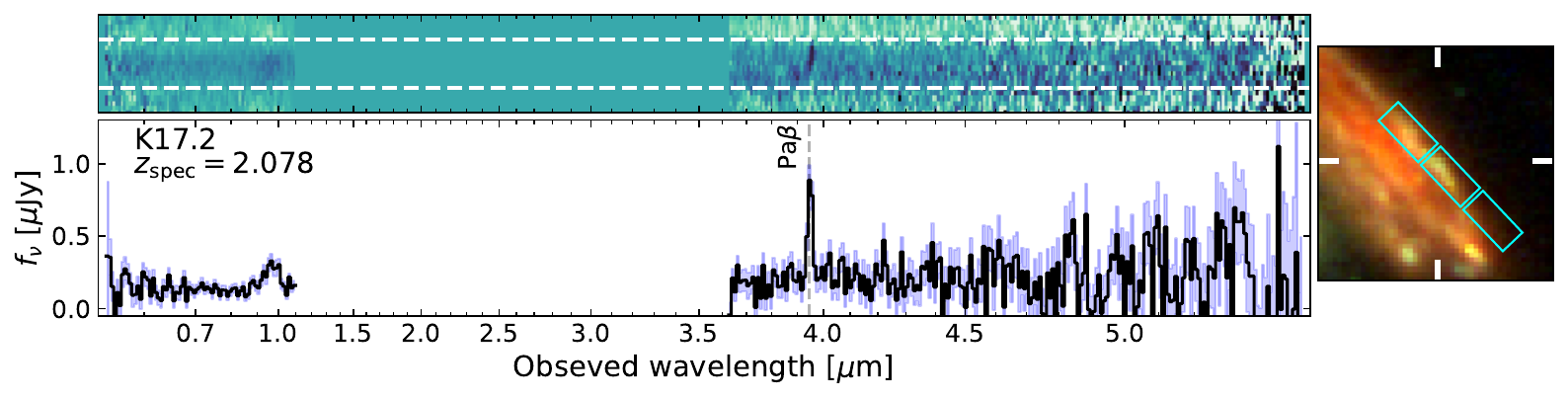}\\ 
\includegraphics[width=\linewidth]{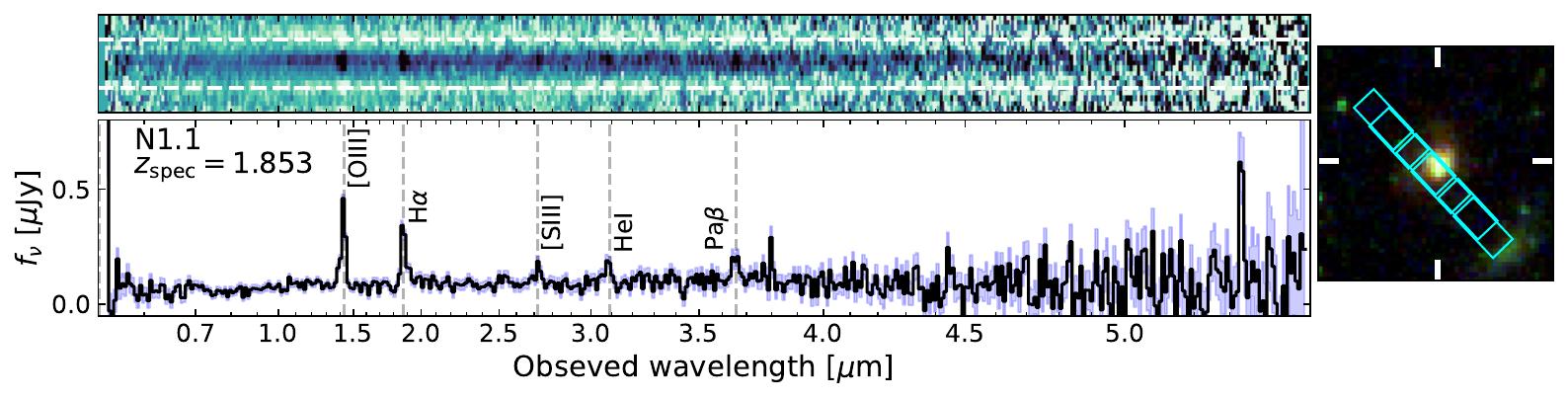}\\
\includegraphics[width=\linewidth]{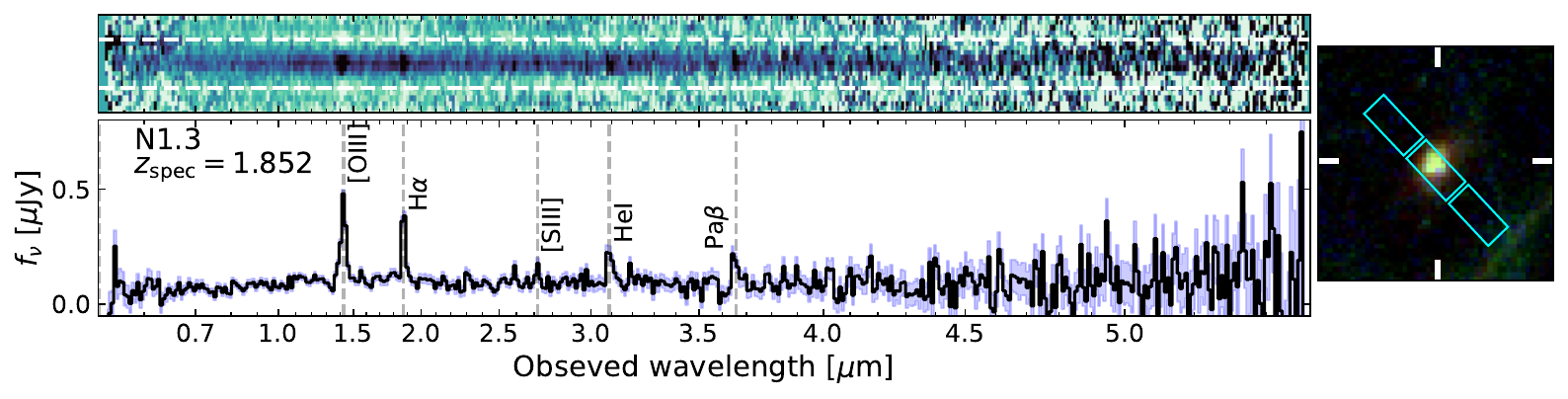}\\
\includegraphics[width=\linewidth]{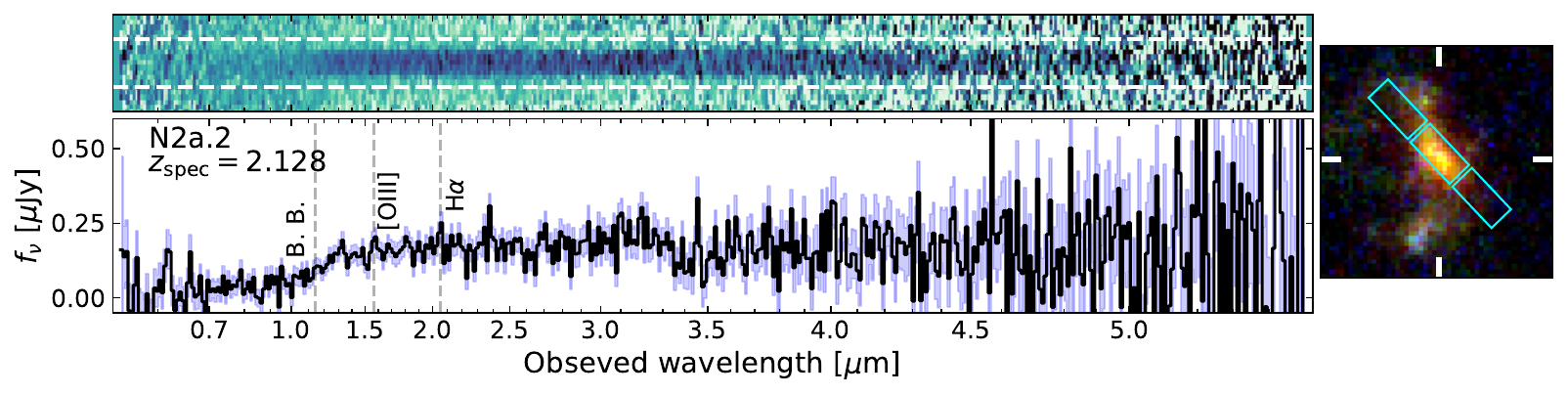}\\
\caption{continued}
\end{figure}

\begin{figure}[h]
\ContinuedFloat
\centering
\includegraphics[width=\linewidth]{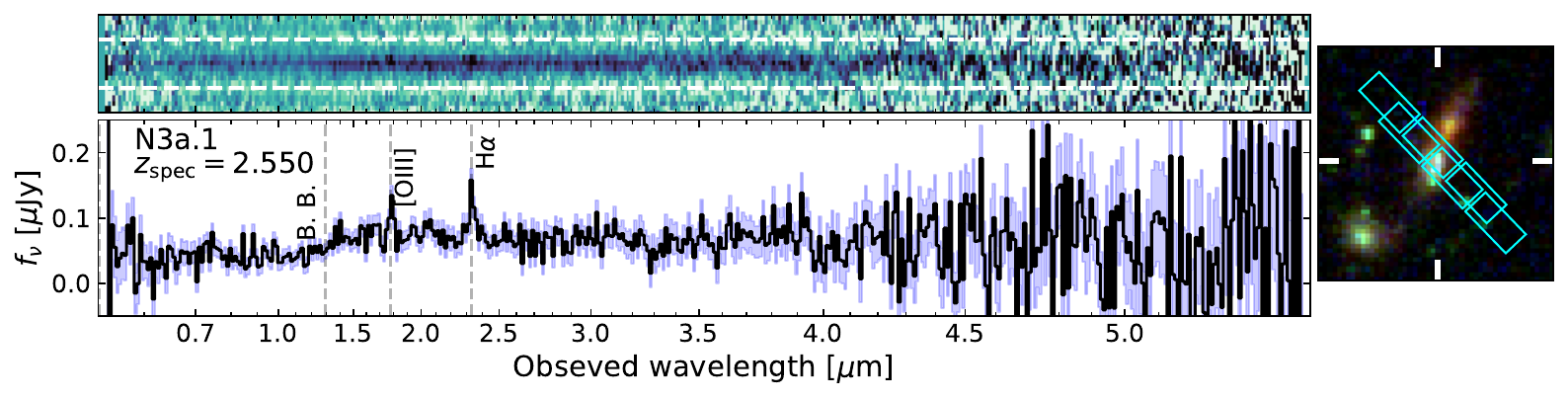}\\ 
\includegraphics[width=\linewidth]{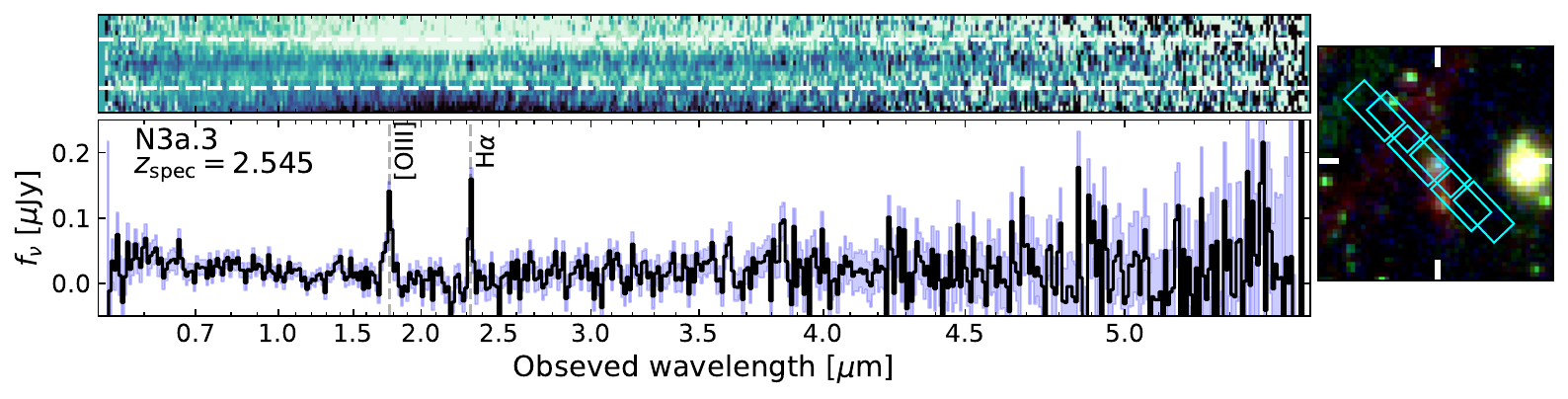}\\ 
\includegraphics[width=\linewidth]{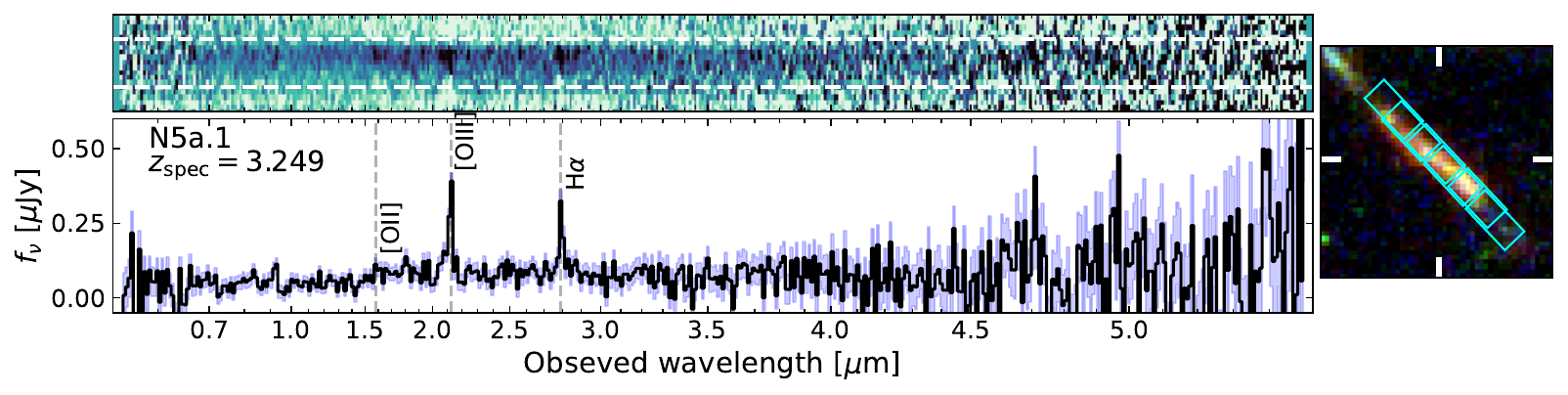}\\ 
\includegraphics[width=\linewidth]{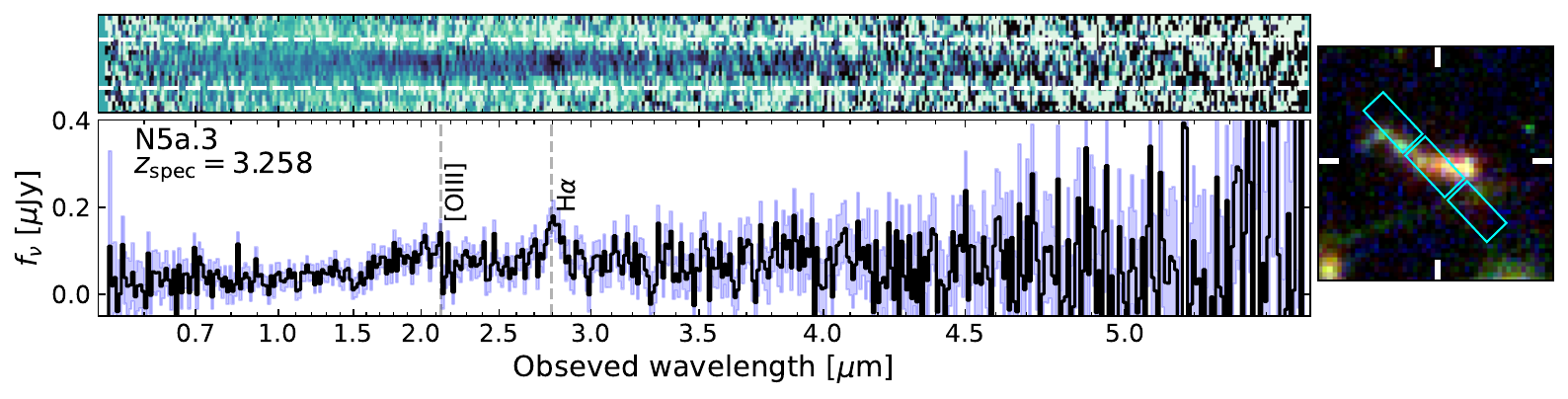}\\ 
\includegraphics[width=\linewidth]{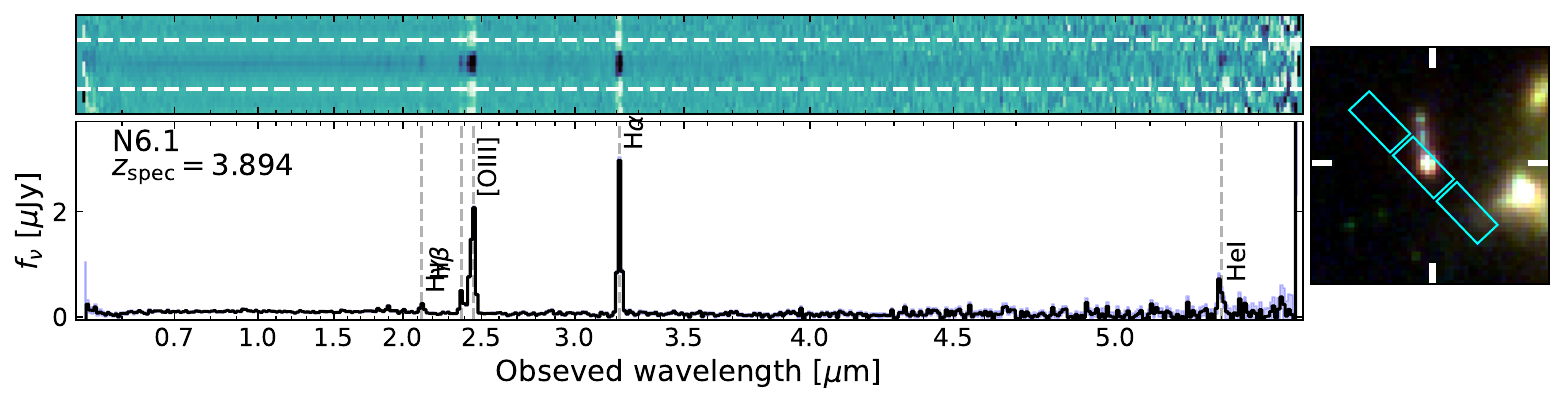}\\ 
\includegraphics[width=\linewidth]{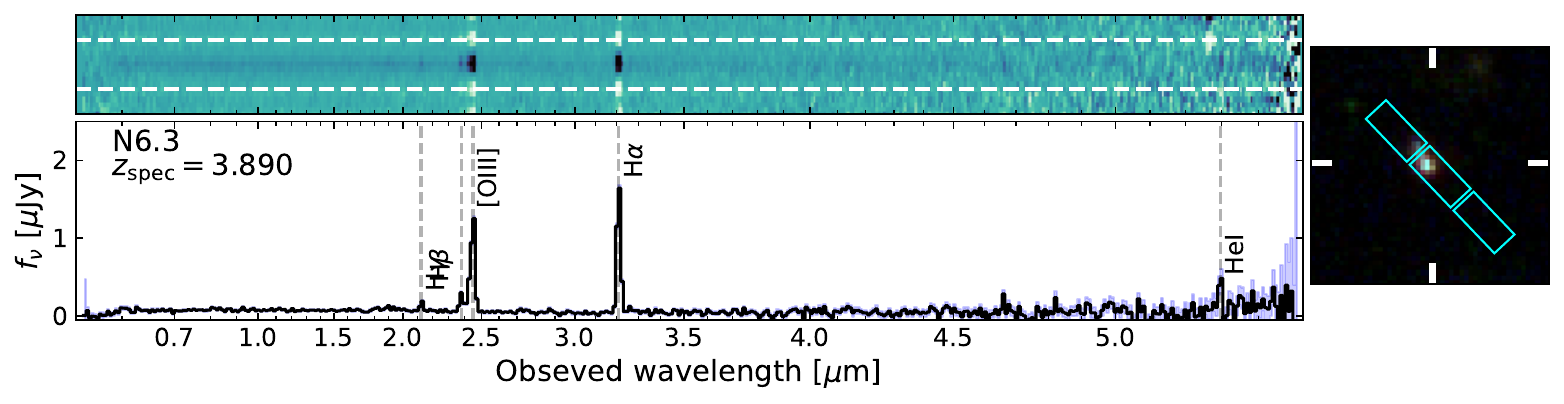}\\ 
\includegraphics[width=\linewidth]{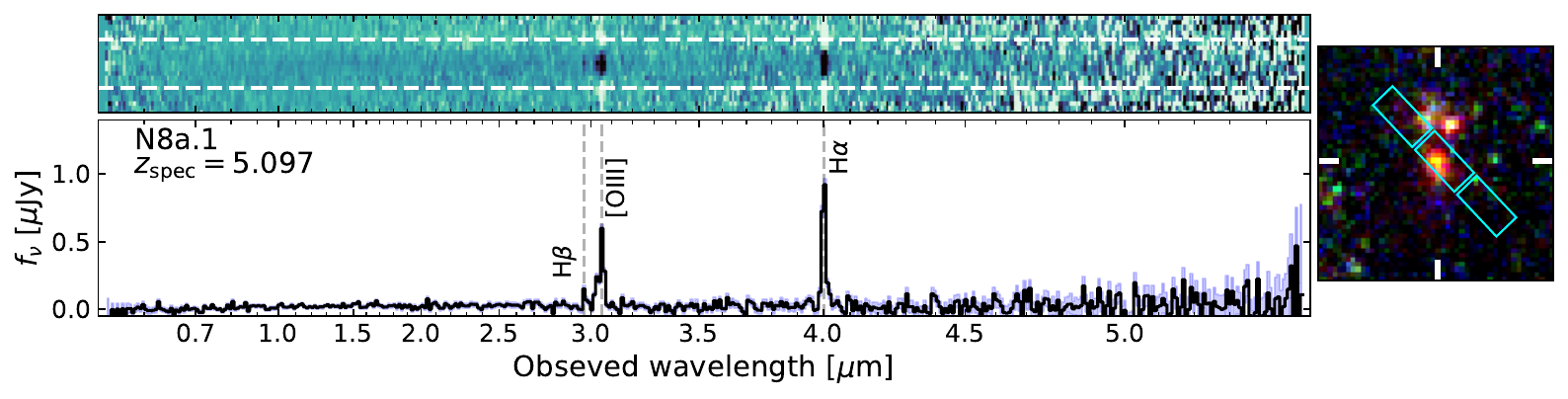}\\ 
\includegraphics[width=\linewidth]{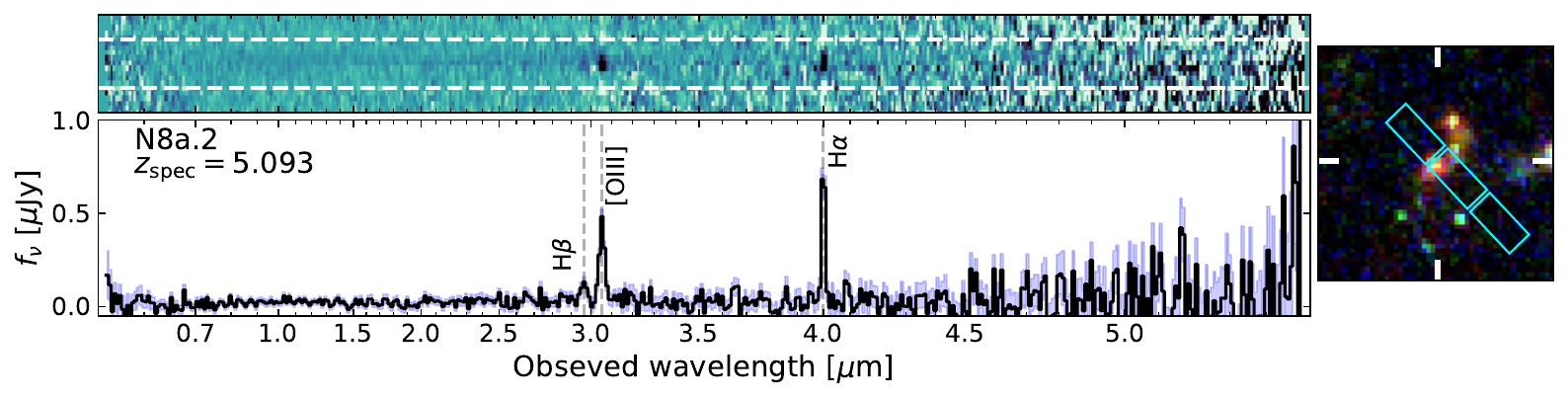}\\
\includegraphics[width=\linewidth]{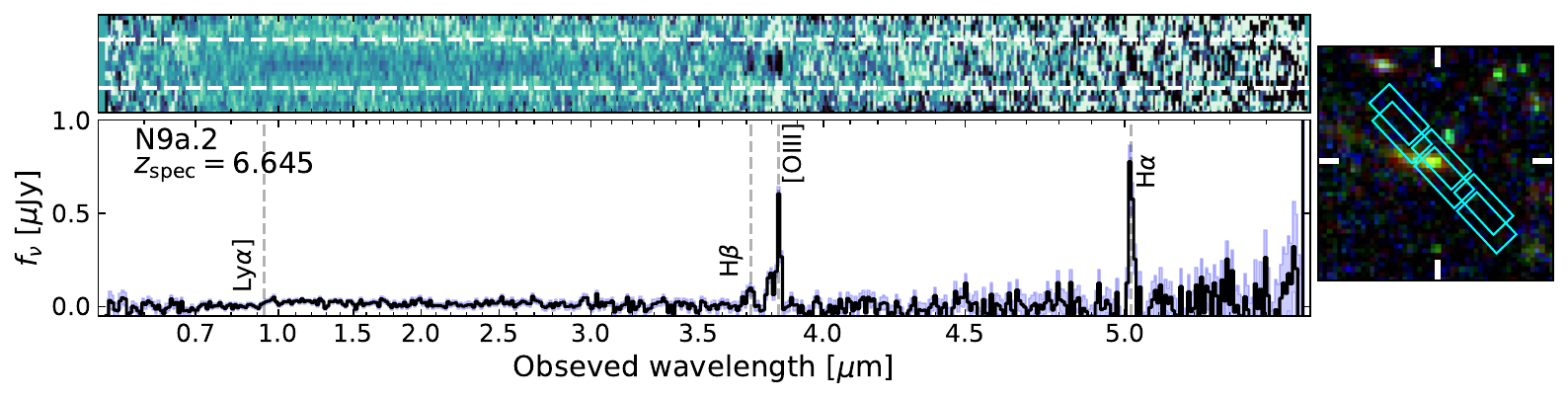}\\
\includegraphics[width=\linewidth]{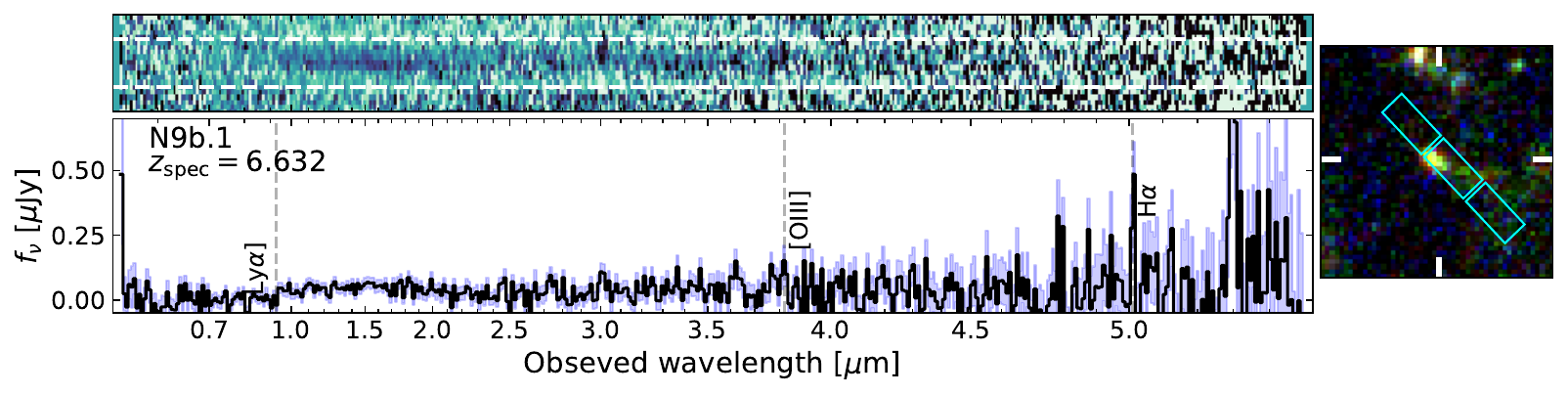}\\
\caption{continued}
\end{figure}

\begin{figure}[h]
\ContinuedFloat
\centering
\includegraphics[width=\linewidth]{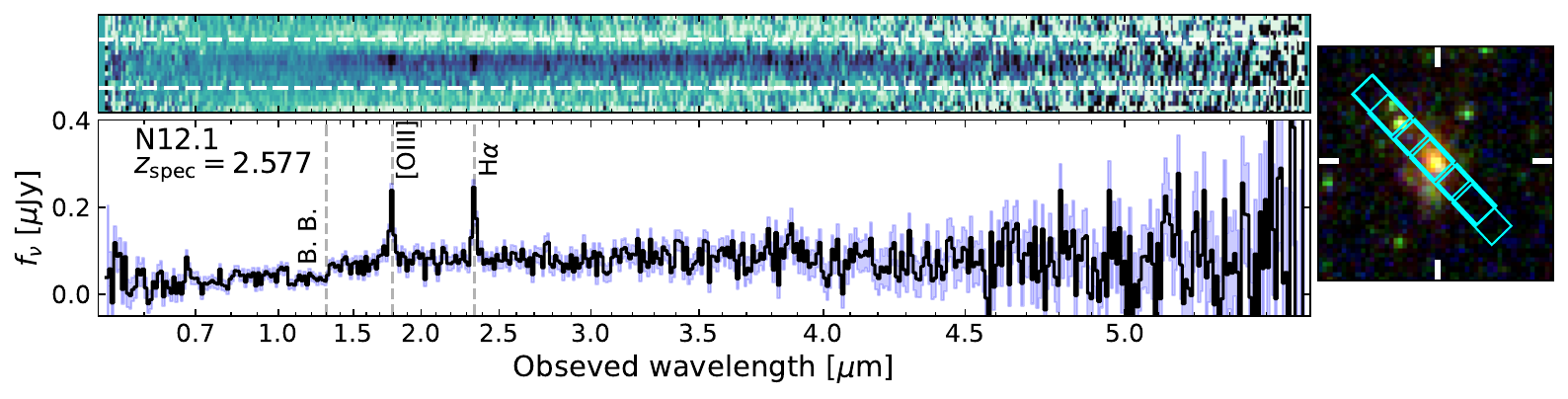}\\ 
\includegraphics[width=\linewidth]{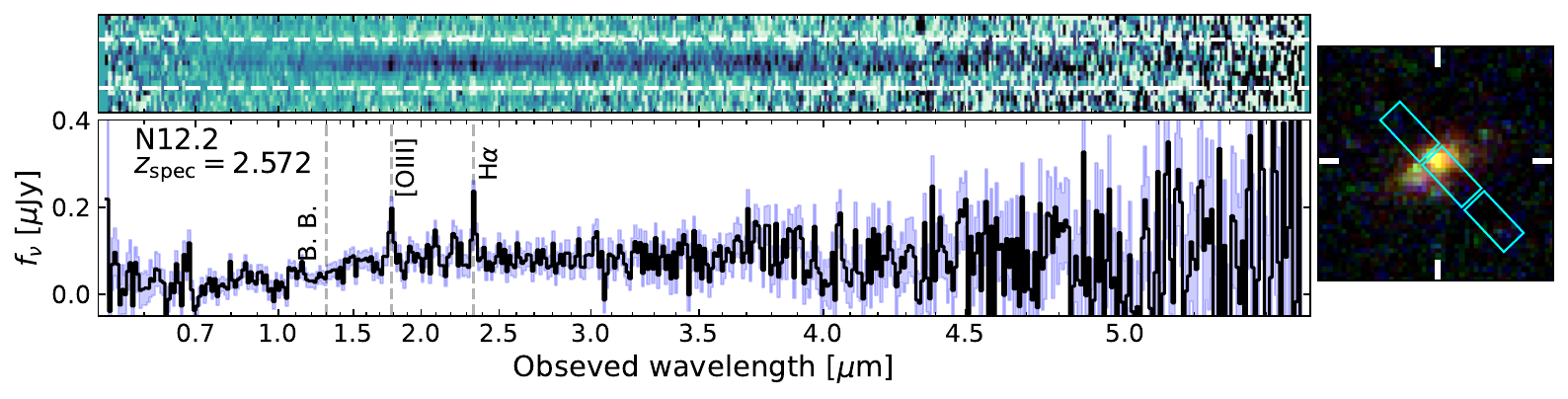}\\ 
\includegraphics[width=\linewidth]{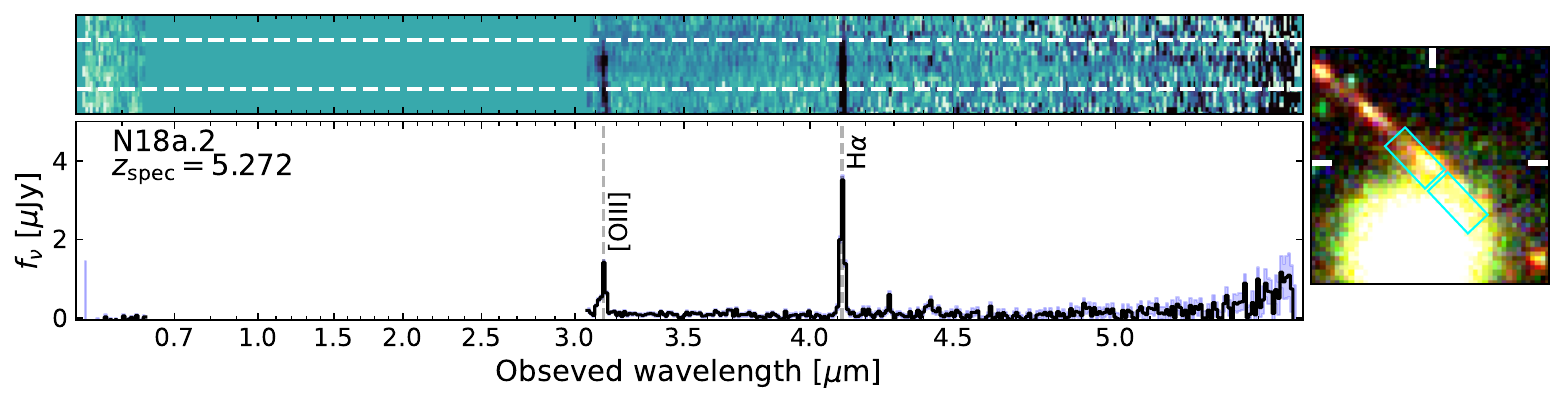}\\ 
\includegraphics[width=\linewidth]{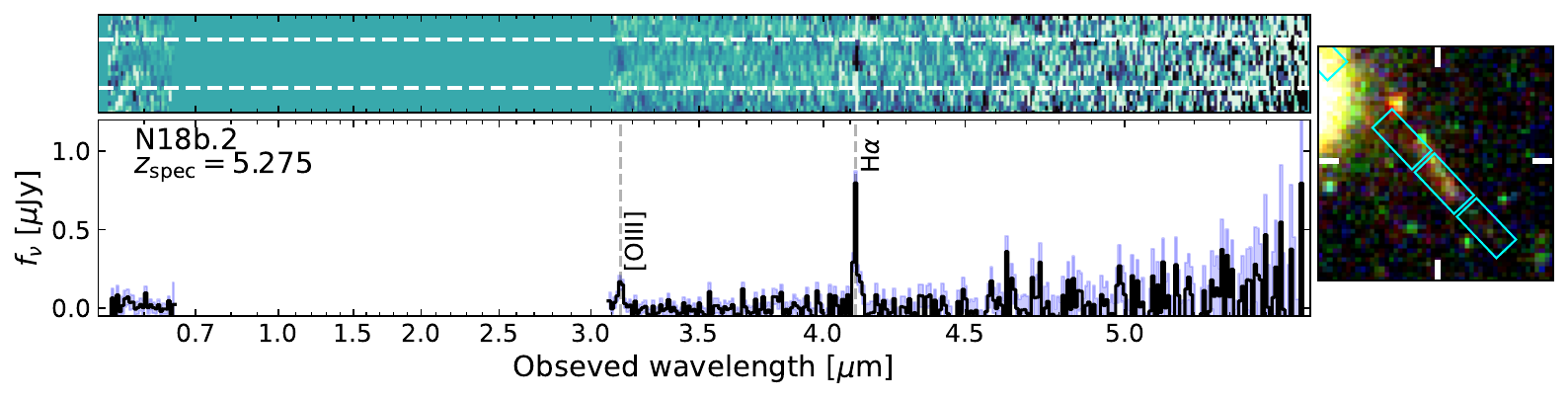}\\ 
\caption{continued}
\end{figure}

\begin{figure}[h]
\centering
\includegraphics[width=\linewidth]{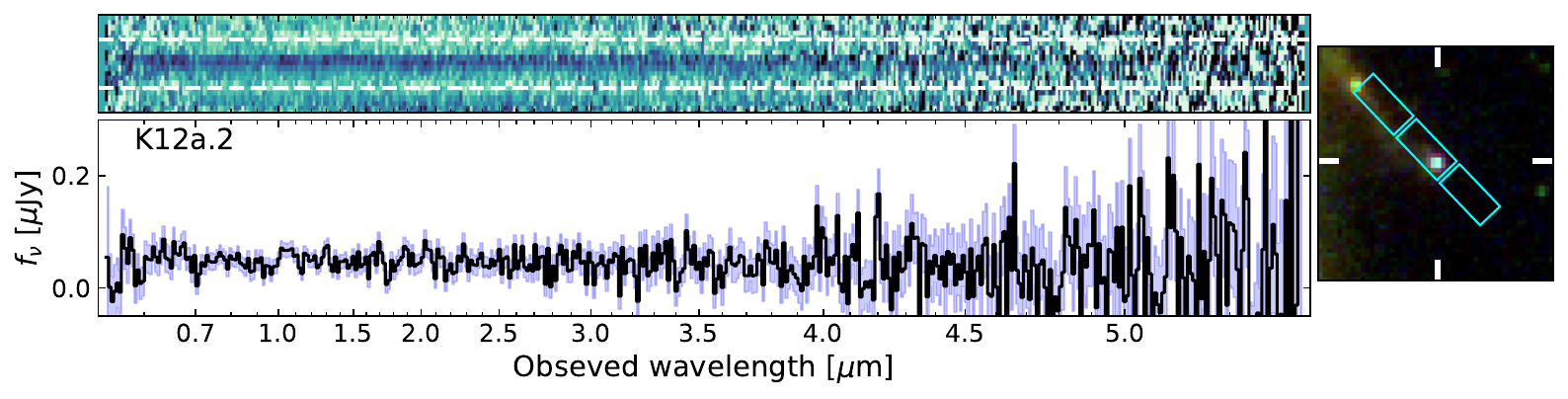}
\includegraphics[width=\linewidth]{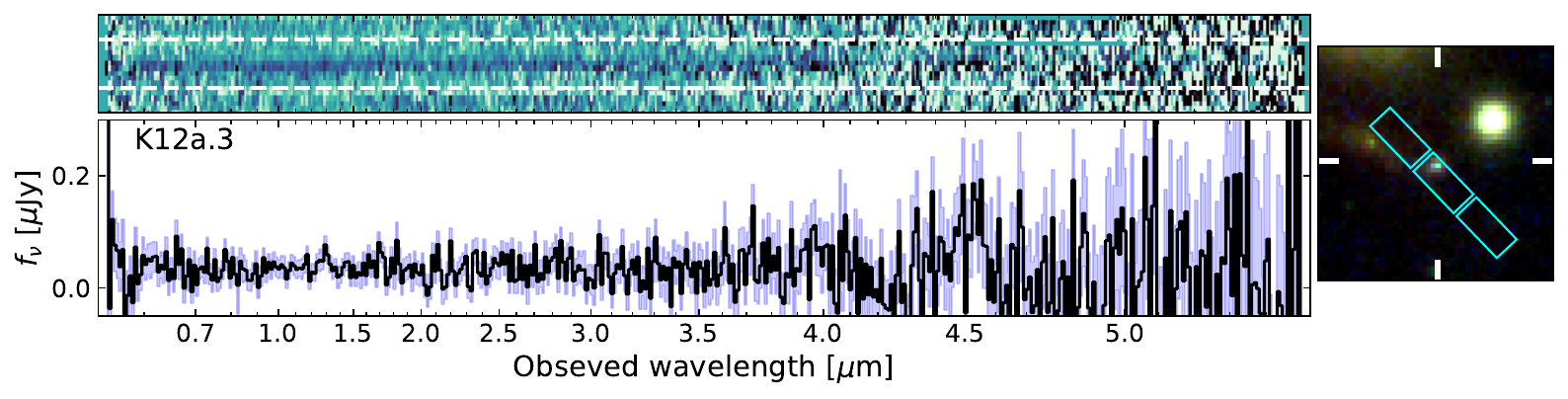}
\includegraphics[width=\linewidth]{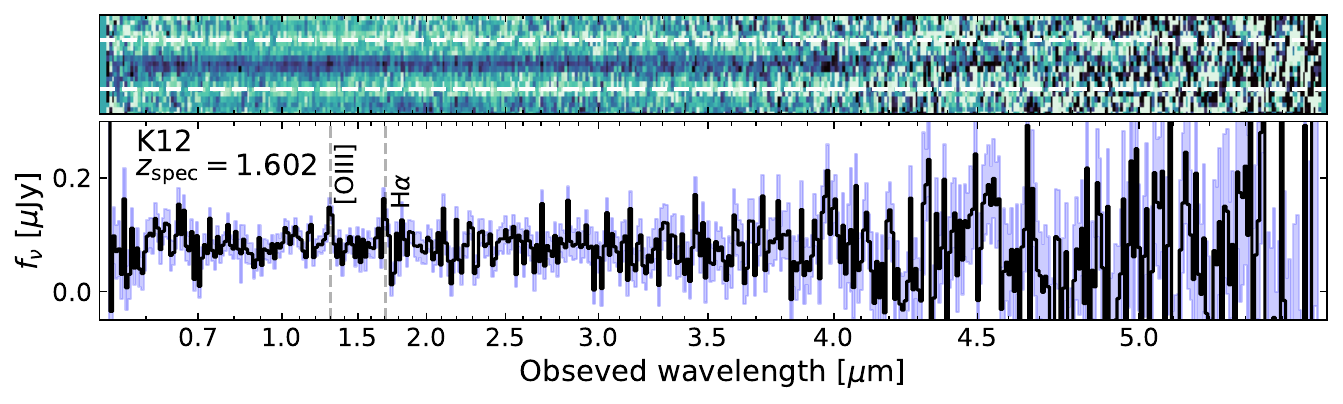}
\includegraphics[width=\linewidth]{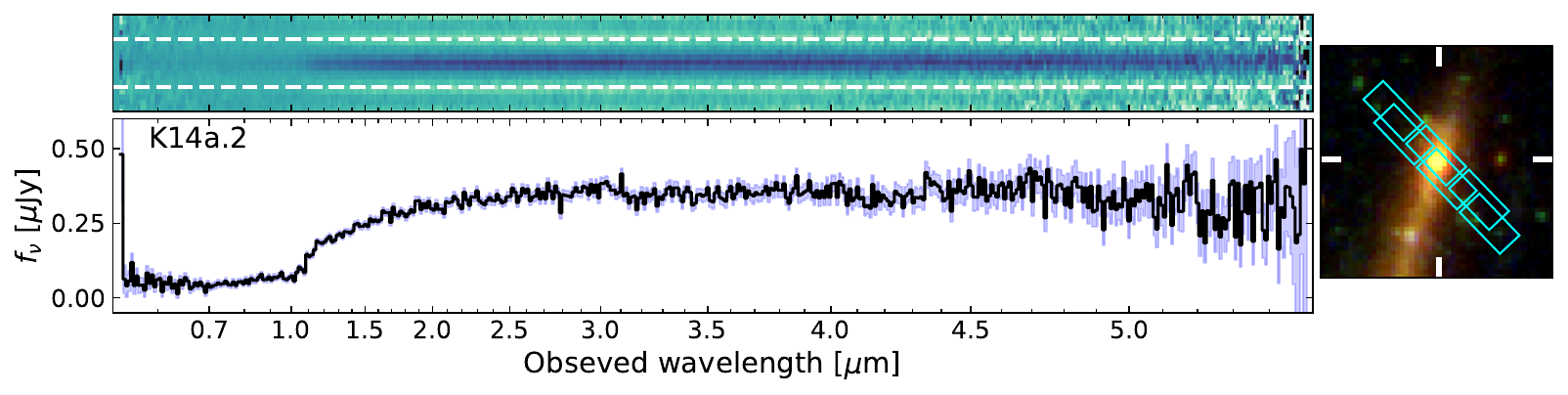}
\includegraphics[width=\linewidth]{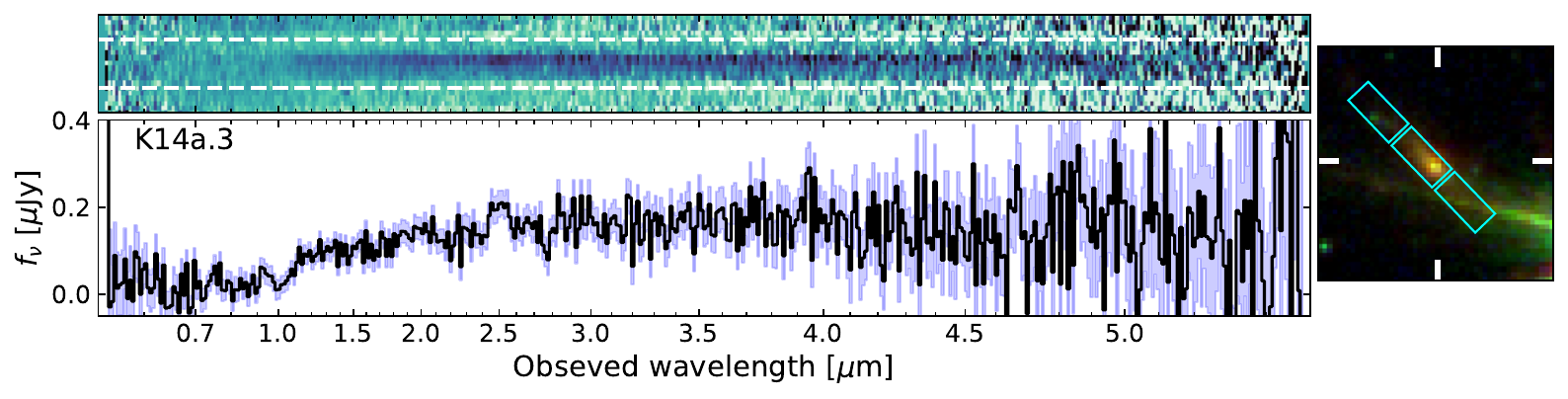}
\includegraphics[width=\linewidth]{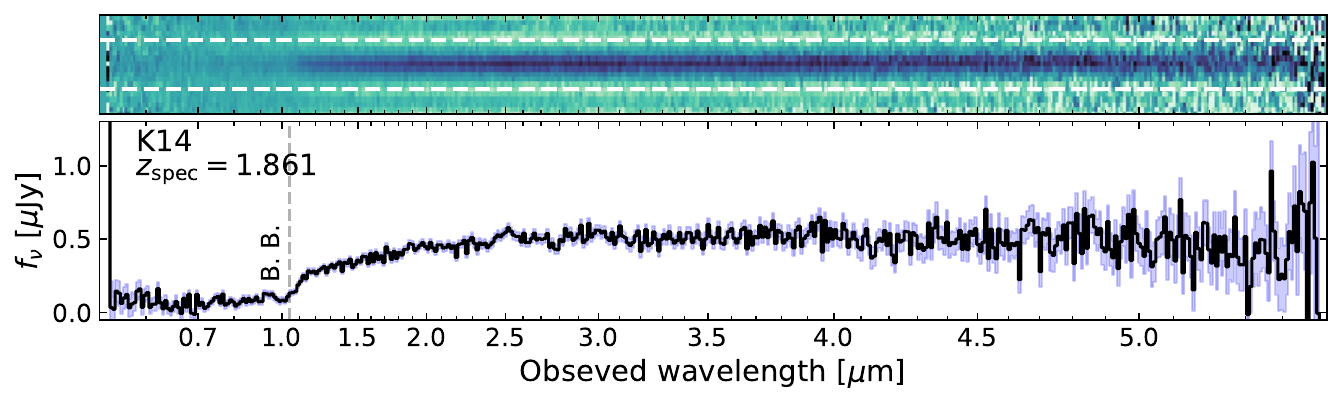}
\includegraphics[width=\linewidth]{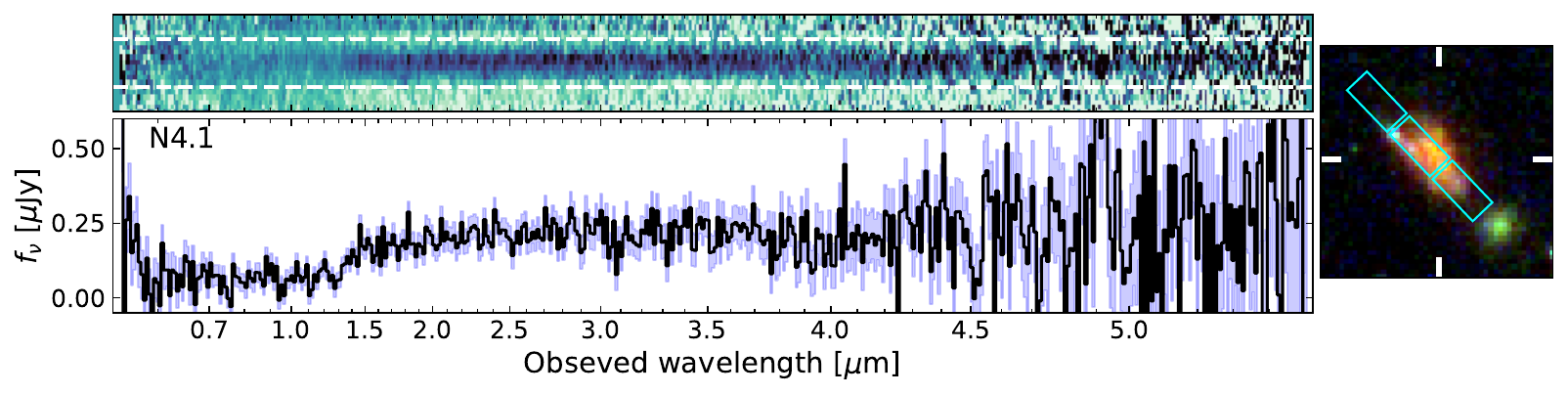}
\includegraphics[width=\linewidth]{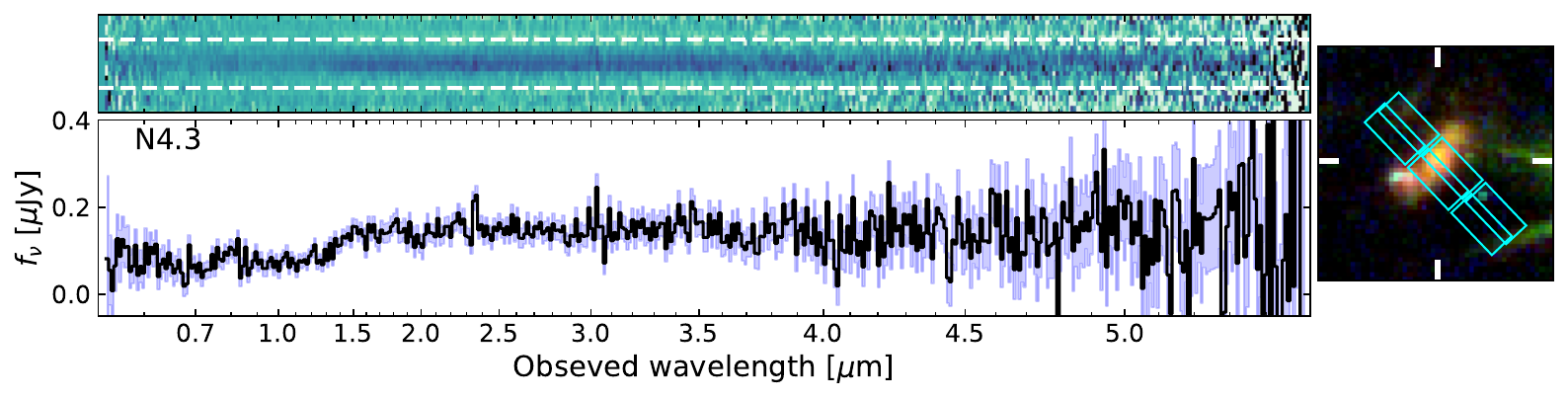}
\includegraphics[width=\linewidth]{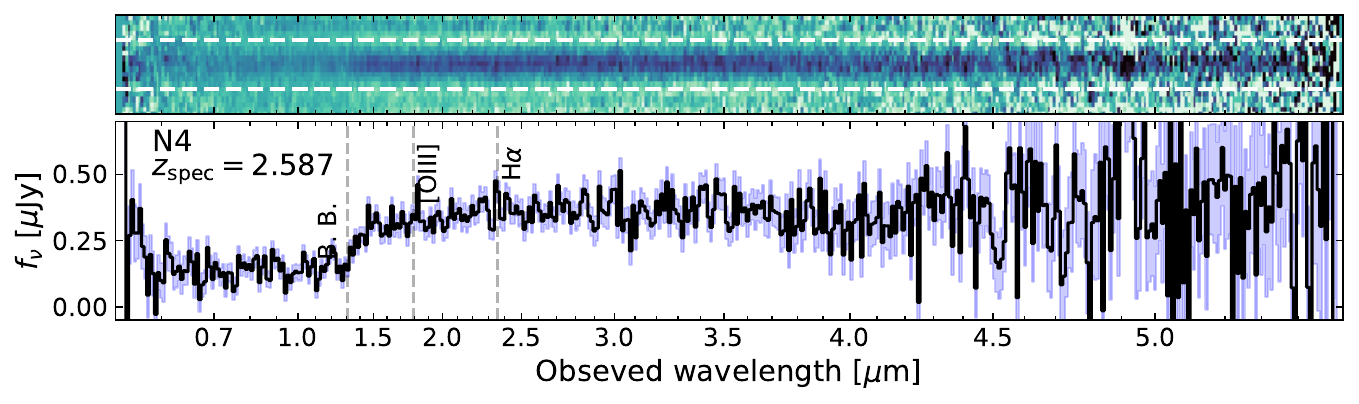}
\caption{NIRSpec prism spectra, where redshift was obtained by stacking the spectra of several multiple images of the system to increase SNR. The combined spectrum of each system is shown without the RGB cutout.}
\label{fig:stackedspectra}
\end{figure}

\begin{figure}[h]
\ContinuedFloat
\centering
\includegraphics[width=\linewidth]{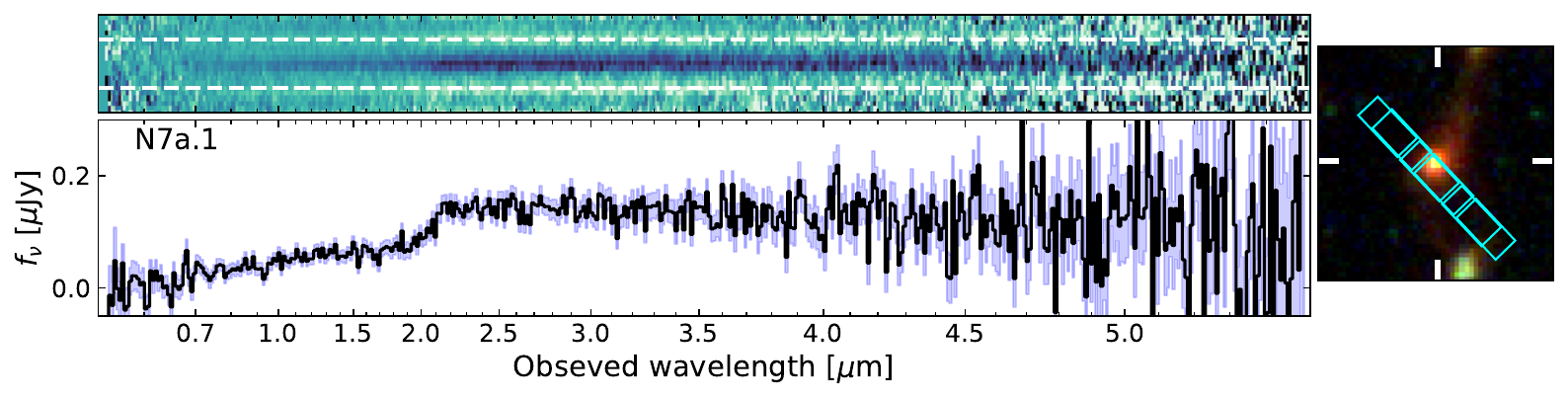}
\includegraphics[width=\linewidth]{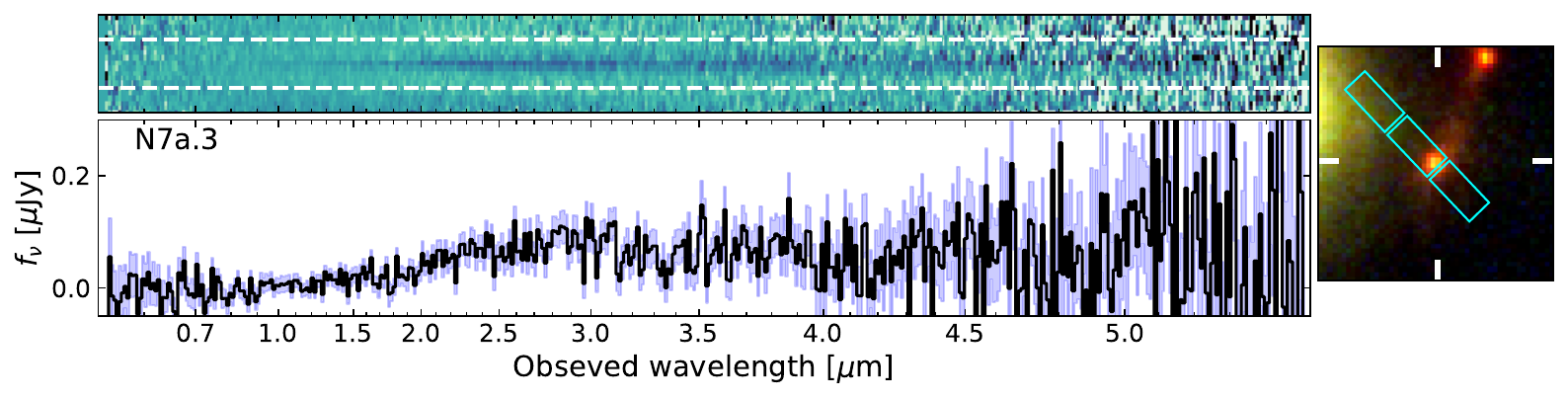}
\includegraphics[width=\linewidth]{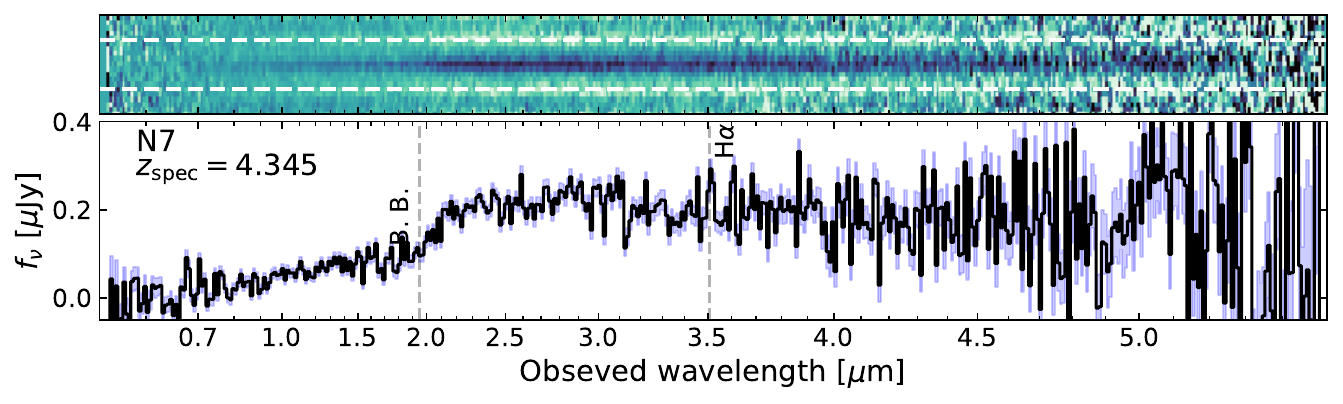}
\includegraphics[width=\linewidth]{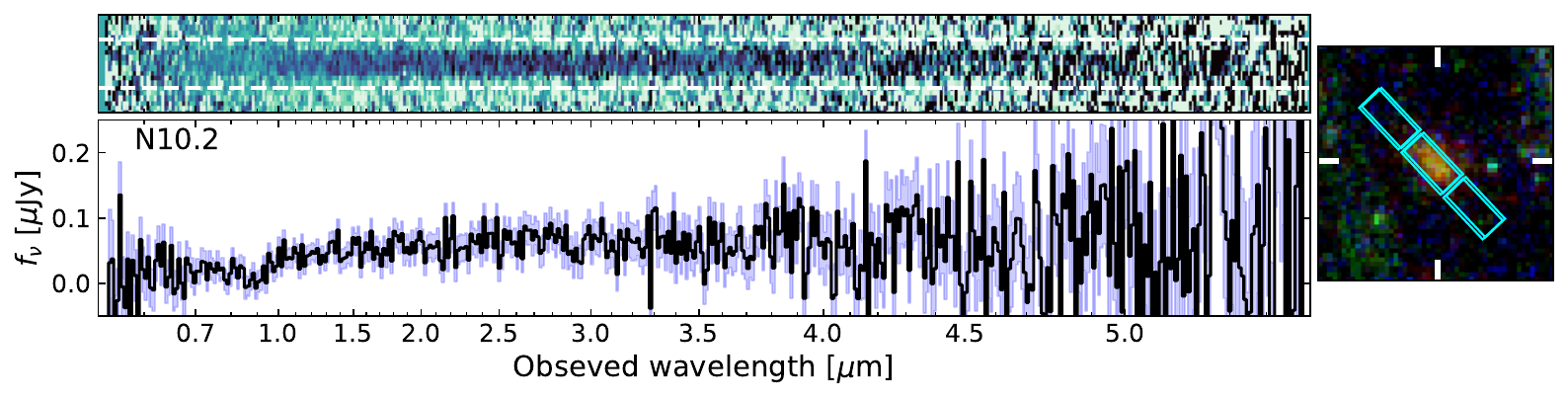}
\includegraphics[width=\linewidth]{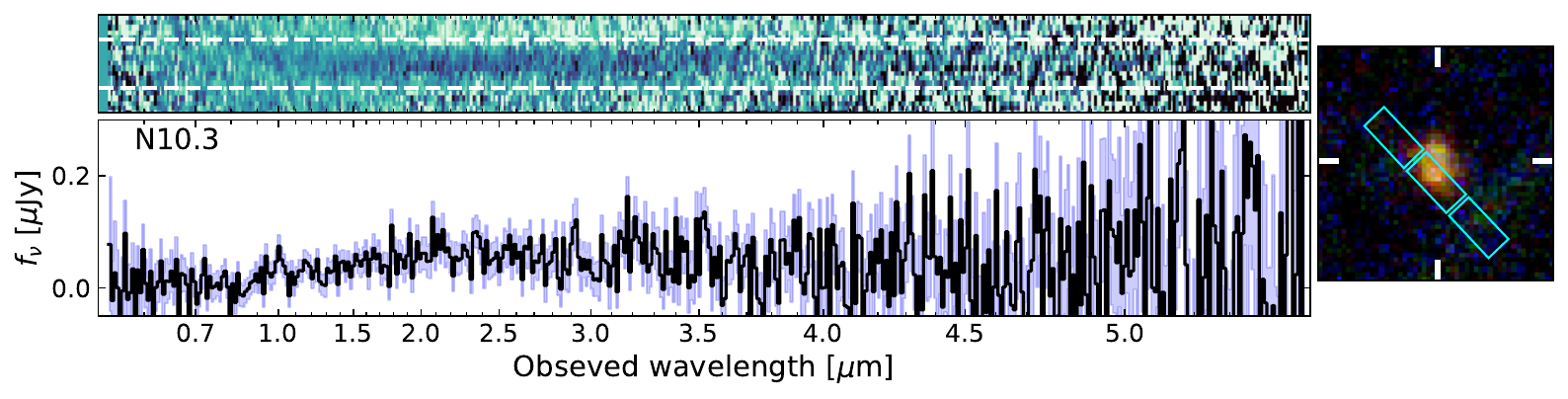}
\includegraphics[width=\linewidth]{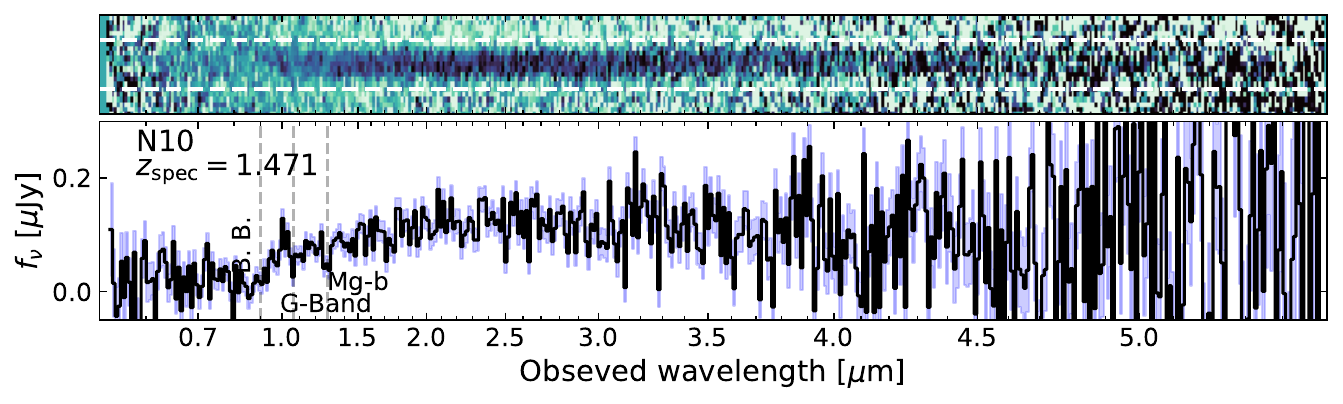}
\caption{continued}
\end{figure}

\FloatBarrier

\begin{table*}[h]
    \centering
    \renewcommand{\arraystretch}{1.05}
    \caption{Cluster members for which we obtained spectroscopic redshifts $z_{\mathrm{NIRSpec}}$ ordered by magnitude $m_{\rm F277W}$. }
    \begin{tabular}{c c c c c }
    
      \hline
      \hline
         ID & RA (ICRS) & DEC (ICRS) &$z_{\mathrm{NIRSpec}}$&mag (F277W) \\
      \hline
      \hline

24 & 104.620054 & $-$55.947492 & 0.293 & 17.92 \\
37 & 104.623280	& $-$55.964413 & 0.295 & 18.26 \\
67 & 104.603309 & $-$55.967364 & 0.292 & 18.94 \\
77 & 104.622469 & $-$55.932222 & 0.286 & 19.22 \\
80 & 104.623707 & $-$55.931473 & 0.293 & 19.26 \\
101& 104.628115 & $-$55.937880 & 0.298 & 19.53 \\
107 & 104.661242 & $-$55.963057 & 0.309 & 19.65 \\
113	& 104.652838 & $-$55.938851 & 0.302 & 19.71	\\
124 & 104.661848 & $-$55.965049 & 0.292 & 19.90 \\
142 & 104.650172 & $-$55.956542 & 0.300 & 20.31 \\
150	& 104.674586 & $-$55.944796 & 0.296 & 20.39 \\
174	& 104.654520 & $-$55.949101	& 0.301 & 20.67 \\
178 & 104.640859 & $-$55.936452 & 0.312 & 20.72 \\
185 & 104.624063 & $-$55.936961 & 0.300 & 20.84 \\
192 & 104.555272 & $-$55.954081 & 0.284 & 20.91 \\
203 & 104.646245 & $-$55.941056 & 0.301 & 21.05 \\

\hline
9197 & 104.585473 & $-$55.957941 & 0.299 & 21.47 \\
8772 & 104.611413 & $-$55.965685 & 0.299 & 21.47 \\
10793 & 104.656191 & $-$55.935241 & 0.309 & 21.53 \\
9887 & 104.643000 & $-$55.947458 & 0.304 & 21.56 \\ 
10440 & 104.581551 & $-$55.931446 & 0.296 & 21.84 \\
10333 & 104.601635 & $-$55.941478 & 0.302 & 21.94 \\
11360 & 104.570285 & $-$55.921061 & 0.294 & 22.16 \\
10113 & 104.572386 & $-$55.943793 & 0.292 & 22.23 \\
1220 & 104.606805 & $-$55.966671 & 0.299 & 22.36 \\
8975 & 104.664755 & $-$55.960275 & 0.300 & 22.46 \\
9498 & 104.575392 & $-$55.953001 & 0.264 & 22.89 \\
9959 & 104.668263 & $-$55.945702 & 0.315 & 23.13 \\
2809 & 104.674357 & $-$55.955599 & 0.337 & 23.51 \\
829 & 104.618893 & $-$55.970494 & 0.304 & 23.59 \\
6012 & 104.568577 & $-$55.937357 & 0.298 & 23.61 \\
4199 & 104.571534 & $-$55.948041 & 0.293 & 23.64 \\
8749 & 104.657913 & $-$55.963921 & 0.300 & 23.68 \\
4816 & 104.537840 & $-$55.944510 & 0.292 & 24.56 \\
    \end{tabular}
\tablefoot{Cluster members above the horizontal line are sufficiently bright to be included in our lens model. The uncertainty of $z_{\rm NIRSpec}$can be estimated as $0.002(1+z_{\rm NIRSpec})\sim0.003$.}
    \label{tab:newnirspecclusterm}
\end{table*}

\clearpage
\onecolumn
\section{Lens model parameters}
\label{app:modelparams}

\begin{table*}[h] 
    \centering
    \renewcommand{\arraystretch}{1.}
    \caption{\label{tab:parameterpriors}Values of fixed parameters and the prior ranges (indicated with $\div$) of free parameters of our fiducial model. }
\begin{tabular}{c|ccccccc}
&$x_0\left[{ }^{\prime \prime}\right]$ & $y_0\left[{ }^{\prime \prime}\right]$ & $e$ & $\theta\left[{ }^{\circ}\right]$ & $\sigma_{\text {lt }}\left[\mathrm{km} \mathrm{s}^{-1}\right]$ & $r_{\text {core }}\left[\mathrm{kpc}\right]$ & $r_{\text {cut }}\left[\mathrm{kpc}\right]$ \\
\hline 
\hline
H1 & -5 $\div$ 14 & -5 $\div$ 19 & 0 $\div$ 0.7 & 0 $\div$ 180 & 0 $\div$ 2000 & 5 $\div$ 200 & 2000\\
H2 & 20 $\div$ 56 & 25 $\div$ 57 & 0 $\div$ 0.7 & 0 $\div$ 180 & 0 $\div$ 2000 & 5 $\div$ 300 & 2000\\
H3 & 175 $\div$ 195 & 40 $\div$ 60 & 0 $\div$ 0.6 & 0 $\div$ 180 & 0 $\div$ 2000 & 5 $\div$ 200 & 2000\\
H4 & 3.70 & -21.33 & 0 $\div$ 0.7 & 0 $\div$ 180 & 0 $\div$ 2000 & 5 $\div$ 200 & 2000\\
\hline
BCG1 & -0.02 & -0.03 & 0.13 & 78 & 50 $\div$ 500 & 0.1 & 144\\
BCG2 & 24.04 & 29.13 & 0.06 & 29 & 50 $\div$ 500 & 0.1 & 126\\
BCG3 & 186.08 & 49.62 & 0.07 & 173 & 50 $\div$ 500 & 0.1 & 113\\
G9 & 51.96 & 48.92 & 0.10 & 10 & 0 $\div$ 400 & 0.1 & 71\\
G10 & 5.21 & 23.02 & 0 & 0 & 0 $\div$ 400 & 0.1 & 70\\
G5 & 42.76 & 50.69 & 0.03 & 149 & 0 $\div$ 400 & 0.1 & 89\\
\hline

S1 & 67.21 & 92.87 & 0 & 0 & 0 $\div$ 1292 & 50 & 1286\\
S2 & -82.28 & -170.95 & 0 & 0 & 378 & 50 & 777\\
S3 & -168.87 & 73.35 & 0 & 0 & 563 & 50 & 1164\\
S5 & 158.47 & -120.29 & 0 & 0 & 435 & 50 & 896\\
S7 & 4.45 & -84.11 & 0 & 0 & 0 $\div$ 1058 & 50 & 1088\\
S9 & 328.97 & 177.08 & 0 & 0 & 508 & 50 & 1048\\
S10 & -23.19 & 66.77 & 0 & 0 & 385 & 50 & 791\\
\hline
Scaling relations & $N_{\mathrm{gal}}=213$ & $m_{\mathrm{F} 277 \mathrm{~W}}^{\text {ref }}=15.77$ & $\alpha=0.25$ & $\beta_{\text {cut }}=0.5$  &$\sigma_{\text {lt }}^{\text {ref }}=0.25\div350$ && $r_{\text {cut }}=1.0 \div 200$
\end{tabular}
\end{table*}

\begin{table*}[h] 
    \centering
    \renewcommand{\arraystretch}{1.4}
    \caption{\label{tab:bestfitparam}Median values of the main halo and scaling relation parameters with uncertainties and best-fit values in parentheses.}
 \begin{tabular}{c|ccccccc}
 &$x_0\left[{ }^{\prime \prime}\right]$ & $y_0\left[{ }^{\prime \prime}\right]$ & $e$ & $\theta\left[{ }^{\circ}\right]$ & $\sigma_{\text {lt }}\left[\mathrm{km} \mathrm{s}^{-1}\right]$ & $r_{\text {core }}\left[''\right]$ & $r_{\text {cut }}\left[''\right]$ \\
\hline 
\hline

H1&
$6.9_{-1.0}^{+1.0}$ &	$6.9_{-1.3}^{+1.5}$  &	$0.69_{-0.01}^{+0.01}$ &	$61_{-3}^{+2}$ &	
$579_{-48}^{+49}$&
$11_{-2}^{+2}$ &	 \\
&[7.6] &
[8.2] &
[0.70] &
[63] &
[589] &
[12] &\\

\hline

H2&
$40.6_{-2.8}^{+4.3}$ &	$41.0_{-4.5}^{+5.7}$ &	$0.53_{-0.12}^{+0.11}$ &	$68_{-6}^{+16}$ &
$567_{-159}^{+97}$ &
$22_{-8}^{+5}$ &	 \\
&[37.0] &
[36.4] &
[0.44] &
[75] &
[423] &
[13] &\\
\hline

H3&
$185.2_{-0.6}^{+0.6}$  &	$49.7_{-0.2}^{+0.1}$ &	$0.45_{-0.06}^{+0.07}$ &	$1_{-1}^{+1}$ & 
$690_{-20}^{+23}$ &
$11_{-2}^{+2}$&	 \\
&[185.2] &
[49.7] &
[0.47] &
[1] &
[673] &
[10] &\\
\hline

H4 & & &
$0.65_{-0.08}^{+0.04}$ &	$30_{-10}^{+10}$ &	
$518_{-95}^{+102}$ &
$34_{-9}^{+7}$ &	 \\
& & &
[0.70] &
[30] &
[619] &
[44] &\\

\hline
\hline

 Scaling relations & & & & &

 $238_{-16}^{+17}$ & & $27_{-7}^{+8}$  \\

 & & & & &
 [252]
& &
 [22]
 \end{tabular}
\end{table*}

\begin{table*}[h] 
    \centering
    \renewcommand{\arraystretch}{1.1}
    \caption{\label{tab:gold_canucs}Median values of galaxy and group scale halo velocity dispersions with uncertainties and best-fit values in parentheses. }
 \begin{tabular}{c|cc}
&$\sigma_{\text {lt }}\left[\mathrm{km} \mathrm{s}^{-1}\right]$&\\
\hline
\hline

BCG1 & 
$215_{-15}^{+13}$
&[ 223] \\

BCG2 & 
$240_{-17}^{+17}$
&[216]  \\

BCG3 & 
$202_{-60}^{+37}$
&[193]  \\

G9 & 
$141_{-11}^{+11}$ 
&[157] \\

G10 & 
$130_{-10}^{+10}$  
& [132] \\

G5 & 
$158_{-13}^{+14}$
& [170] \\

\hline

S1 &
$268_{-191}^{+140}$
& [392]\\

S7 & 
$538_{-55}^{+44}$
& [565] \\
\end{tabular}
\end{table*}

\end{document}